\documentclass{emulateapj}

\newcommand{\ts}{\thinspace}
\newcommand{\etal}{\mbox{et\ts al.\ts}}

\shorttitle{70\ts${\mu}$m Sources: SEDs and $L_{\rm IR}$}
\shortauthors{Kartaltepe \etal}

\begin{document}

\title{A Multiwavelength Study of a Sample of \lowercase{70\ts${\mu}$m} Selected Galaxies in the COSMOS Field I:  Spectral Energy Distributions and Luminosities }

\author{Jeyhan S. Kartaltepe\altaffilmark{1,2}, D. B. Sanders\altaffilmark{1}, E. Le Floc'h\altaffilmark{1}, 
D. T. Frayer\altaffilmark{3},
H. Aussel\altaffilmark{4},
S. Arnouts\altaffilmark{5},
O. Ilbert\altaffilmark{1,6},
M. Salvato\altaffilmark{7},
N. Z. Scoville\altaffilmark{7},
J. Surace\altaffilmark{8},
L. Yan\altaffilmark{8} ,
M. Brusa\altaffilmark{9}, 
P. Capak\altaffilmark{7},
K. Caputi\altaffilmark{10},
C. M. Carollo\altaffilmark{10},
F. Civano\altaffilmark{11},
M. Elvis\altaffilmark{11},
C. Faure\altaffilmark{12} , 
G. Hasinger\altaffilmark{9}, 
A. M. Koekemoer\altaffilmark{13},
N. Lee\altaffilmark{1},
S. Lilly\altaffilmark{10},
C. T. Liu\altaffilmark{14},
H. J. McCracken\altaffilmark{15},
E. Schinnerer\altaffilmark{16},
V. Smol{\v c}i{\' c}\altaffilmark{7}, 
Y. Taniguchi\altaffilmark{17},
D. J. Thompson\altaffilmark{18}, and
J. Trump\altaffilmark{19}}

\altaffiltext{$\star$}{Based on observations with the NASA/ESA {\em Hubble Space Telescope}, obtained at the Space Telescope Science Institute, which is operated by AURA Inc, under NASA contract NAS 5-26555; also based on data collected at: the Subaru Telescope, which is operated by the National Astronomical Observatory of Japan; the XMM-Newton, an ESA science mission with instruments and contributions directly funded by ESA Member States and NASA; the European Southern Observatory under Large Program 175.A-0839, Chile; the National Radio Astronomy Observatory which is a facility of the National Science Foundation operated under cooperative agreement by Associated Universities, Inc ;  and and the Canada-France-Hawaii Telescope with MegaPrime/MegaCam operated as a joint project by the CFHT Corporation, CEA/DAPNIA, the National Research Council of Canada, the Canadian Astronomy Data Centre, the Centre National de la Recherche Scientifique de France, TERAPIX and the University of Hawaii.}  

\altaffiltext{1}{Institute for Astronomy, 2680 Woodlawn Dr., University of Hawaii, Honolulu, HI, 96822, email: jeyhan@ifa.hawaii.edu}

\altaffiltext{2}{Current address: National Optical Astronomy Observatory, 950 N. Cherry Ave., Tucson, AZ, 85719, email: jeyhan@noao.edu}

\altaffiltext{3}{Infrared Processing and Analysis Center, California Institute of Technology 100-22, Pasadena, CA 91125, USA }

\altaffiltext{4}{CNRS, AIM-Unit«e Mixte de Recherche CEA-CNRS-Universit«e Paris VII-UMR 7158, F-91191 Gif-sur-Yvette, France. }

\altaffiltext{5}{Canada France Hawaii telescope corporation, 65-1238 Mamalahoa Hwy, Kamuela, Hawaii 96743, USA }

\altaffiltext{6}{Laboratoire d'Astrophysique de Marseille, BP 8, Traverse du Siphon, 13376 Marseille Cedex 12, France}

\altaffiltext{7}{California Institute of Technology, MC 105-24, 1200 East California Boulevard, Pasadena, CA 91125}

\altaffiltext{8}{Spitzer Science Center, California Institute of Technology, Pasadena, CA 91125}

\altaffiltext{9}{Max Planck Institut f\"ur Extraterrestrische Physik,  D-85478 Garching, Germany}

\altaffiltext{10}{Department of Physics, ETH Zurich, CH-8093 Zurich, Switzerland}

\altaffiltext{11}{Harvard-Smithsonian Center for Astrophysics, 60 Garden Street, Cambridge, MA 02138}

\altaffiltext{12}{Laboratoire d'Astrophysique, Ecole Polytechnique F\'ed\'erale de Lausanne (EPFL), Observatoire de Sauverny, 1290 Versoix, Switzerland}

\altaffiltext{13}{Space Telescope Science Institute, 3700 San Martin Drive, Baltimore, MD 21218}

\altaffiltext{14}{Astrophysical Observatory, City University of New York, College of Staten Island, 2800 Victory Blvd, Staten Island, NY  10314}

\altaffiltext{15}{Institut d'Astrophysique de Paris, UMR7095 CNRS, Universit\`e Pierre et Marie Curie, 98 bis Boulevard Arago, 75014 Paris, France}

\altaffiltext{16}{Max Planck Institut f\"ur Astronomie, K\"onigstuhl 17, Heidelberg, D-69117, Germany}

\altaffiltext{17}{Physics Department, Graduate School of Science, Ehime University, 2-5 Bunkyou, Matsuyama, 790-8577, Japan}

\altaffiltext{18}{Large Binocular Telescope Observatory, University of Arizona, 933 N. Cherry Ave. Tucson, AZ  85721-0065,   USA}

\altaffiltext{19}{Steward Observatory, University of Arizona, 933 North Cherry Avenue, Tucson, AZ 85721}

\begin{abstract}

We present a large robust sample of 1503 reliable and unconfused 70\ts${\mu}$m selected sources from the multiwavelength data set of the Cosmic Evolution Survey (COSMOS). Using the {\it Spitzer} IRAC and MIPS photometry, we estimate the total infrared luminosity, $L_{\rm IR}$ (8--1000\ts${\mu}$m), by finding the best fit template from several different template libraries. The long wavelength 70 and 160\ts${\mu}$m data allow us to obtain a reliable estimate of  $L_{\rm IR}$, accurate to within 0.2 and 0.05\ts dex, respectively. The 70\ts$\mu$m data point enables a significant improvement over the luminosity estimates possible with only a 24\ts$\mu$m detection. The full sample spans a wide range in $L_{\rm IR}$, $L_{\rm IR}\approx 10^{8}-10^{14}\ts L_{\odot}$, with a median luminosity of $10^{11.4}\ts L_{\odot}$. We identify a total of 687 luminous, 303 ultraluminous, and 31 hyperluminous infrared galaxies (LIRGs, ULIRGs, and HyLIRGs) over the redshift range $0.01<z<3.5$ with a median redshift of 0.5. Presented here are the full spectral energy distributions for each of the sources compiled from the extensive multiwavelength data set from the ultraviolet (UV) to the far-infrared (FIR). A catalog of the general properties of the sample (including the photometry, redshifts, and $L_{\rm IR}$) are included with this paper. We find that the overall shape of the spectral energy distribution (SED) and trends with $L_{\rm IR}$ (e.g., IR color temperatures and optical-IR ratios) are similar to what has been seen in studies of local objects, however, our large sample allows us to see the extreme spread in UV to near-infrared (NIR) colors spanning nearly three orders of magnitude. In addition, using SED fits we find possible evidence for a subset of cooler ultraluminous objects than observed locally. However, until direct observations at longer wavelengths are obtained, the peak of emission and the dust temperature cannot be well constrained. We use these SEDs, along with the deep radio and X-ray coverage of the field, to identify a large sample of candidate active galactic nuclei (AGN). We find that the fraction of AGN increases strongly with $L_{\rm IR}$, as it does in the local universe, and that nearly 70\% of ULIRGs and all HyLIRGs likely host a powerful AGN. 

\end{abstract}

\keywords{cosmology: observations --- galaxies: active  --- galaxies: evolution --- galaxies: high-redshift --- infrared: galaxies --- surveys }

\section{Introduction}

Luminous and ultraluminous infrared galaxies (LIRGs, $L_{\rm IR} = 10^{11}-10^{12}\ts L_{\odot}$ and ULIRGs, $L_{\rm IR} = 10^{12}-10^{13}\ts L_{\odot}$) have played a major role in the study of galaxy formation and evolution since their initial discovery. They were first identified in large numbers in the local universe by the {\it Infrared Astronomical Satellite} (IRAS) and follow-up redshift surveys revealed that though they are rare in the local universe, their number density increases strongly with redshift (e.g., the Bright Galaxy Survey (BGS): \citealt{Soifer:1989p2523}; the Revised Bright Galaxy Survey (RBGS): \citealt{Sanders:2003p1575}; 1\ts Jy ULIRG Survey: \citealt{Kim:1998p3280}).

Since the launch of the {\it Spitzer Space Telescope},  many deep infrared surveys (particularly at 24\ts${\mu}$m where the MIPS detector is most sensitive) have been undertaken and have shaped our view of the significance of IR galaxies out to $z\sim1$. We now know that the cosmic star formation rate is dominated by LIRGs beyond $z>0.7$ \citep{LeFloch:2005p2544} and ULIRGs start to dominate by $z\sim2$ \citep{Caputi:2007p2597,Magnelli:2009p2619}. However, one limitation in all 24\ts${\mu}$m selected studies is the difficulty in obtaining an accurate and reliable measurement of the total infrared luminosity of these sources since the 24\ts${\mu}$m selection wavelength represents rest-frame 12\ts${\mu}$m by $z=1$ and 8\ts${\mu}$m by $z=2$. Since the peak of the infrared emission for star-forming galaxies and galaxies with AGN is typically at $\sim50-200$\ts${\mu}$m, at higher redshifts the selection wavelength moves further from the peak and becomes heavily influenced by the presence of polycyclic aromatic hydrocarbon (PAH) features. 

In order to study how various galaxy properties are related to luminosity, an accurate estimate of the total infrared luminosity is needed. Previous studies \citep{Bavouzet:2008p2620, Symeonidis:2008p2625} have shown that rest-frame monochromatic luminosities in the mid-infrared (MIR) correlate very well with the total $L_{\rm IR}$, but they show a considerable amount of scatter. Of course, this problem becomes worse at higher redshifts as observed MIR probes shorter wavelengths. Ideally, one would use observations in the far-infrared (FIR) and submillimeter in order to sample the peak of the infrared emission in these sources, but current IR and submillimeter detectors lack the sensitivity to reach the necessary depths and only the most luminous sources are detected. Studies at long wavelengths are also hampered by confusion due to the large beam size of most detectors.

Previous studies \citep[e.g.,][]{Huynh:2007p62, Symeonidis:2008p2625, Magnelli:2009p2619} based on 70\ts${\mu}$m selected sources have been hampered by this sensitivity problem and have lacked the depth or area necessary to detect a large sample of objects over a wide range of luminosities and redshifts. Deep surveys over small areas allow the detection of fainter, lower-luminosity sources, but they suffer from cosmic variance and poor statistics. Shallow surveys over wide areas only sample the bright end of the luminosity function. In order to overcome these problems, we have obtained deep imaging at 24, 70, and 160\ts${\mu}$m over the entire $\sim 2 \rm\ts deg^2$ of the COSMOS field (Le Floc'h \etal 2009, Frayer \etal 2009).

The Cosmic Evolution Survey (COSMOS: \citealt{Scoville:2007p1776}) was designed to study galaxy evolution over a large range of redshifts and environments, particularly with respect to large scale structure. The initial HST imaging \citep{Koekemoer:2007p2299} has been complemented with wide multiwavelength coverage from space and the ground.  Along with many spectroscopic redshifts, the deep multiwavelength photometry has allowed us to obtain photometric redshifts with unprecedented accuracy, both for normal galaxies \citep{Ilbert:2009p2146} and AGN \citep{Salvato:2009p2142}. This unique data set provides the first opportunity to study the properties of a large sample of 70\ts${\mu}$m selected sources in detail over a wide range in redshift, luminosity, and environment.

This paper is the first of a two part series on the properties of a large sample of $\sim$1500 70\ts${\mu}$m selected galaxies in the COSMOS field. \S2 summarizes the COSMOS data sets used for this study and our sample selection is described in \S3 along with a summary of general properties of the sources. We present the full UV--FIR SEDs and describe the estimate of their total infrared luminosity in \S4. We discuss our results and their implications in \S5 and present our conclusions in \S6. Paper II will discuss the morphological and optical color properties of these sources as a function of redshift and luminosity.  Throughout this paper we assume a  $\Lambda$CDM cosmology with $\rm H_0=70\ts \rm km\ts s^{-1} \ts Mpc^{-1}$, $\Omega_{\Lambda}=0.7$, and $\Omega_{m}=0.3$. All magnitudes are in the AB system unless otherwise stated.

\section{The Data sets}

The imaging data used for this study were obtained as part of the HST-COSMOS project \citep{Scoville:2007p1769}. COSMOS originated as an HST Treasury program imaging an $\sim$ 2 deg$^{2}$ equatorial field with the Advanced Camera for Surveys (ACS), using the F814W filter (I-band). This is the largest contiguous field ever observed by HST. To supplement the ACS coverage, follow-up observations have been obtained across the entire spectrum using both space and ground based facilities. A complete description of the COSMOS data sets can be found in \cite{Scoville:2007p1776}. In this section we provide a brief description of each data set and refer the reader to the relevant papers. These data sets are summarized in Table~\ref{data}.

\subsection{S-COSMOS}
Spitzer-COSMOS (S-COSMOS: \citealt{Sanders:2007p11}) is a Spitzer legacy survey designed to cover the entire 2 deg$^2$ COSMOS field with both the MIPS and IRAC instruments.  The deep IRAC data for the full COSMOS field was taken during cycle 2 (Jan 2006) with a total of 166 hrs to cover the field in all 4 bands (3.6, 4.5, 5.6, and 8\ts${\mu}$m) down to a total depth of 1200\ts s. The $5\ts\sigma$ sensitivity at 3.6\ts${\mu}$m is 0.9\ts${\mu}$Jy. For a detailed description of the data reduction of all S-COSMOS data, see \cite{Sanders:2007p11}.

The MIPS coverage of the COSMOS field was taken during cycle 2 and cycle 3. The aim of the cycle 2 observations was to obtain shallow (80, 40, and 8\ts s per pixel at 24, 70, and 160\ts${\mu}$m, respectively) imaging over a large $\sim4\ts \rm deg^2$ area centered on the COSMOS field. In addition to the shallow imaging, deeper imaging was obtained for a small test region covering $\sim8\%$ of the field. The total depths obtained over this test region were 3200, 1560, and 320\ts s at 24, 70, and 160\ts${\mu}$m, respectively, with a 5\ts$\sigma$ sensitivity of 0.071, 7.5, and 70 mJy. Deep MIPS imaging was obtained for the entire field during cycle 3 (Jan--May 2007 and Jan 2008) with a depth of 3400, 1350, and 273\ts s at 24, 70, and 160\ts${\mu}$m with a 5\ts$\sigma$ sensitivity of 0.08, 8.5, and 65 mJy, respectively.

The cycle 2 and cycle 3 MIPS 24\ts${\mu}$m data were combined and reduced using the MOPEX package \citep{Makovoz:2005p2643} to produce a final mosaic. The source catalog was produced using SExtractor \citep{Bertin:1996p322} for source detection and the DAOPHOT package \citep{Stetson:1987p2648} for measuring flux densities using the Point Spread Function (PSF) fitting technique. This final catalog is $>90\%$ complete above 80\ts${\mu}$Jy and $\sim80\%$ complete above 60\ts${\mu}$Jy. Details on the 24\ts${\mu}$m observations, data reduction, and source properties are presented in Le Floc'h \etal (2009, in press).

The raw 70 and 160\ts${\mu}$m data were reduced using the Germanium Reprocessing Tools (GeRT) and then combined using the MOPEX package. The source detection was done using the Astronomical Point-Source Extraction (APEX) tools within MOPEX. The systematic photometric errors are $5\%$ and $12\%$ at 70 and 160\ts${\mu}$m, respectively. Details on the 70 and 160\ts${\mu}$m observations, data reduction, catalogs, and source counts are presented in \cite{Frayer:2009p2141}.

\subsection{{\it Hubble Space Telescope} images}

The HST/ACS images were taken during HST cycles 12 and 13 as part of the COSMOS treasury survey. A total of 583 orbits were used to obtain a coverage of 1.64\ts deg$^2$ in the F814W filter.  The ACS images were taken between Oct 2003 and June 2005 and reach a depth of 27.2 mag (5\ts$\sigma$). Details  of the ACS images, including their calibration and reduction, are given in \cite{Koekemoer:2007p2299}. In addition to the ACS images, NICMOS parallel images were taken over $\sim 6\%$ of the ACS area in the F160W (H-band) filter down to a depth of 25.9 mag (5\ts$\sigma$). The high resolution of the ACS images ($FWHM = 0.09\arcsec$) is necessary for the detailed morphological analysis presented in Paper II. The NICMOS parallel images ($FWHM = 0.16\arcsec$) are particularly useful for studying the effects of morphological $k$-corrections. Due to the requirement of morphological information for this study, we limited our sample to sources within the ACS area.

\subsection{Ground based UV/Optical/NIR Data}

The groundbased UV, optical, and near-infrared (NIR) coverage consists of data taken with Subaru-SuprimeCam ($B_J,V_J,g^+,r^+,i^+,z^+$), CFHT-Megacam ($u^*,i^*$), UKIRT-WFCAM ($J$), and CFHT-WIRCAM ($K_{\rm S}$). The complete description of the data reduction of the optical and near-infrared (NIR) ground based data can be found in Capak \etal (2009, in preparation). 

The broadband optical data ($B_J,V_J,g^+,r^+,i^+,z^+$) were taken in 2004--2005 using Suprime-Cam on the 8\ts m {\it Subaru Telescope} on Mauna Kea in Hawaii. The seeing ranged from 0.4--2.0\arcsec and the depth at $i^+$ is 26.2 mag. Deep $u^*$ and $i^*$ images were obtained using the Megacam camera at the 3.6\ts m {\it Canada France Hawaii Telescope} (CFHT) with a depth of 26.5 mag at $u^*$.

The J-band images were taken in several observing runs between 2004--2007 with the Wide-Field Camera on the 3.8\ts m {\it United Kingdom Infrared Telescpe} (UKIRT) and the K$_{\rm S}$-band images were taken in queue mode between 2005--2007 with the Wide-field Infrared Camera (WIRCAM) at CFHT. The data were reduced using custom scripts in IRAF\footnote{http://iraf.noao.edu/}. The TERAPIX software Scamp \citep{Bertin:2006p2670} was used to calculate the astrometric solution and the final images were combined using Swarp \citep{Bertin:2002p2651}. The final catalogs are complete to $J=23.7$ and $K_{\rm S}=23.5$ mag. For details on the reduction of the $K_{\rm S}$ images, see McCracken \etal (2009, submitted to ApJ). 

Photometry for all of these data sets was produced using SExtractor in dual mode \citep{Bertin:1996p322} using the I-band image as the detection image and have been combined into a master catalog.

\subsection{GALEX UV Data}
Ultraviolet data were taken using the {\it Galaxy Evolution Explorer} (GALEX) in 2004. Four pointings of 50\ts ks each were used to observe the entire COSMOS field \citep{Zamojski:2007p28} in the near and far-ultraviolet bands (NUV and FUV). The $u^*$ images from CFHT were used as a prior for the flux measurements of the GALEX images. These data reach a limiting magnitude of 25.7 and 26.0 in the FUV and NUV, respectively.

\subsection{X-ray Data}

The COSMOS field was observed with the {\it XMM-Newton} satellite in 2004--2005 for a total of $\sim1.5\ts$ Ms  \citep{Hasinger:2007p2291}. A total of 1887 point sources were detected in at least one of the bands (soft: 0.5--2\ts keV, hard: 2--10\ts keV, and ultra-hard: 5--10\ts keV). The limiting fluxes in each of these bands is $5\times 10^{-16}$, $2\times 10^{-15}$, and $5\times 10^{-15}\ts \rm erg\ts cm^{-2}\ts s^{-1}$ for the soft, hard, and ultra-hard bands, respectively \citep{Cappelluti:2009p2159}. The Chandra-COSMOS survey (C-COSMOS) observed the central 0.5 deg$^2$ of the COSMOS field to a depth of 160\ts ks and an outer 0.4 deg$^2$ to a depth of 80\ts ks with the {\it Chandra X-ray Observatory} \citep{Elvis:2009p2154}. The flux limits obtained are $1.9\times 10^{-16}$ and $7.3\times 10^{-16}\ts \rm erg\ts cm^{-2}\ts s^{-1}$  for the soft (0.5--2\ts keV) and hard (2--10\ts keV) bands, respectively. A total of 1761 point-sources have been detected and are presented in the catalog of \cite{Elvis:2009p2154}. In this paper, we use the optical identification catalogs (Brusa \etal 2009, Civano \etal 2009, in preparation) from both of these data sets to identify AGN among our 70\ts${\mu}$m selected sample of galaxies.

\subsection{Radio Data}

The COSMOS field was observed with the {\it Very Large Array} (VLA) at 1.4\ts GHz as part of a VLA Large Project in 2004-2005 for a total of 280 hours in A and C configuration to cover the entire 2 deg$^2$ \citep{Schinnerer:2007p2300}. The data has  a mean rms noise of 10.5\ts${\mu}$Jy/beam and a $5\ts\sigma$ sensitivity of 55\ts${\mu}$Jy. The catalog used here consists of $\sim 2417$ sources with $S/N>5$ \citep{Bondi:2008p68}. 

\subsection {Spectroscopy}
Spectra of the 70\ts${\mu}$m sources were obtained from several different sources. The first is a follow-up survey of 24\ts${\mu}$m sources in the deep MIPS test region of the field using DEIMOS on {\it Keck II}. The complete survey details will be described in Kartaltepe \etal (2009, in preparation). The observations were taken on 5 nights between 2007--2009 using the 600 l/mm grating with a 1\arcsec \ts wide slit and typical slit lengths of $\sim 6$\arcsec. The typical wavelength coverage with these settings is 4000--10000\ts\AA\ts with a dispersion of {$0.65\ts\rm\AA$}  per pixel and a spectral resolution of 4.6\ts\AA. The total integration times were between 1 and 3 hours, depending on the magnitudes of the sources. The data were reduced using the DEEP2 pipeline\footnote{http$://$astro.berkeley.edu/$\sim$cooper/deep/spec2d/} and redshifts were measured individually for each source and a quality flag 1--4 was assigned, with 4 being a very secure redshift based on multiple lines, 3 being a likely redshift based on multiple lines, 2 being a less reliable redshift often based on only one line and a noisy spectrum, and 1 being a best-guess redshift based on a noisy spectrum. A total of 62 70\ts${\mu}$m detected sources have DEIMOS spectroscopy.

Spectra were also drawn from the zCOSMOS-bright 10k sample \citep{Lilly:2007p2297} observed with the Visible Multi-Object Spectrograph (VIMOS: \citealt{LeFevre:2003p2676}) on the 8\ts m {\it Very Large Telescope} (VLT) in Paranal, Chile.  This sample consists of the first 10,000 galaxies from a large flux limited ($I<22.5$) survey of galaxies over the entire COSMOS field. The zCOSMOS-bright survey used the $R\sim 600$ MR grism with a velocity resolution of $<100\ts\rm km\ts s^{-1}$ and a wavelength range of 5550--9650\ts\AA.

Additional spectra have been obtained from a survey of 677 {\it XMM-Newton} selected AGN candidates with the Inamori Magellan Areal Camera and Spectrograph (IMACS) on the 6.5\ts m {\it Magellan (Baade) Telescope} from 2005--2008 on a total of 18 clear nights (\citealt{Trump:2007p2229}, \citeyear{Trump:2009p5204}). The 200 and 150 l/mm gratings were used. The wavelength range of these spectra is 5600--9200 \AA\ with a resolution of 10\ts\AA\ for the lower resolution 150 l/mm grating. While AGN were the main targets of this survey, additional sources were observed when possible. Similarly, \cite{Prescott:2006p2333} conducted a spectrocopic survey of quasar candidates using the Hectospec multiobject spectrograph on the 6.5\ts m {\it Multi Mirror Telescope} (MMT) on Mt.\ts Hopkins using the 270 l/mm grating and covering a wavelength range of 3100--9000\ts\AA. Four sources were also observed using the Focal Reducer and low dispersion Spectrograph (FORS1) on the VLT (PI: C. Faure) covering a wavelength range of 3300--11000\ts\AA\ (Anguita \etal 2009, submitted).

In addition to the surveys described above, spectra of the 70\ts${\mu}$m sources were found on the {\it NASA Extragalactic Database} (NED) public archive from the Sloan Digital Sky Survey (SDSS: \citealt{Abazajian:2009p5536}) and the 2dF Galaxy Redshift Survey (2dFGRS: \citealt{Colless:2001p5237}). See Table~\ref{spectra} for a summary of the spectroscopic surveys and a breakdown of the number of sources that were observed by each.

\section{The 70\ts${\mu}$m Sample}

\begin{figure}[t]
\plotone{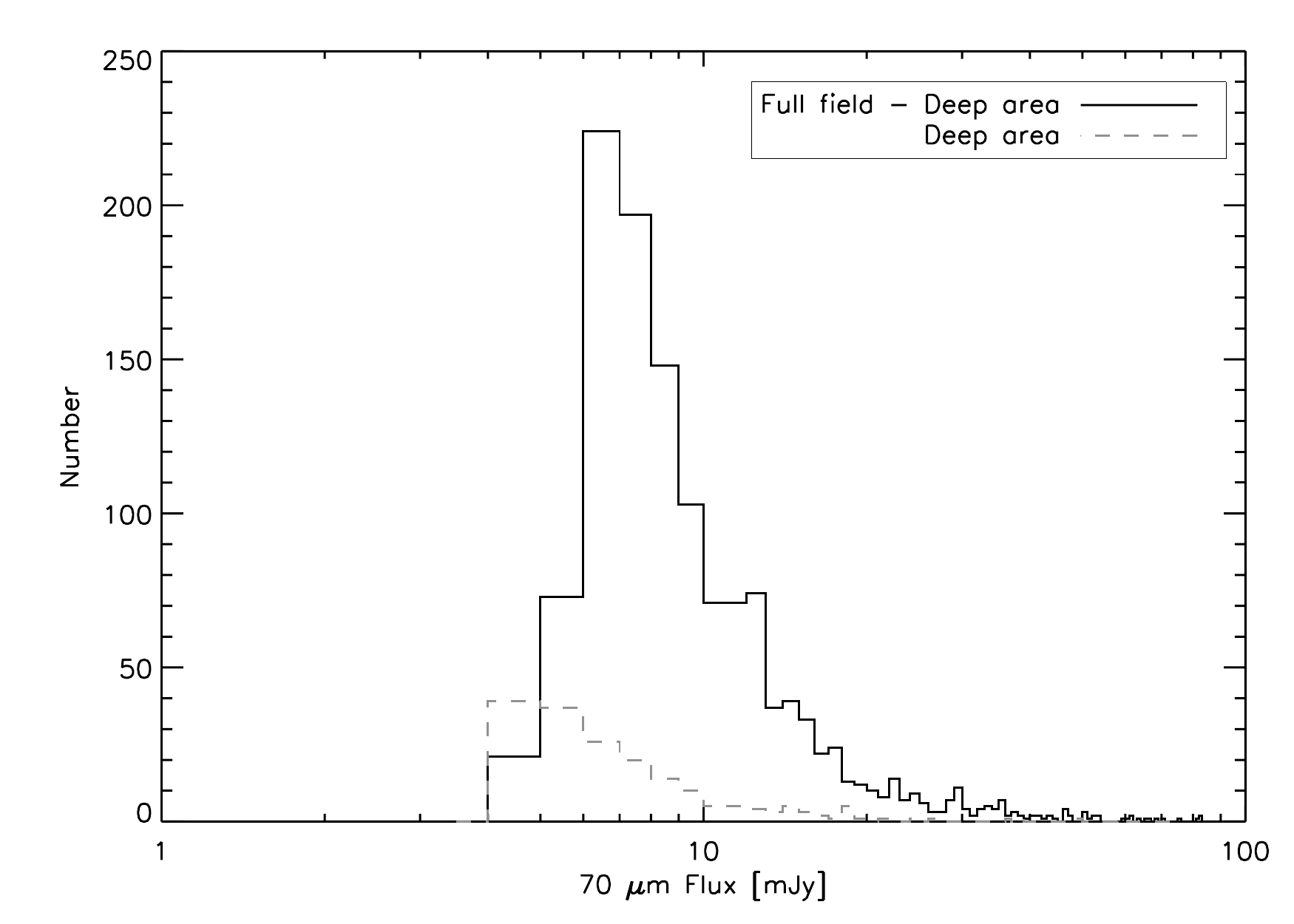}
\caption{Histogram of 70\ts$\mu$m source fluxes in the deep MIPS test area (covering $\sim 0.16$ deg$^{2}$: gray dashed) and the full COSMOS field (covering 1.64 deg$^{2}$) excluding the deep area (black). The test area represents $\sim 9$\% of the COSMOS field and contains $\sim12\%$ of the 70\ts$\mu$m sources (includes more sources since this region is deeper). The flux limit for the full field is $\sim$ 6.5\ts mJy and the flux limit for the deep area is $\sim$ 5\ts mJy for sources with $S/N>3$.}
\label{flux}
\end{figure}

We restricted our study to the area of the field with HST/ACS coverage and selected sources at 70\ts${\mu}$m with $S/N >3$. After removing spurious sources detected around bright objects and removing sources in areas around bright stars masked out in the optical or IRAC photometry, the number of sources remaining is 1743. These 1743 sources represent the initial catalog of 70\ts${\mu}$m selected sources discussed below. This catalog includes sources within the deeper cycle 2 MIPS coverage area as well as the full field cycle 3 coverage, therefore, the deeper area has a slightly lower flux limit ($\sim$ 5\ts mJy). The distribution of 70\ts${\mu}$m fluxes for the deep area and the rest of the field is shown in Figure~\ref{flux}.

\subsection{Identification of Counterparts}

\begin{figure}[t]
\plotone{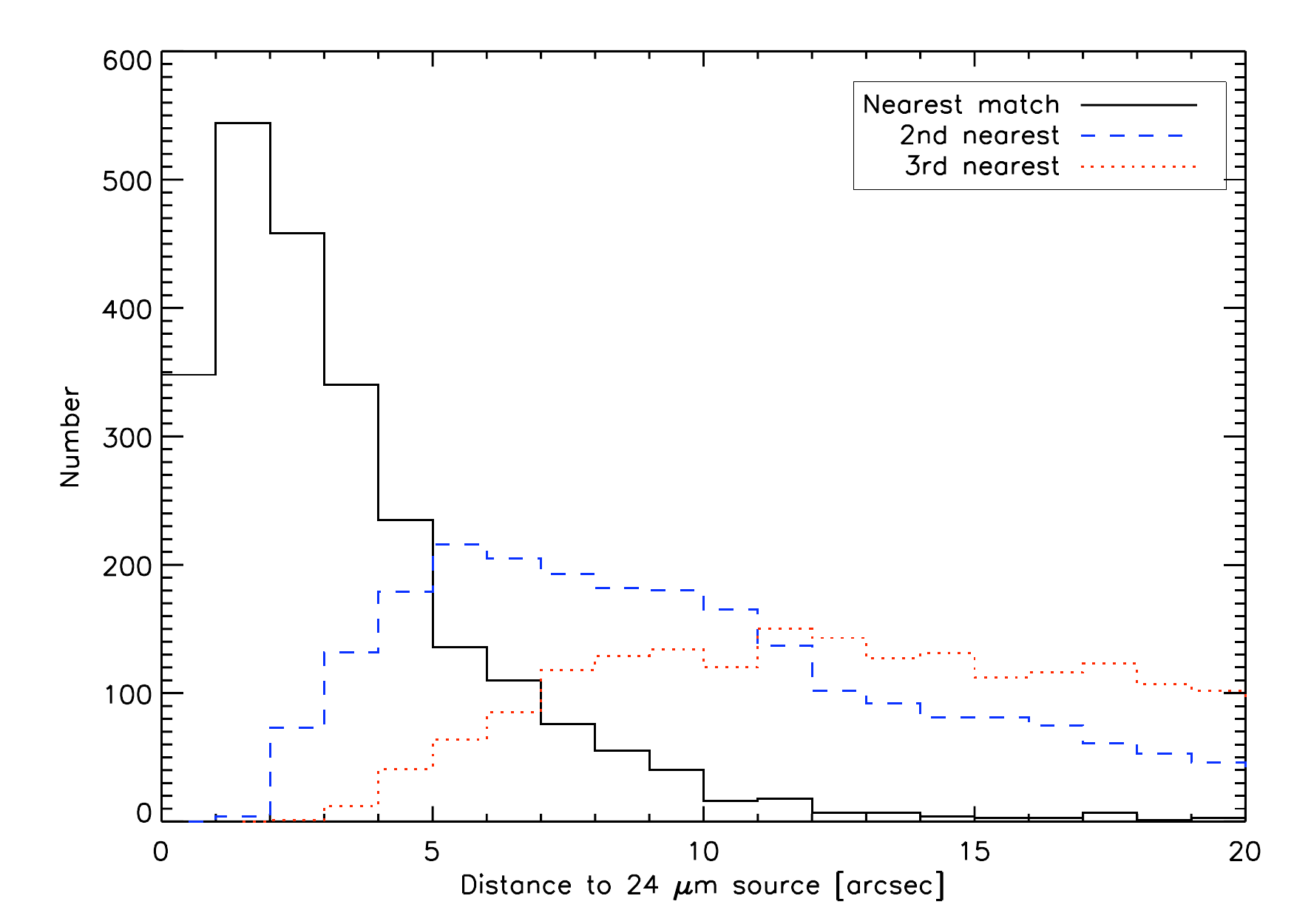}
\plotone{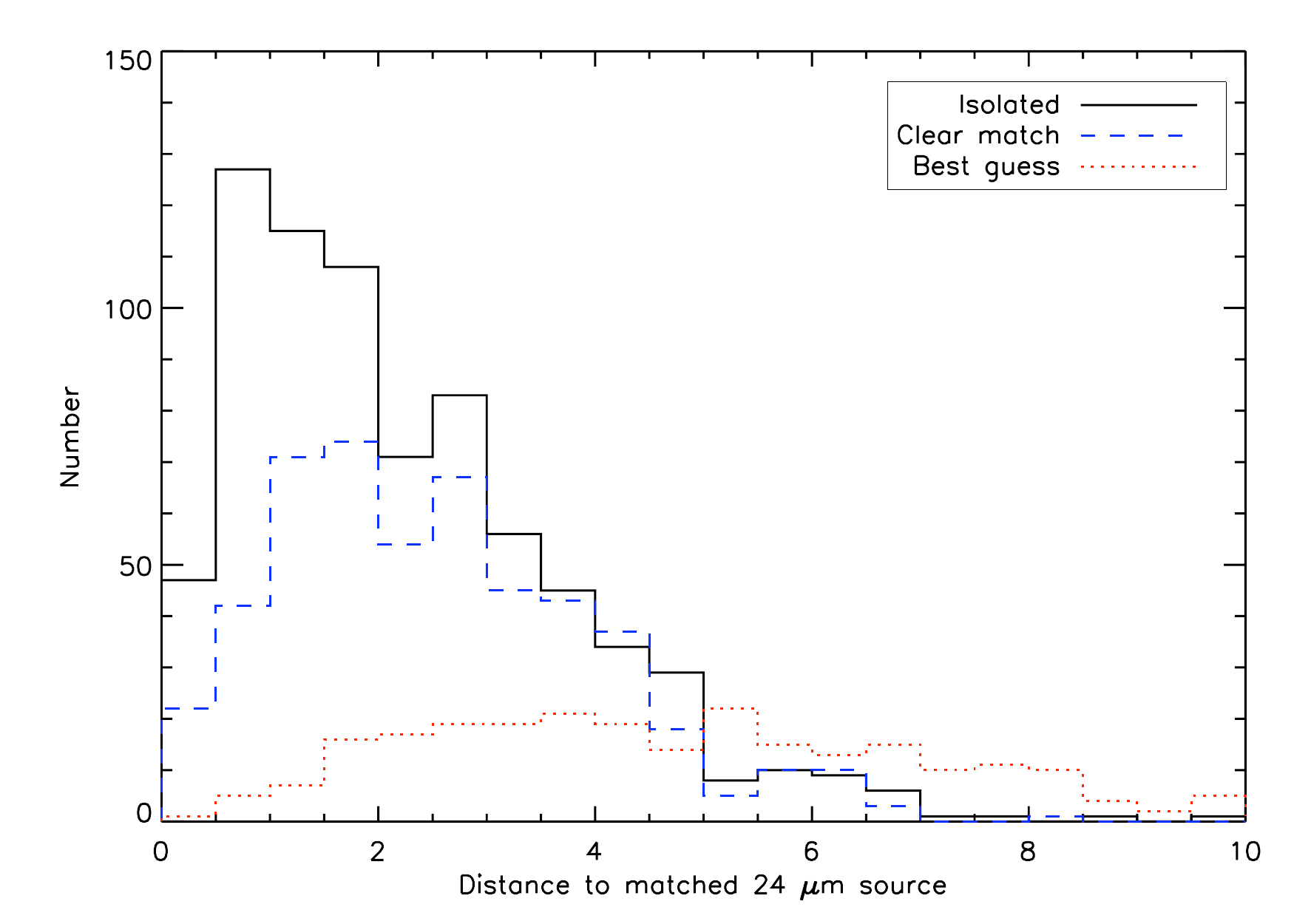}
\caption{Top: Angular separation between each 70\ts$\mu$m source position and the nearest (as well as the second and third nearest) 24\ts$\mu$m source. Bottom: Separation between matched source 70 and 24\ts$\mu$m positions for the three different cases - those that are isolated (black), those that have a clear match (blue dashed), and those with a best-guess match (red dotted).}
\label{sep}
\end{figure}

In order to ensure the most reliable identification of the optical counterpart for each source in our sample, we first matched the cleaned 70\ts${\mu}$m catalog to the 24\ts${\mu}$m catalog. Figure~\ref{sep} shows the separation between the 70\ts${\mu}$m source positions and the nearest (as well as the second and third nearest) 24\ts${\mu}$m source. 79 of the 70\ts${\mu}$m  sources (4.5\%) do not have a 24\ts ${\mu}$m counterpart within the 18$\arcsec$  70\ts${\mu}$m beam or there is one on the very edge that does not appear to be a likely counterpart. The vast majority of these sources (92\%) are at the low $S/N$ end ($3<S/N<4$).  In order to not be detected at 24\ts${\mu}$m, these sources would have to have a 70 to 24\ts${\mu}$m flux ratio of $> 100$. Sources with this extreme flux ratio are very rare in the rest of the sample with 24\ts${\mu}$m counterparts (at the 1\% level). Physically, such a large flux ratio is possible at $z>2$ where 24\ts${\mu}$m could potentially pass through a silicate absorption feature at rest frame 8\ts${\mu}$m, however, given the large number of sources not detected it is unlikely that all of them are such extreme high redshift sources. Furthermore, there are no sources in the local universe with such extreme ratios between rest-frame 8 and 24\ts${\mu}$m. This, combined with their low $S/N$, indicates that they are most likely spurious detections and we therefore remove them from further analysis.

Of the remaining 1664, 1528 sources were successfully matched to a 24\ts${\mu}$m counterpart. Of the 1528 successfully matched sources, 759 ($\sim$ 50\%) have a single isolated 24\ts${\mu}$m source within the 70\ts${\mu}$m  beam making the identification of the correct counterpart straightforward. For 513 sources ($\sim$ 34\%) there is more than one 24\ts ${\mu}$m source in the beam but the most likely counterpart is clear (i.e., the source closest to the center of the beam is also the brightest at 24\ts${\mu}$m). 256 sources ($\sim$ 17\%) have more than one 24\ts ${\mu}$m source in the beam but the likely counterpart is not as clear as in the previous case. For these sources we chose the source with the highest 24\ts${\mu}$m flux as the best-guess match to the counterpart. Although it is possible that some fraction of these ``best-guess" matches are the wrong counterpart, we find that the results of our analysis do not change if they are left out. We estimate that our false match rate is on the order of 2.5\%. For the 136 sources that cannot be matched there either appears to be more than one equally likely counterpart or the 70\ts${\mu}$m source appears to be a blend of multiple 24\ts${\mu}$m sources.  It is highly probable that the multiple 24\ts${\mu}$m sources within the beam each contribute to the 70\ts${\mu}$m flux and therefore it is not possible to determine an accurate 70\ts${\mu}$m flux for each source. Leaving these sources in our sample will only add an additional uncertainty, therefore we omit these ambiguous sources from further discussion.

We then matched the remaining 1528 sources with believable 24\ts${\mu}$m counterparts to the IRAC source positions. All but 9 of them have a clear match within 2$\arcsec$ of the 24\ts${\mu}$m source position. For these 9, the 24\ts${\mu}$m source appears to be a blend of two IRAC sources so we remove them from the sample for the same reasons described above. For the remaining 1520 sources we then matched the IRAC counterparts to the optical I-band positions and found clear counterparts for all but 11. Five of these appear to be two optical sources with blended IRAC photometry so we also exclude these 5 sources from our sample.  The remaining six sources (0.3\%) do not have an optical counterpart above the flux limit of the catalog. Without an optical counterpart we have no redshift measurement for these six sources and thus cannot obtain an estimate of the total infrared luminosity so we exclude them from our final sample but note that their IRAC/MIPS properties do not appear to differ from the rest of the sample. Of the remaining 1508 sources, 5 are stars and so we remove them from the final sample as well. This results in a final sample size of 1503 70\ts${\mu}$m selected sources with reliable counterparts at 24\ts${\mu}$m, the IRAC bands, and the optical.  We believe that the 9\% of sources removed, mostly due to blending, will not affect our final result and removing them leaves us with a robust sample.

\begin{figure}[t]
\plotone{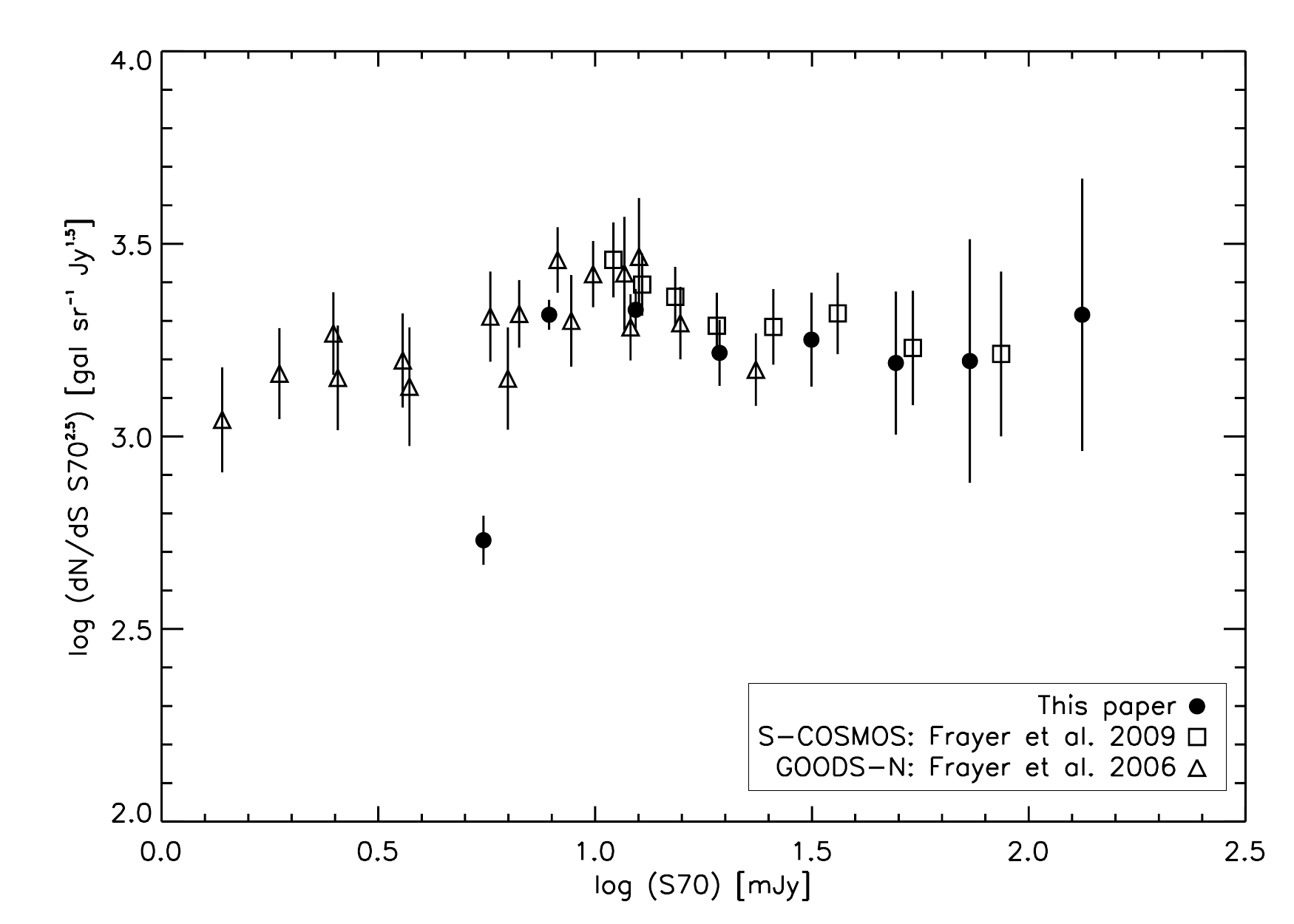}
\caption{Differential source counts ($dN/dS \times S^{2.5}$) for the final sample of 1503 sources compared to the published S-COSMOS (Frayer et al. 2009) and GOODS-N (Frayer et al. 2006) source counts. Note that the counts agree except at the lowest flux bin ($S70 <$ 6.5 $\mu$Jy).}
\label{counts}
\end{figure}

We plot the differential source counts ($dN/dS \times S^{2.5}$) for the final sample of 1503 sources in Figure~\ref{counts} along with the published source counts from S-COSMOS (Frayer et al. 2009) and GOODS-N (Frayer et al. 2006). The source counts for our sample match the published source counts very well except at the lowest flux bin ($S70 <$ 6.5 $\mu$Jy). This indicates that spurious sources at low $S/N$ do not have a significant contribution and that the sample is therefore reliable though incomplete at these low flux levels.

The cross-matching at 160\ts$\mu$m was done by looking for 160\ts$\mu$m sources near each of the 70\ts$\mu$m sources. For the vast majority of the cases (89\%), only one 70\ts$\mu$m source was identified as a possible match to a 160\ts$\mu$m source. For most of the cases where more than one 70\ts$\mu$m source could have been identified as a possible match, one was much closer to the center of the beam and the other was far enough away that the choice was unambiguous (9\%).  For the handful of ambiguous cases (remaining 2\%), we chose the brighter 70 micron source as the match. In total, we find that 463 of the 1503 70\ts${\mu}$m selected sources ($\sim31\%$) are detected at 160\ts${\mu}$m.

\subsection{Photometric Redshifts}
All of the UV--NIR (including GALEX, IRAC, and all of the ground based optical and NIR) multiwavelength photometry was used to produce a photometric redshift catalog for all galaxies \citep{Ilbert:2009p2146} and X-ray detected AGN \citep{Salvato:2009p2142} in the COSMOS field. Thanks to the depth and broad range of this multiwavelength coverage, the  COSMOS photometric redshifts have an unprecedented accuracy of $dz/(1+z) = 0.007$ at $i^+< 22.5$ and 0.012 at $i^+< 24$ out to $z<1.25$. The AGN photometric redshifts have an accuracy of $dz/(1+z) = 0.02$ and 0.03 for $i^+ < 22.5$ and $22.5 < i^+ < 25$, respectively. The number of outliers for both galaxies and AGN are negligible. The accuracy and high precision of these photometric redshifts allows us to use our entire sample for the analysis of the properties of 70\ts$\mu$m sources instead of limiting our sample to those with spectroscopy as has generally been done in the past.

\begin{figure}[t]
\plotone{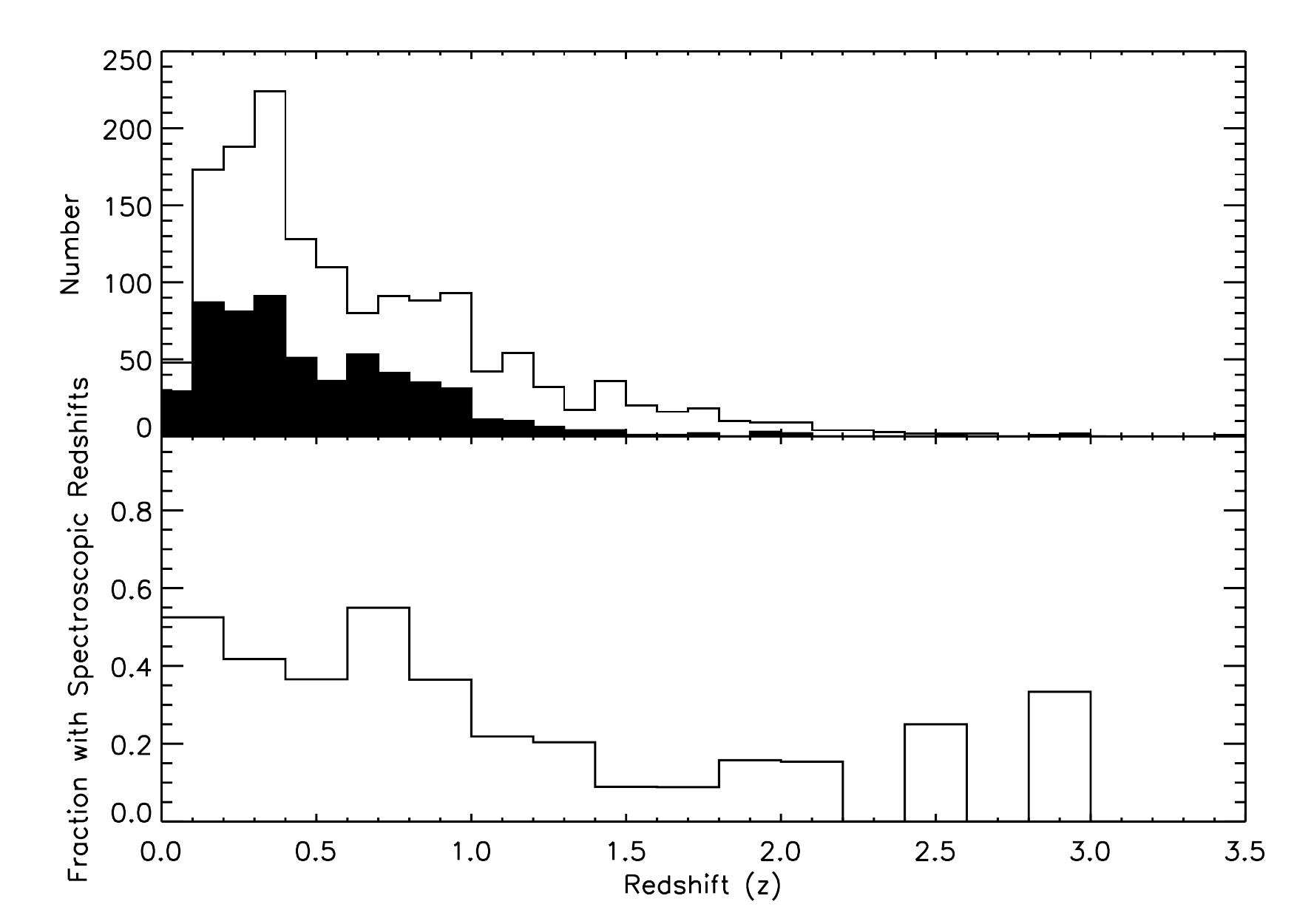}
\caption{Top: Redshift distribution of the entire 70\ts$\mu$m selected sample (white: photometic redshifts) and those with spectroscopy (black). Bottom: Fraction of sample with reliable spectroscopic redshifts as a function of redshift. Note that spectroscopic redshifts are available for $\sim 40\%$ of the sample and spread across the entire redshift range.}
\label{redshift}
\end{figure}

The redshift distribution of the entire 70\ts$\mu$m sample is shown in Figure~\ref{redshift}. Photometric redshifts are available for all of the 1503 galaxies in the sample.  154 of these sources are detected in the X-ray by {\it XMM-Newton} and {\it Chandra} and so we use the photometric redshifts determined by \cite{Salvato:2009p2142}. Our 70\ts$\mu$m selected sample peaks at $z\sim 0.35$ and the number of sources drops beyond ($z\sim 1$). However, given our large sample size, a significant number (273: 18\%) are detected at $1< z < 3$

\subsection{Spectroscopic Redshifts}

\begin{figure}[t]
\plotone{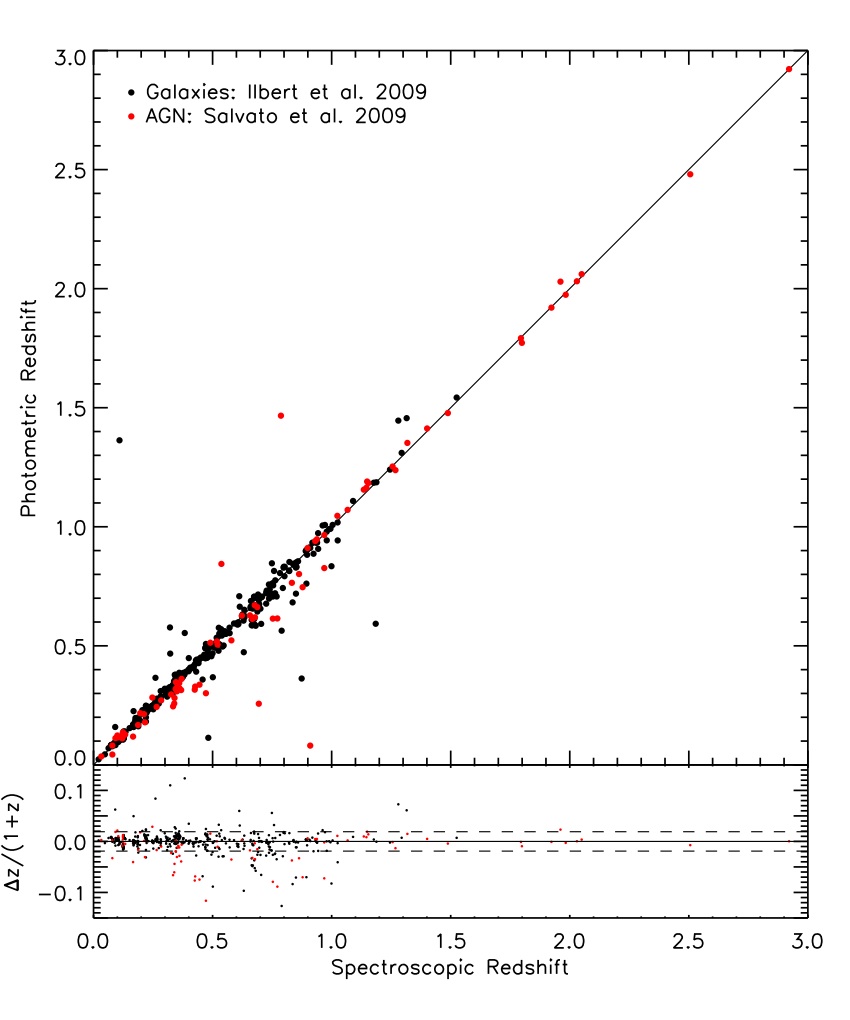}
\caption{Comparison of photometric and spectroscopic redshifts. Photometric redshifts are from Ilbert \etal 2009 (for normal galaxies) and Salvato \etal 2009 (for X-ray detected AGN marked with red boxes). Note that the agreement is excellent for the AGN as well as the normal galaxies. The overall dispersion is $0.02\times(1+z)$ (shown as dashed lines in the bottom panel) and the fraction of catastrophic failures is 1.7\%. }
\label{phot_spec}
\end{figure}

Reliable spectroscopic redshifts are available for 602 sources in the sample ($\sim40\%$) and are represented by the filled histogram in Figure~\ref{redshift}. Column 5 in Table~\ref{spectra} lists the number of sources in our sample that have spectra from each of the different redshift surveys. Several sources have redshifts available from more than one source. For these objects we compared each available redshift to check their reliability and chose the best one. A comparison between the photometric and spectroscopic redshifts for those sources with very secure redshifts (quality flags 3 and 4) is shown in Figure~\ref{phot_spec}. The XMM sources are shown as the red points and use the photometric redshifts of \cite{Salvato:2009p2142}. The overall dispersion in this infrared selected sample is $0.02\times(1+z)$. There are only 9 catastrophic failures ($|z_{\rm p}-z_{\rm s}|/(1+z_{\rm s}) > 0.15$) in the sample (1.7\%). For the rest of the paper, we adopt the spectroscopic redshift for those sources with reliable spectroscopy and the photometric redshift for the rest of the sample. 

\subsection{UV-to-FIR Spectral Energy Distributions}

\begin{figure*}
\plotone{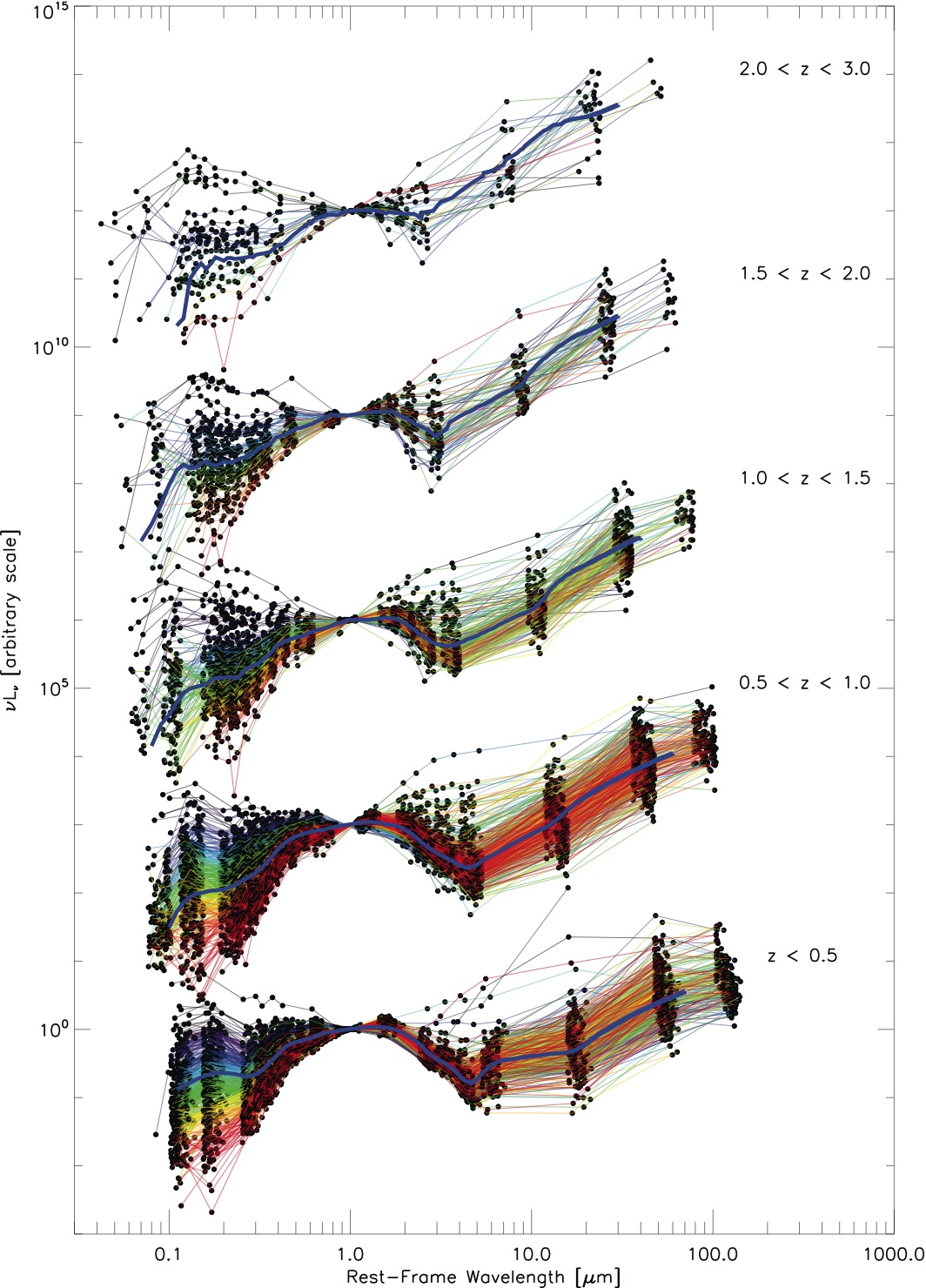}
\caption{Spectral energy distributions of all 1503 70\ts$\mu$m sources binned by redshift and normalized at 1\ts$\mu$m. The SEDs are color coded such that the ones with the bluest slope in the UV-optical are purple/blue and the reddest are red and the spectrum in between shows the range of SED colors. The median SED for each bin is overplotted in blue. Note that at higher redshifts the SEDs extend further into the UV while the FIR points probe shorter wavelengths}
\label{sed_z}
\end{figure*}

Figure~\ref{sed_z} shows a compilation of all 1503 spectral energy distributions ranging from the UV through the FIR binned by redshift and normalized at 1\ts${\mu}$m.  Over plotted in blue is the median SED for each luminosity bin. All of the photometry used to construct these SEDs, the redshifts, and the quantities derived in this paper (such as $L_{\rm IR}$) are listed in Table~\ref{catalog}.

\subsection{X-ray Properties}

\begin{figure}
\plotone{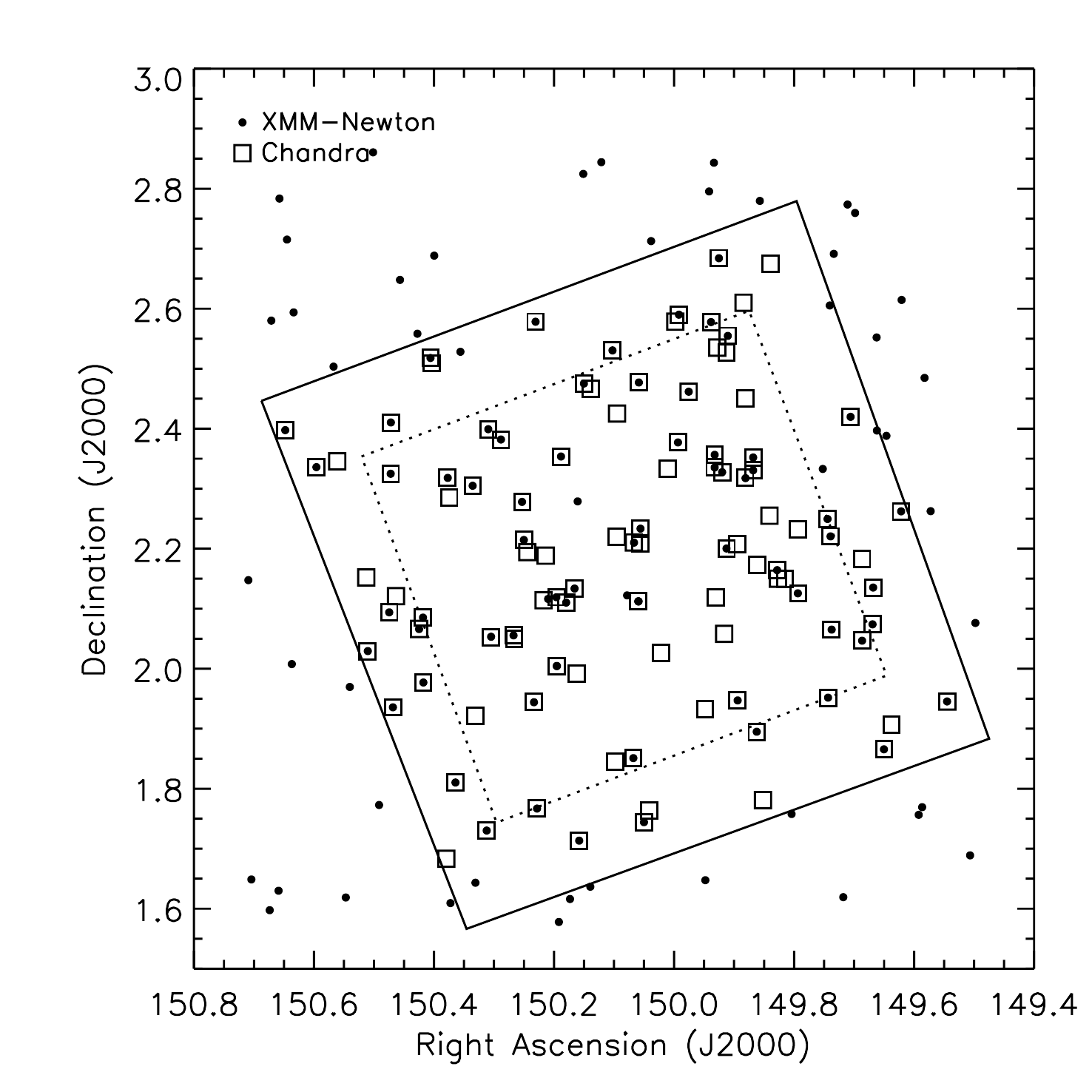}
\caption{Spatial distribution of all of the X-ray sources detected by {\it XMM-Newton} (dots) and {\it Chandra} (boxes). Note that a smaller area of the field was observed by {\it Chandra}. The two rectangles illustrate the different depths of {\it Chandra} coverage. The outer rectangle (solid box) is $\sim 0.9$ deg$^{2}$ (covering 55\% of the ACS area) representing the entire area observed by {\it Chandra}. The inner rectangle (dotted box) is $\sim 0.5$ deg$^{2}$ (covering 30\% of the ACS area) observed to an effective exposure time of $\sim 160$\ts ks. The area between the two rectangles was observed to an effective exposure time of $\sim 80$\ts ks.}
\label{xray_dist}
\end{figure}

\begin{figure}
\plotone{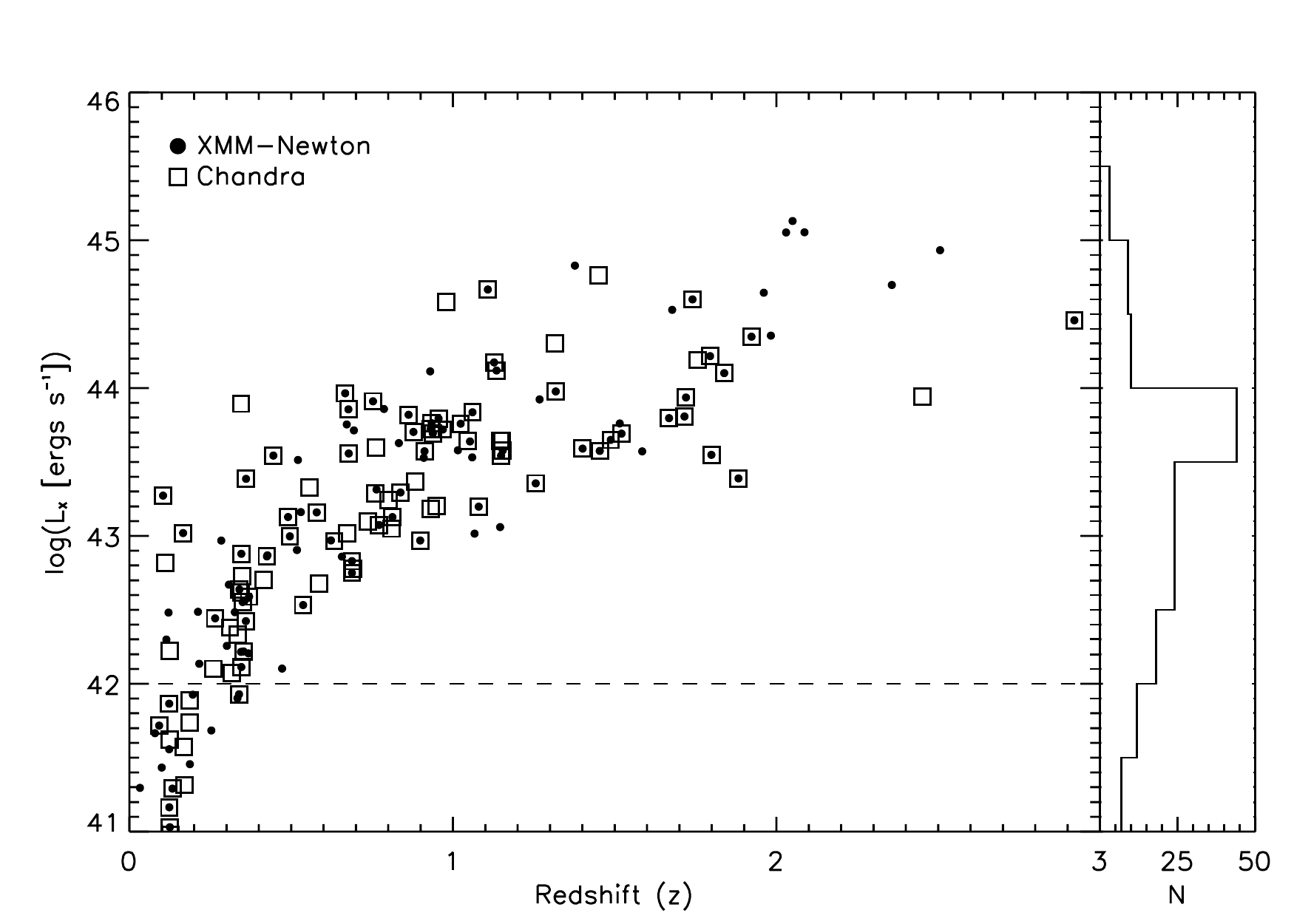}
\plotone{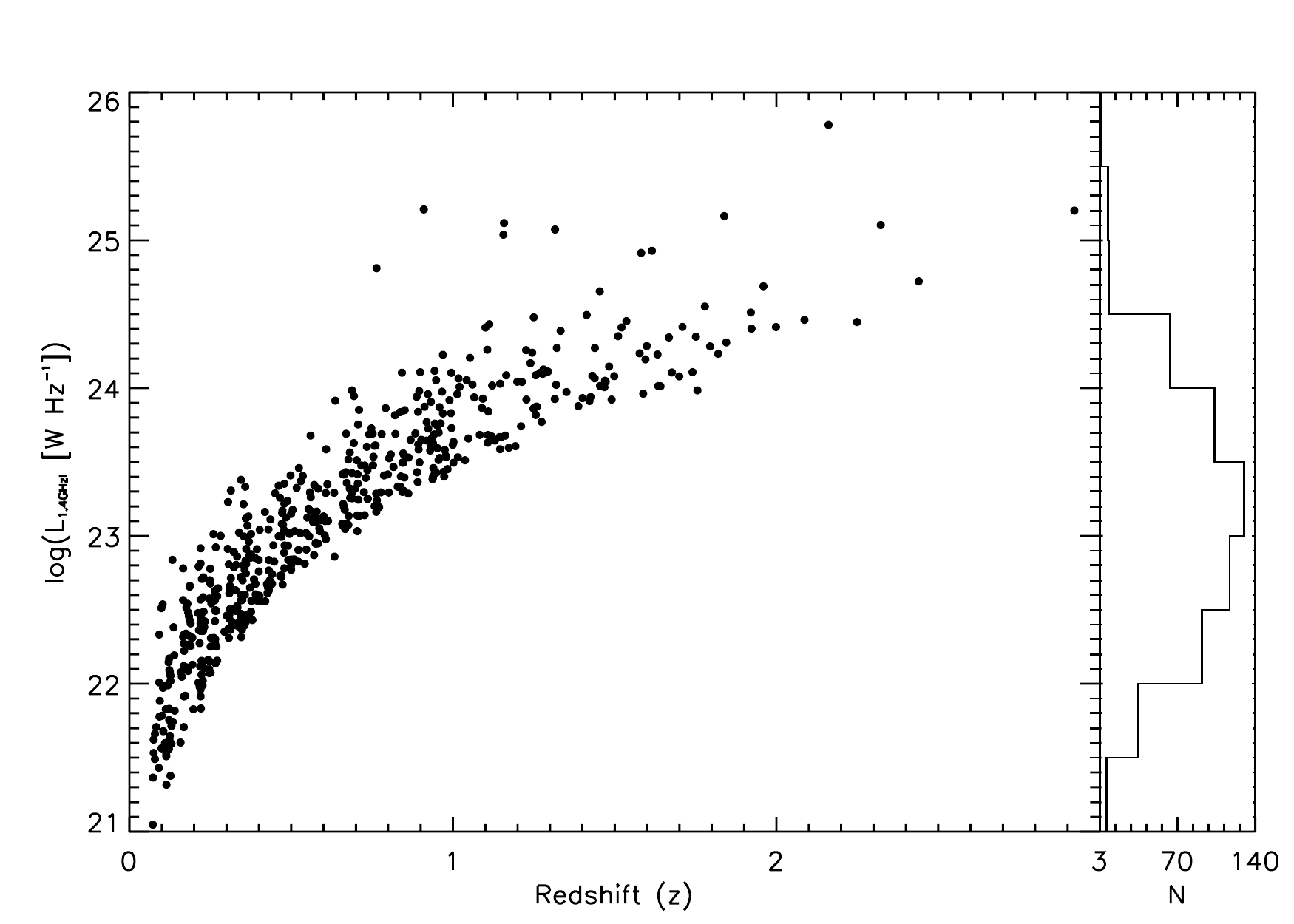}
\caption{Top: Distribution of the rest-frame X-ray luminosities for the 154 sources detected by either {\it Chandra} or {\it XMM} as a function of redshift. The dashed horizontal line represents the luminosity limit above which objects are considered to be AGN ($L_{\rm X}>10^{42}\ts\rm ergs\ts\rm s^{-1}$). Bottom: Distribution of rest-frame 1.4 GHz radio luminosities for the 562 sources detected by the VLA as a function of redshift. }
\label{lxr}
\end{figure}

Using the {\it XMM-Newton} and {\it Chandra} X-ray catalogs (\citealt{Brusa:2009p2165}; Civano \etal 2009, in preparation) of the COSMOS field we identified which of the 70\ts${\mu}$m selected sources are detected at X-ray energies. 119 of the sources were detected by XMM and 103 by {\it Chandra} with 68 detected by both, for a total of 154 X-ray detected sources. The spatial distribution of the sources detected by both XMM and {\it Chandra} is shown in Figure~\ref{xray_dist}. This plot illustrates the relative areas observed with both telescopes within the ACS coverage area. The rest-frame X-ray luminosities for each of these sources was calculated from the hard band (0.5--10\ts keV) fluxes with an assumed power law photon index of $\Gamma=1.7$ and is shown in the top panel of Figure~\ref{lxr} as a function of redshift. 132 (85\%) of these sources have $L_{\rm X} > 10^{42} \rm ergs\ts \rm s^{-1}$ (dashed line in Fig.~\ref{lxr}) and so are likely to be AGN. 

\begin{figure}
\plotone{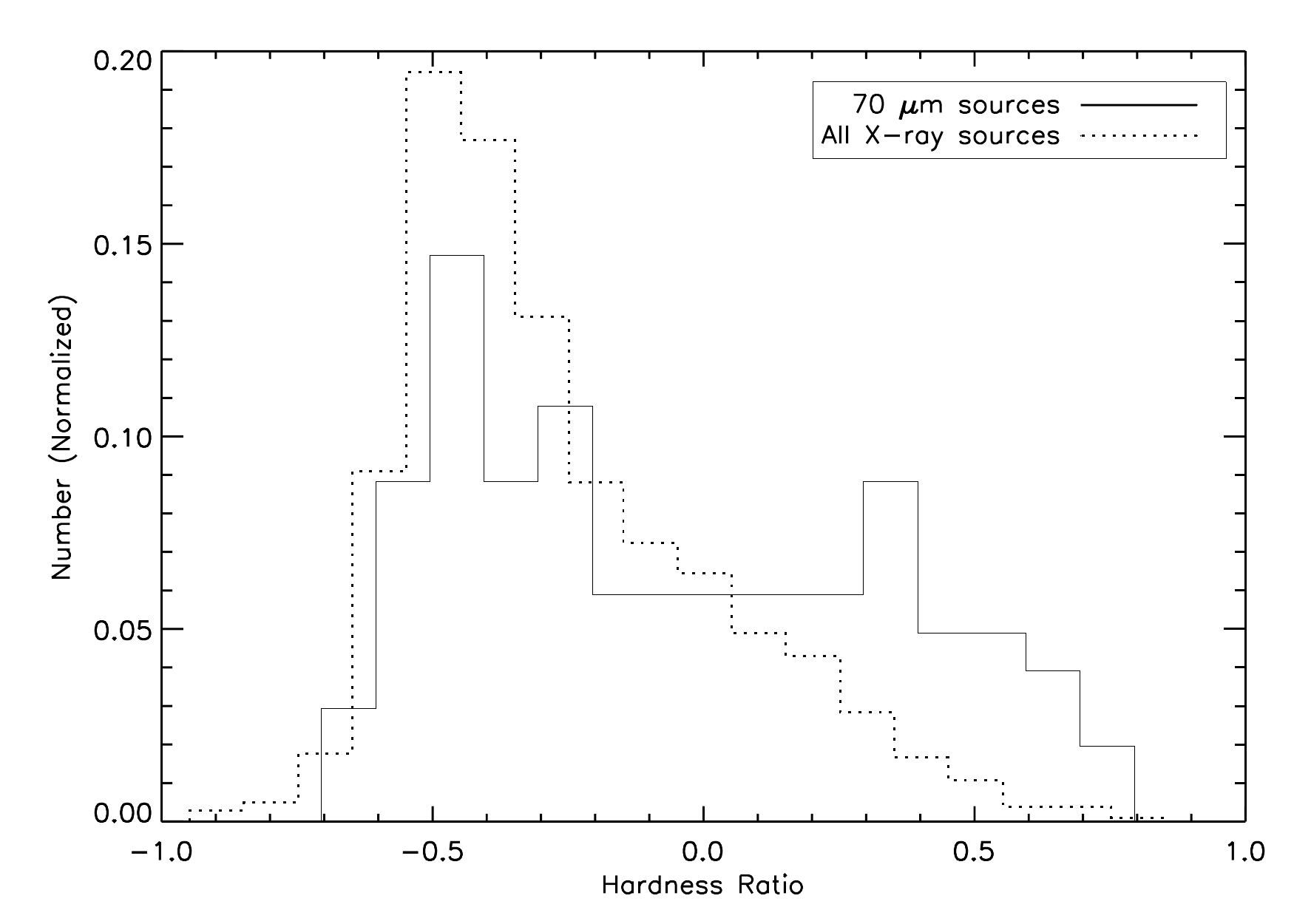}
\caption{Normalized distribution of hardness ratios for 70\ts $\mu$m  selected sources (solid line) and the general X-ray population (dotted line). Note that the 70\ts $\mu$m sample matches the general X-ray population very well at low hardness ratios but that there is an excess of 70\ts $\mu$m sources with larger hardness ratios.}
\label{xray}
\end{figure}

Figure~\ref{xray} shows the distribution of hardness ratios ($(H-S)/(H+S)$, where $H=2-10$\ts keV and $S=0.5-2$\ts keV) for the 102 X-ray sources in the sample that are detected in both the soft and hard X-ray bands. For the remaining objects, 24 are completely hard (i.e., no detection in the soft band), 29 are completely soft (no hard band detection) and only upper limits are available for 3. Over plotted on this histogram (dotted line) is the normalized distribution for the entire X-ray selected population that is detected in both bands. These two populations show striking differences. Whereas the full X-ray population has a hardness ratio that peaks around $-0.5$, the 70\ts${\mu}$m selected population appears bimodal, with peaks at $\sim -0.5$ and 0.35. The 70\ts${\mu}$m selected population of X-ray detected AGN is on average harder than the general X-ray selected population. The large hardness ratios of some of the 70\ts${\mu}$m sources indicates that there is significant obscuration present in these galaxies absorbing UV and soft X-ray emission. 

\subsection{Radio Properties}

562 of the 70\ts${\mu}$m sources were matched to a radio source within 1 arcsecond. For each of these sources, we estimated the rest frame 1.4\ts GHz luminosity ($L_{1.4\ts\rm GHz}$) using equation 5 of \cite{Yang:2007p3462} and applying a $k$-correction assuming a power law of the form $F_{\nu} \propto \nu^{-\alpha}$ where $\alpha$, the radio spectral index, is assumed to be $\sim 0.7$ \citep{Condon:1992p2865}. The radio luminosity is presented in the bottom panel of Figure~\ref{lxr} as a function of redshift for all 562 radio detected sources in our sample. 

Using this subsample of objects detected in both the radio and infrared, we investigated the infrared-radio correlation. Even though this correlation has been shown to hold over a wide range in redshift and luminosity \citep[e.g.,][]{Helou:1985p2892,Condon:1991p1977, Yun:2001p2885,Appleton:2004p5559,Ibar:2008p5546}, there is some evidence that it might break down at extreme infrared luminosities or at high redshift \citep[e.g.,][]{Kovacs:2006p2929, Sajina:2008p2894, Younger:2009p2901}. A complete analysis of the evolution of the infrared-radio correlation with redshift is beyond the scope of this paper, however, a full treatment of this issue will be discussed in Sargent \etal (2009, submitted to ApJS).  \cite{Condon:1991p1982} found the mean value of $q$ for BGS starbursts to be $<q>=2.34$ with a very small dispersion ($\sigma \sim 0.19$). This is in excellent agreement with the results from \cite{Yun:2001p2885} who found $<q>=2.34$ for a sample of 1809 galaxies from the IRAS 2 Jy sample. The \cite{Yun:2001p2885} study also found that there were slight variations in $q$ with IR luminosity. For faint IR sources, they found values of $q$ that were slightly higher than the mean and for bright IR sources they found a higher overall dispersion ($\sigma \sim 0.33$).

\begin{figure}
\plotone{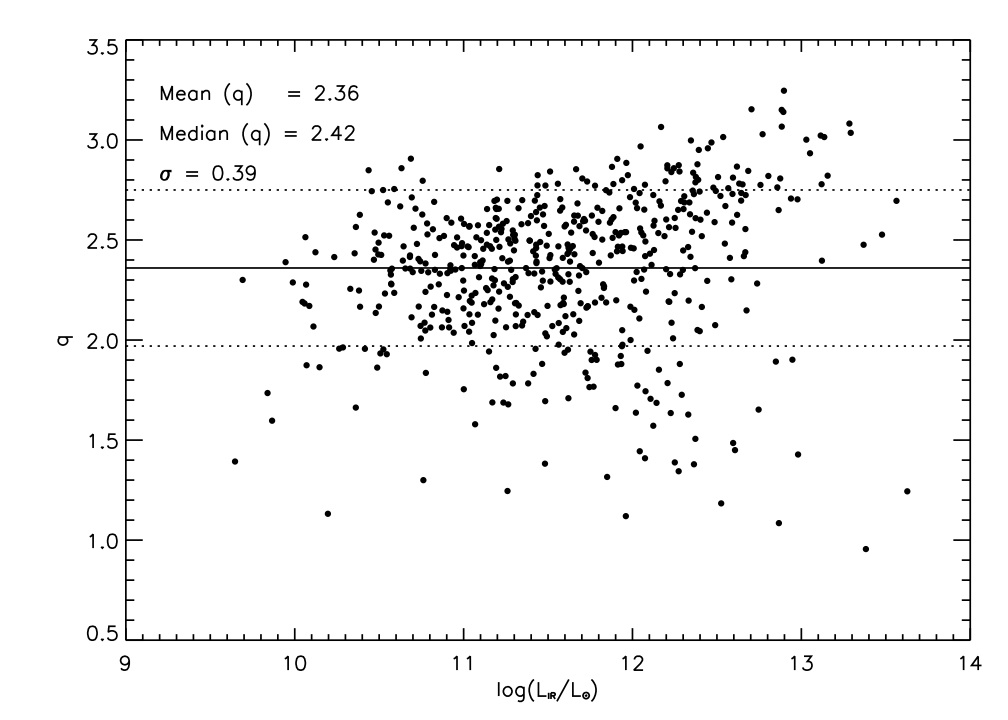}
\plotone{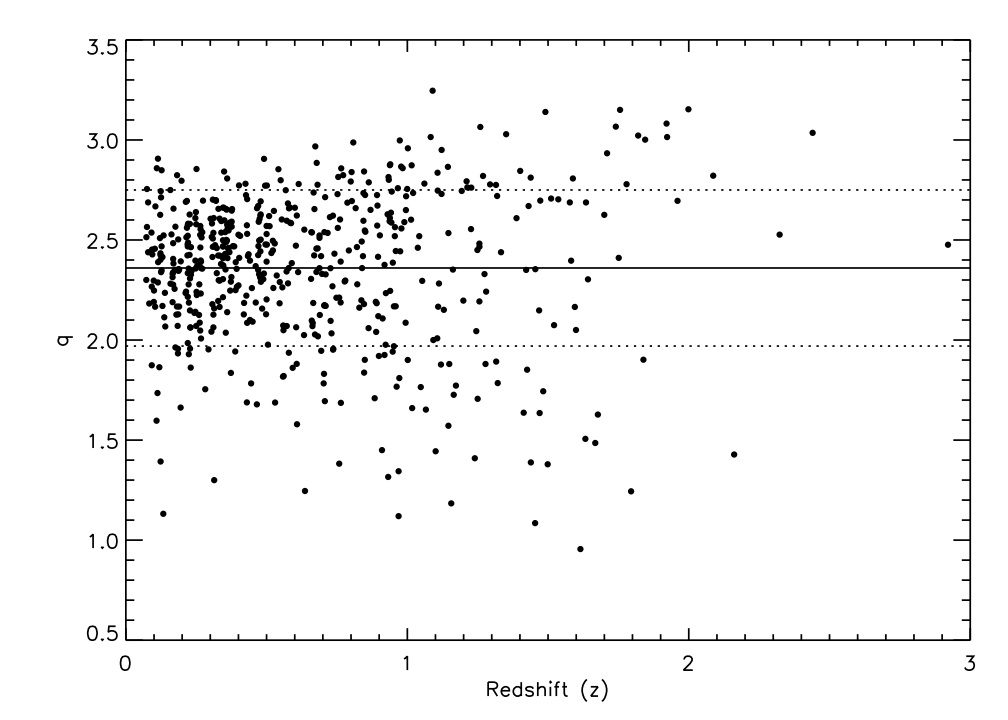}
\caption{Radio/infrared flux ratio ($q$) as a function of total infrared luminosity (top) and redshift (bottom). The horizontal line represents the mean value and the dotted lines are the values $\pm 1\ts\sigma$ from the mean. The scatter increases with $L_{\rm IR}$ and the distribution appears to bifurcate at $\sim L_{\rm IR} > 10^{12}\ts L_{\odot}$. The large scatter at high luminosities is observed even when a more stringent (i.e., $S/N > 5$) cut is used.}
\label{q}
\end{figure}

We computed the value of $q$ as defined originally by \cite{Helou:1985p2892} as the ratio between the FIR (a combination of infrared flux measurements from 60 and 100\ts${\mu}$m) and the flux at 1.4\ts GHz. We adopted this definition to allow for a direct comparison with previous work. For our sample of 562 radio detected 70\ts${\mu}$m selected sources we compute $q$ using the rest frame 60 and 100\ts${\mu}$m flux densities obtained from the best-fit SED template and the $k$-corrected 1.4\ts GHz flux density. The resulting $q$ distribution for our sample is shown in Figure~\ref{q} as a function of $L_{\rm IR}$ (top) and redshift (bottom). We find a wide range in $q$ (0.65--3.7) with an average $q$ of 2.36 and a dispersion of 0.39 overall, in agreement with previous studies. However, we find that this dispersion increases dramatically with $L_{\rm IR}$ with an increase in both radio-excess (small values of $q$) and infrared-excess (large values of $q$) galaxies. For the LIRGs and ULIRGs in the sample, we find $<q>=2.34\pm0.31$ and $<q>=2.42\pm 0.50$, respectively. It is thought that the slightly higher mean $q$ at high luminosities is the result of incredibly high radio free-free opacity \citep{Condon:1991p1982}. However, we note that this effect should decrease with $L_{\rm IR}$ in our sample as the rest frame radio frequency probed increases, and therefore, the infrared excess observed in these sources is not likely to be caused by this opacity effect. 

\section{Total Infrared Luminosities}
\subsection{SED Fits}

We obtained the total infrared luminosity, $L_{\rm IR}$ (8--1000\ts${\mu}$m), for each source in our sample by fitting the 8, 24, 70, and 160\ts${\mu}$m data points to various SED templates using the SED fitting code {\it Le Phare}\footnote{http$://$www.cfht.hawaii.edu/$\sim$arnouts/LEPHARE/cfht$\_$lephare/lephare.html} written by S. Arnouts and O. Ilbert.  We weighted each of the four points equally to avoid over-weighting the points with the smallest error bars, which are typically those at shorter wavelengths (i.e., 8 and 24\ts${\mu}$m). The 8\ts${\mu}$m  data point was only used at low redshifts. Once it shifted below 7\ts$\mu$m, it was no longer used due to confusion with PAH features and since the 24, 70, and 160\ts${\mu}$m  data points well represent the portion of the SED that contributes the bulk of the infrared luminosity. Beyond $z\sim2.4$ the 24\ts${\mu}$m point is no longer used for the same reason.

\begin{figure*}
\epsscale{1.0}
\hspace*{-0.3in}
\plotone{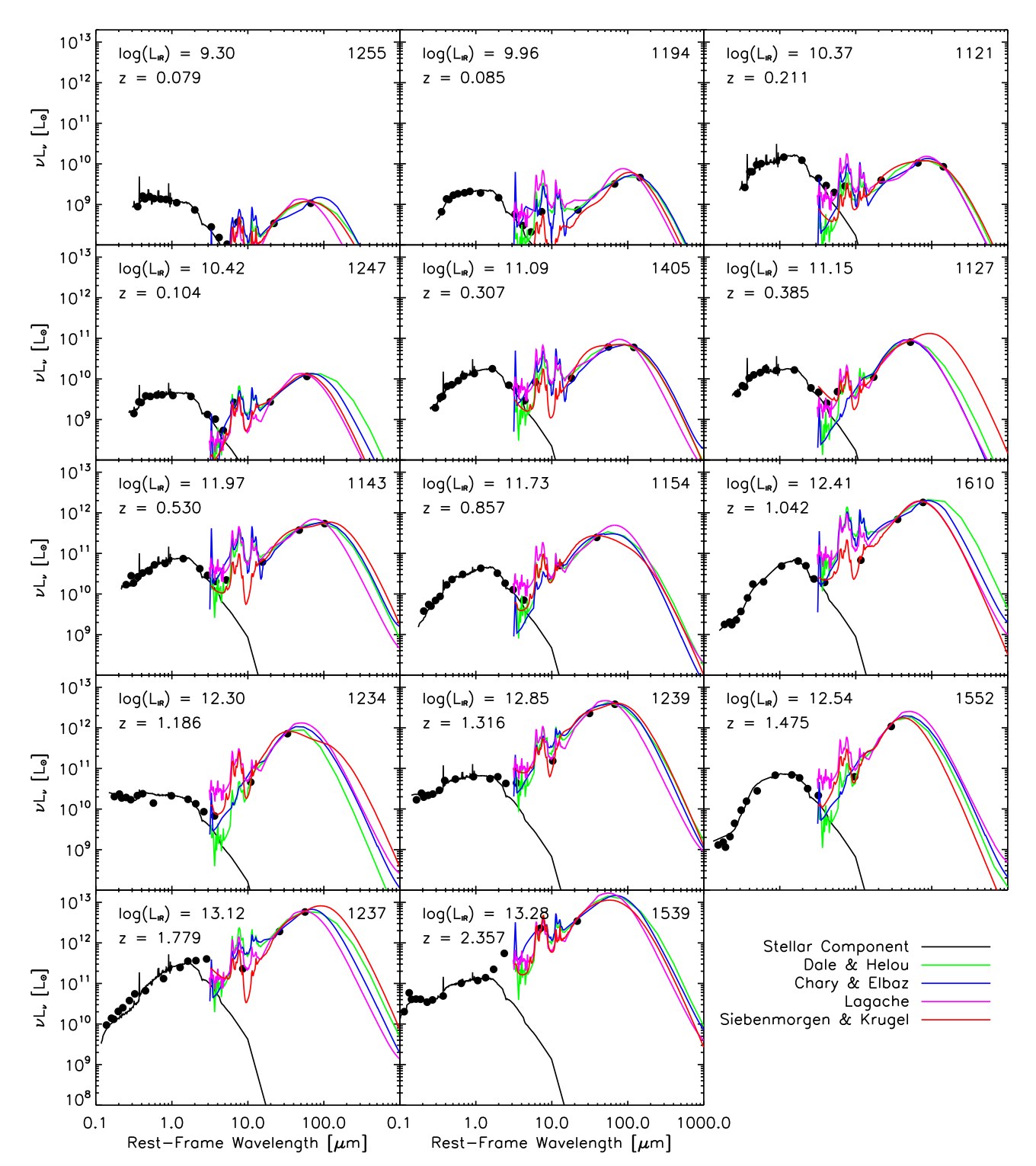}
\caption{Spectral energy distributions for a sample of 14 sources at a variety of luminosities and redshifts (both photometric and spectroscopic). Over plotted are the best fit templates from each of the four infrared template libraries as well as the best fit to the stellar component using empirical SEDs from Coleman et al. (1980). In order to illustrate the range of fits, half of the sources here have 160\ts${\mu}$m detections and the other half do not. Note that at high redshift both the 70 and 160\ts${\mu}$m points are on the low wavelength side of the peak and that the fits get progressively worse. Also, the fit to the stellar component becomes worse for sources with a significant contribution from AGN activity in the NIR (e.g., sources 1237 and 1539).}
\label{lir4}
\end{figure*}

\subsection{Template Libraries}
We used four different template libraries for the SED fitting -- \cite{Chary:2001p2083}, \citealt{Dale:2002p2130}, \cite{Lagache:2003p1825}, and \cite{Siebenmorgen:2007p2697}. We chose to use these template libraries since they represent a wide range of SED shapes and luminosities and they are widely used in the literature. Here, we briefly describe how each of these libraries was derived and discuss the differences between them. For more information on these template libraries, see the original papers and the detailed description in \cite{Symeonidis:2008p2625}.

{\it  \cite{Chary:2001p2083}:} This template library consists of 105 templates based on the SEDs of four ``prototypical" galaxies (Arp220: ULIRG, NGC 6090: LIRG, M82: starburst, and M51: normal star forming galaxy). They were derived using the \cite{Silva:1998p2723} models with the MIR region replaced with ISOCAM observations.  These templates were then divided into two portions (4--20\ts${\mu}$m and 20--1000\ts${\mu}$m) and interpolated between to arrive at a set of libraries of varying shapes and luminosities. The  \cite{Dale:2001p2742} templates are also included in this set to extend the range of shapes.

{\it \cite{Dale:2002p2130}:}  These templates are updated versions of the \cite{Dale:2001p2742} templates which used the model of \cite{Desert:1990p2747}. This model involves three components, large dust grains in thermal equilibrium, small grains semistochastically heated, and stochastically heated PAHs. They are based on IRAS/ISO observations of 69 normal star-forming galaxies in the wavelength range 3--100\ts${\mu}$m. \cite{Dale:2002p2130} improved upon these models at longer wavelengths using SCUBA observations of 114 BGS (\citealt{Soifer:1989p2523}) galaxies, 228 galaxies with ISOLWS (52--170\ts${\mu}$m; Brauher 2002) and 170\ts${\mu}$m observations for 115 galaxies from the ISOPHOT Serendipity Survey \citep{Stickel:2000p2754}. All together, these 64 templates span the IR luminosity range $10^{8}-10^{12} L_{\odot}$.

{\it \cite{Lagache:2003p1825}:} This library of starburst templates was built on the model of Maffei (1994), which uses the dust emission model of \cite{Desert:1990p2747}, with a few modifications applied. First they replace the PAH portion of the templates with the one from \cite{Dale:2001p2742}, then they increase the proportions of PAHs and very small grains while adding in extinction, and finally, they broaden the FIR peak and flatten the SED at long wavelengths. This library also includes templates of normal star-forming galaxies derived from the ISOPHOT serendipity survey \citep{Stickel:2000p2754}, the FIRBACK galaxy SEDs, and long wavelength data taken from \cite{Dunne:2000p2799} and  \cite{Dunne:2001p2816}. 

{\it 	:}  This template library consists of over 7000 model SEDs derived using a radiative transfer and dust model described in \cite{Kruegel:2003p2827}. This model uses the assumption of spherical symmetry and divides the starburst luminosity into two classes, that from OB stars in dense clouds ($L_{\rm OB}$) and the total from all other stars ($L_{\rm tot}-L_{\rm OB}$). Five parameters are varied in the calculations to create the grid of template SEDs: 1) The total IR luminosity ranges from $10^{10}-10^{14}\ts L_{\odot}$, 2) The radius of the starburst nucleus: 0.35, 1, 3, 9, and 15 kpc, 3) The visual extinction of the nucleus: $A_{V} = $2.2, 4.5, 7, 9, 18, 35, 70, and 120 mag, 4) The ratio of the luminosity contributed by the OB stars to the total luminosity: $L_{\rm OB}/L_{\rm tot} =$ 0.4, 0.6, and 0.9, and 5) The dust density in the hot spots: $n^{\rm hs} = 10^2, 10^3,$ and $10^4\ts \rm cm^{-3}$. Combinations of parameters that are considered to be physically unlikely are left out of the final grid.

\subsection{$L_{\rm IR}$}
\begin{figure}
\epsscale{1.0}
\plotone{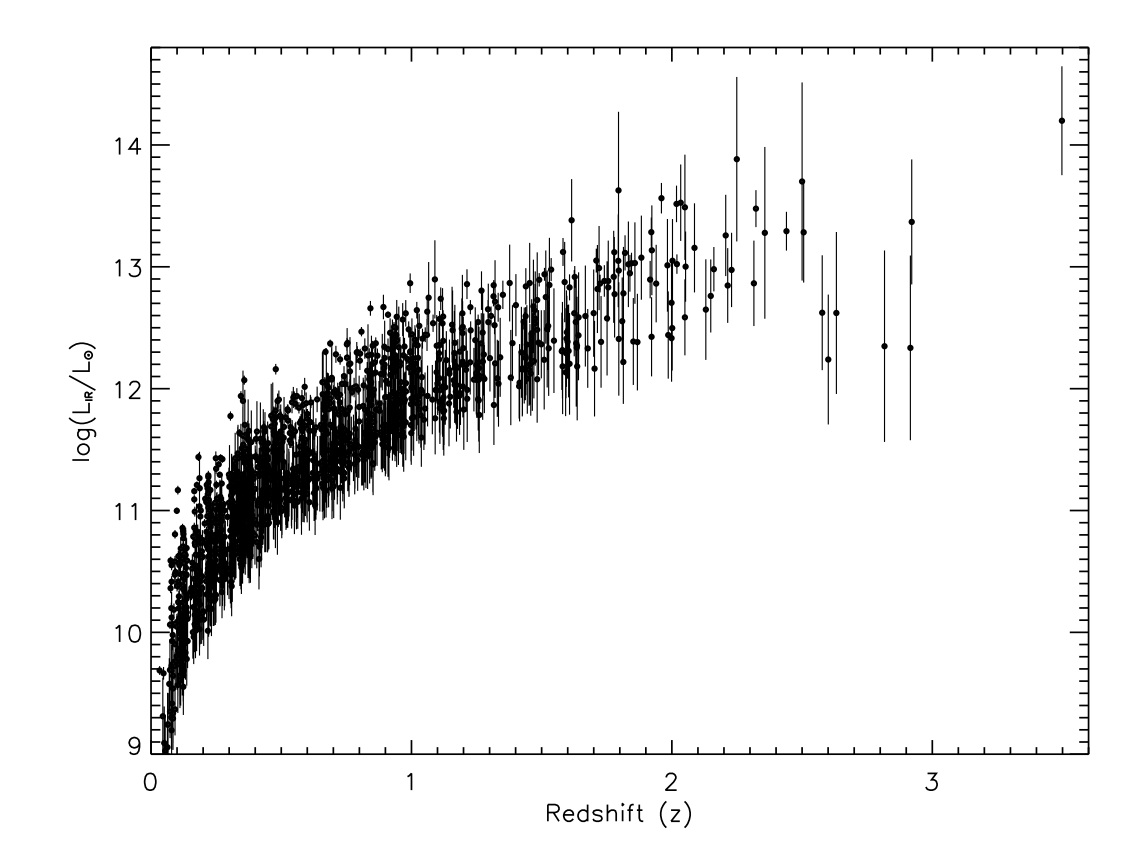}
\caption{Total infrared luminosity ($L_{\rm IR}$) as a function of redshift for all 1503 70\ts${\mu}$m sources in the COSMOS field determined from the best fit template using 8, 24, 70, and 160\ts${\mu}$m where available. The error bars represent the $1\ts\sigma$ uncertainty for each source based on the fit, the redshift errors, and a 0.2 dex systematic uncertainty for the sources without a 160\ts$\mu$m detection.} 
\label{lir_all}
\end{figure}

For each of the above template libraries, the best fit model was chosen by finding the one with the lowest $\chi^{2}$ value and allowing for rescaling of the templates. The total infrared luminosity was then calculated from the best fit template by integrating from 8--1000\ts${\mu}$m. A sample of SEDs and the best fit from each template library is shown in Figure~\ref{lir4} for a range of luminosities. We visually inspected the fits and found that they typically work just as well for AGN as for non-AGN (in the 8--1000\ts${\mu}$m range) since our fit does not include all of the IRAC data points which would be the most affected by an AGN. For the final $L_{\rm IR}$ we adopt the value of the best fit of all the template libraries. This final value is shown in Figure~\ref{lir_all} as a function of redshift and is presented in Table~\ref{catalog}. Since the template library is so large, a range of templates can be used to adequately fit the data points. We use this range, along with the $1\ts\sigma$ uncertainties in the redshift measurement, to calculate the uncertainty on the final derived $L_{\rm IR}$ given in Table~\ref{catalog}. In total we identify 687 LIRGs, 303 ULIRGs, and 31 HyLIRGs.

\subsection{Effect of 160\ts${\mu}$m Detections on $L_{\rm IR}$}

The depth and breadth of the 70\ts${\mu}$m imaging in the COSMOS field allow us to obtain a more accurate measurement of the total infrared luminosity than has previously been possible for samples selected at 24\ts$\mu$m. The long wavelength selection means we can be confident in our estimate even at high redshifts since we are still probing the infrared slope of the SED (35\ts${\mu}$m at $z=1$ and 23\ts${\mu}$m at $z=2$).  In addition, this large sample size of sources, including many that are also detected at 160\ts${\mu}$m, enables us to statistically quantify how well we could have measured the luminosity if we only had detections in a subset of the infrared bands. This information is particularly useful for comparing our results to other studies of sources selected at 24\ts${\mu}$m.

\begin{figure}
\vspace{-0.3in}
\plotone{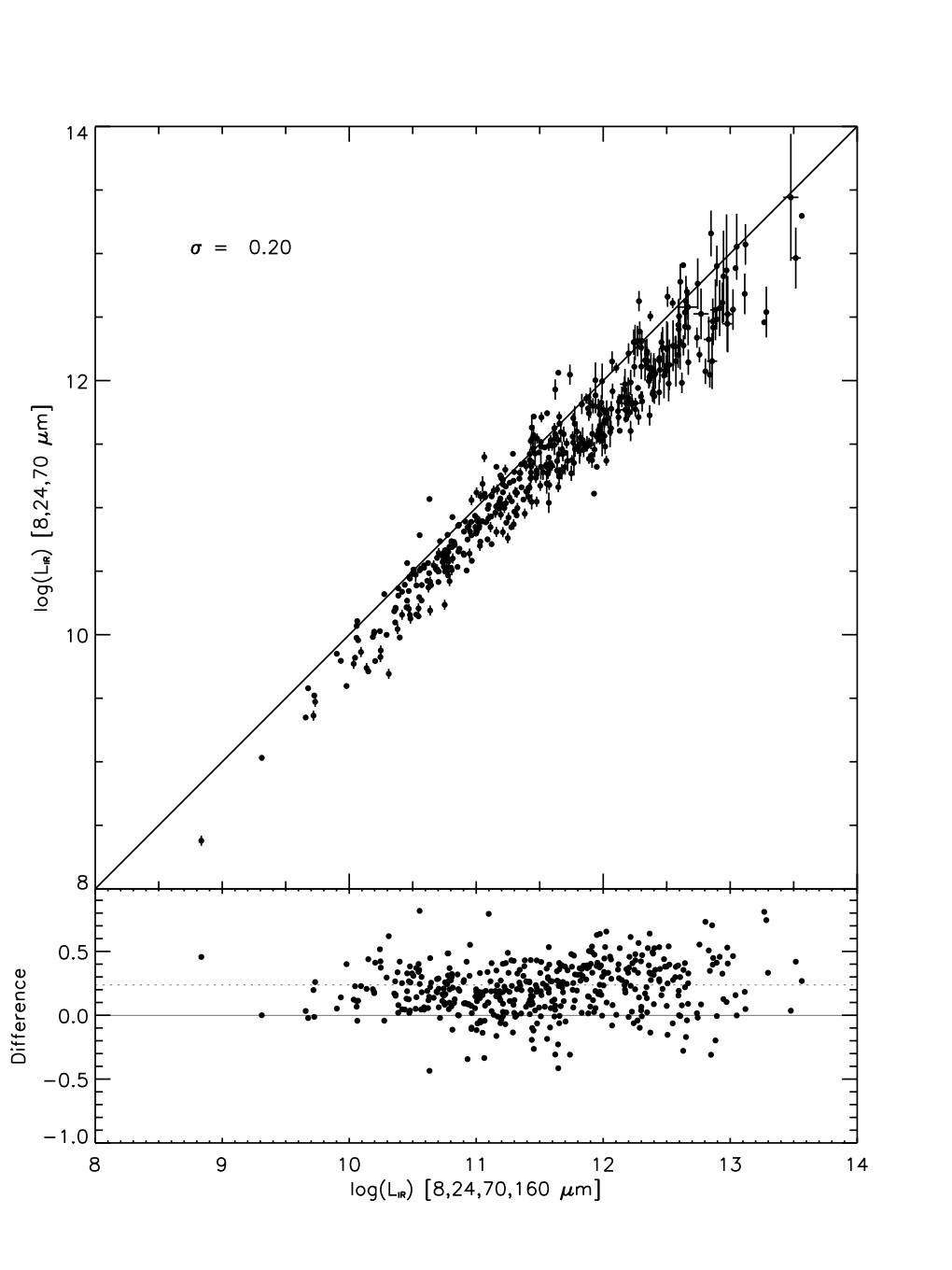}
\caption{Comparison of the total infrared luminosity for the 463 sources detected at 160\ts${\mu}$m obtained with and without the 160\ts${\mu}$m data point. The error bars are the $1\ts\sigma$ uncertainty based on the template fit. The bottom panel shows the difference in $L_{\rm IR}$ between these two measurements. The mean offset is shown as the dotted line and the dispersion about this mean is given.}
\label{lir}
\end{figure}

Figure~\ref{lir} shows a comparison of our estimate of $L_{\rm IR}$ for the subsample of sources detected at 160\ts${\mu}$m (463 sources in total) with the estimate excluding the 160\ts${\mu}$m data point (based only on 8, 24, and 70\ts${\mu}$m). The solid line represents a 1-to-1 relationship and the lower panel shows the difference between these two measurements. Since the majority of 70\ts${\mu}$m selected sources are not detected at 160\ts${\mu}$m, it is important to understand and quantify this relationship. While the internal scatter of the points is quite small ($\sigma = 0.20$ dex),  there appears to be a significant offset (of 0.20 dex on average) between the estimates with and without the 160\ts${\mu}$m data point, which increases with $L_{\rm IR}$. This offset indicates that the $L_{\rm IR}$ determined without the 160\ts${\mu}$m data point is underestimated by 0.20 dex on average.

\begin{figure*}
\vspace{-0.8in}
\plotone{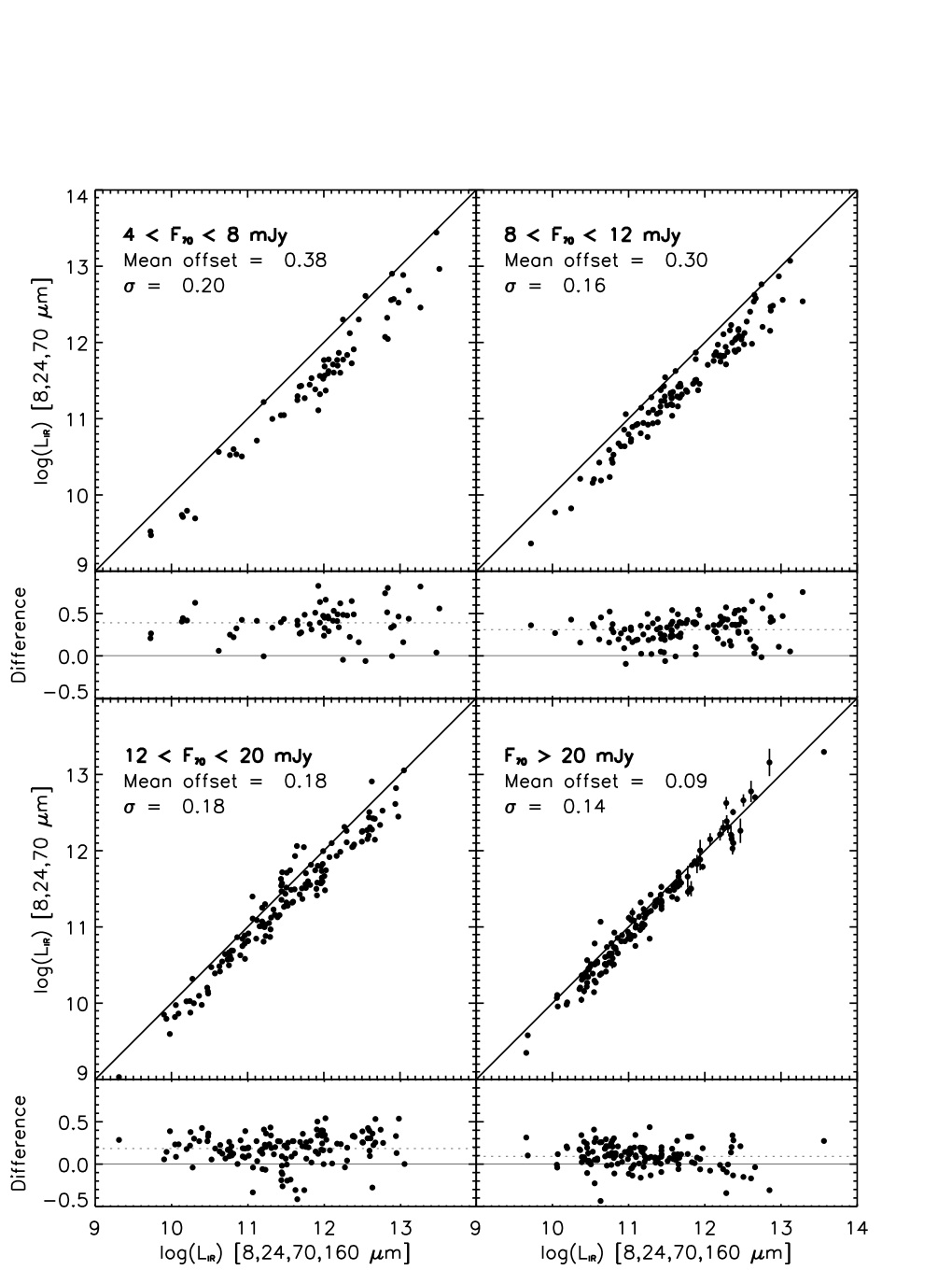}
\vspace{0.5in}
\caption{Same as Figure~\ref{lir} but divided into four 70\ts${\mu}$m flux bins. Note that the mean offset between the two luminosity estimates decreases at higher 70\ts${\mu}$m fluxes while the overall scatter stays about the same.}
\label{L_flux}
\end{figure*}

\begin{figure*}
\vspace{-0.2in}
\plotone{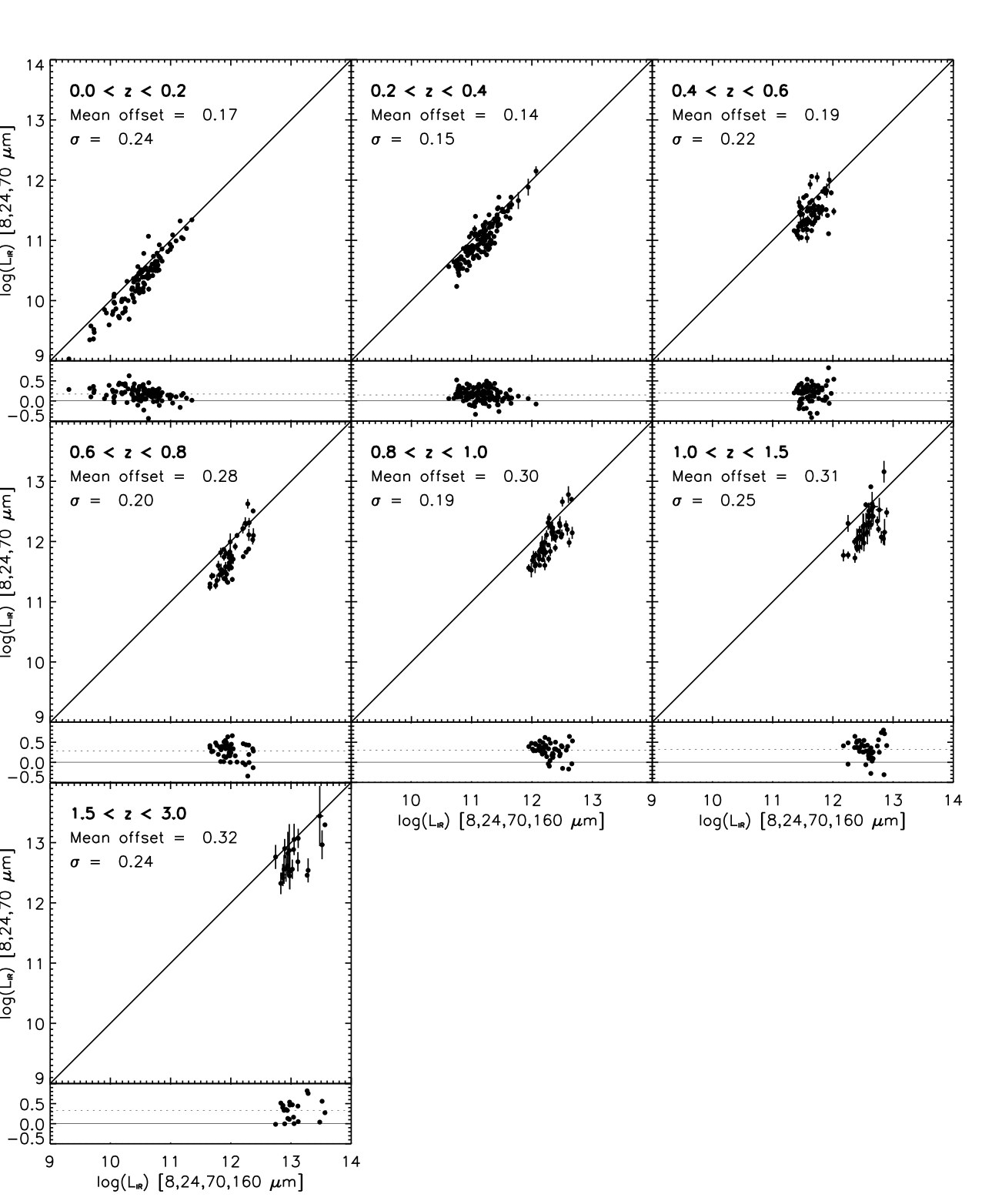}
\caption{Same as Figure~\ref{lir} but divided into seven redshift bins. Note that the mean offset between the two luminosity estimate increases with redshift while the overall scatter stays about the same.}
\label{L_z}
\end{figure*}

In an attempt to isolate the possible cause of this offset, Figures~\ref{L_flux} and \ref{L_z} contain the same plot divided into several flux and redshift bins, respectively. The offset is largest for the sources at the faintest flux levels (0.36 dex at $4<F(70)<8$\ts mJy) and is negligible at the brightest fluxes (0.11 dex at $F(70)>20$\ts mJy). When divided into redshift bins, the largest offsets are seen at the high redshift end. The mean offset and internal dispersion for each redshift and flux bin are summarized in Table~\ref{offsets}. There are a few possible explanations for the observed offset. One is that the intrinsic noise in the 160\ts${\mu}$m measurement at these low flux levels biases the fit toward higher estimates than if the source was not detected at 160\ts${\mu}$m at all. However, this explanation seems unlikely since the offset remains even when using higher $S/N$ cuts at both 70 and 160\ts${\mu}$m. Another possibility is that at higher redshifts the 70\ts${\mu}$m detection samples too short a wavelength to put much of a constraint on the best fit template. This would result in an overall larger scatter but not an offset. And finally, it is possible that the detection at 160\ts${\mu}$m preferentially picks out sources with a colder dust component than the rest of the 70\ts${\mu}$m population. 

\begin{figure}
\plotone{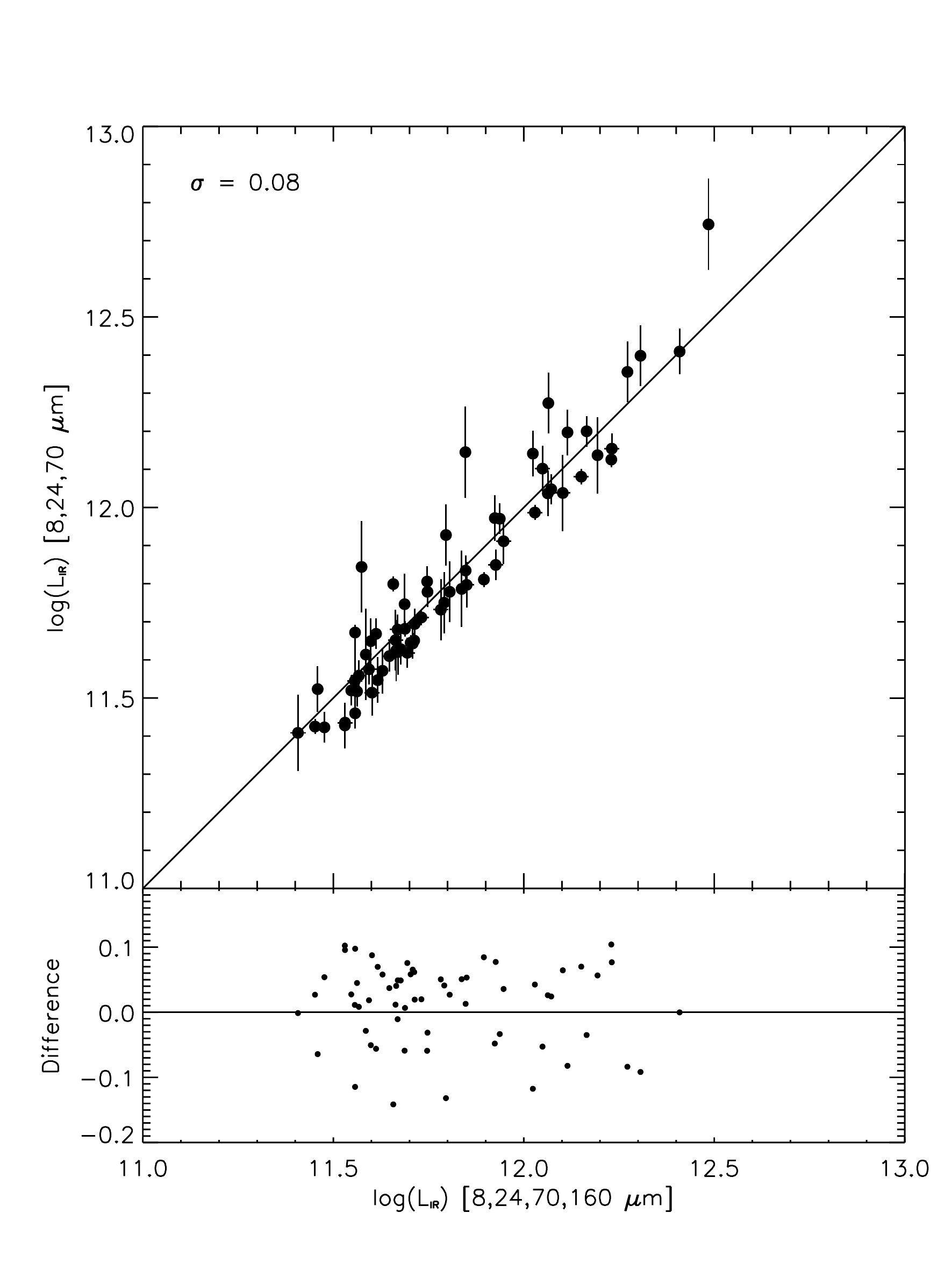}
\caption{Comparison of the total infrared luminosity obtained with and without the 160\ts${\mu}$m data point for a sample of 67 local (U)LIRGs from the GOALS survey. The low dispersion obtained between these two measures indicates that the template fitting method is reliable and that the rest-frame 70\ts$\mu$m data point is long enough to obtain a reliable measure of $L_{\rm IR}$ at low redshift ($z<0.1$).}
\label{goals}
\end{figure}

To test our fitting routine we fit the same data points (8--160\ts${\mu}$m from IRAC and MIPS) for a set of 67 local LIRGs and ULIRGs from the Great Observatory All-Sky LIRG Survey (GOALS: \citealt{Armus:2009p5172}). As for the COSMOS sources, we obtained the best fit $L_{\rm IR}$ with and without the 160\ts${\mu}$m point and plot the comparison in Figure~\ref{goals}. For these local sources we do not see the same offset that we see in the COSMOS sources and the intrinsic scatter is much smaller ($\sigma = 0.08$ dex). In addition, the $L_{\rm IR}$ obtained from this fit agrees very well with $L_{\rm IR}$ in the literature from the IRAS data points. This test gives us confidence that the offset we observe is not due to a problem with the fitting routine and also indicates that the rest-frame 70\ts${\mu}$m data point is at a long enough wavelength to obtain an accurate measurement of $L_{\rm IR}$ at low redshift. We also find that the template libraries based empirically on local sources tend to fit the data points better, as is expected. As an additional test we fit only the 8 and 24\ts${\mu}$m data points for the local sources (roughly corresponding to observed 24 and 70 at $z\sim2$) and find an overall scatter of $\sim0.17$ dex. This result is consistent with the typical uncertainty we obtain for the COSMOS sources.

\begin{figure}
\plotone{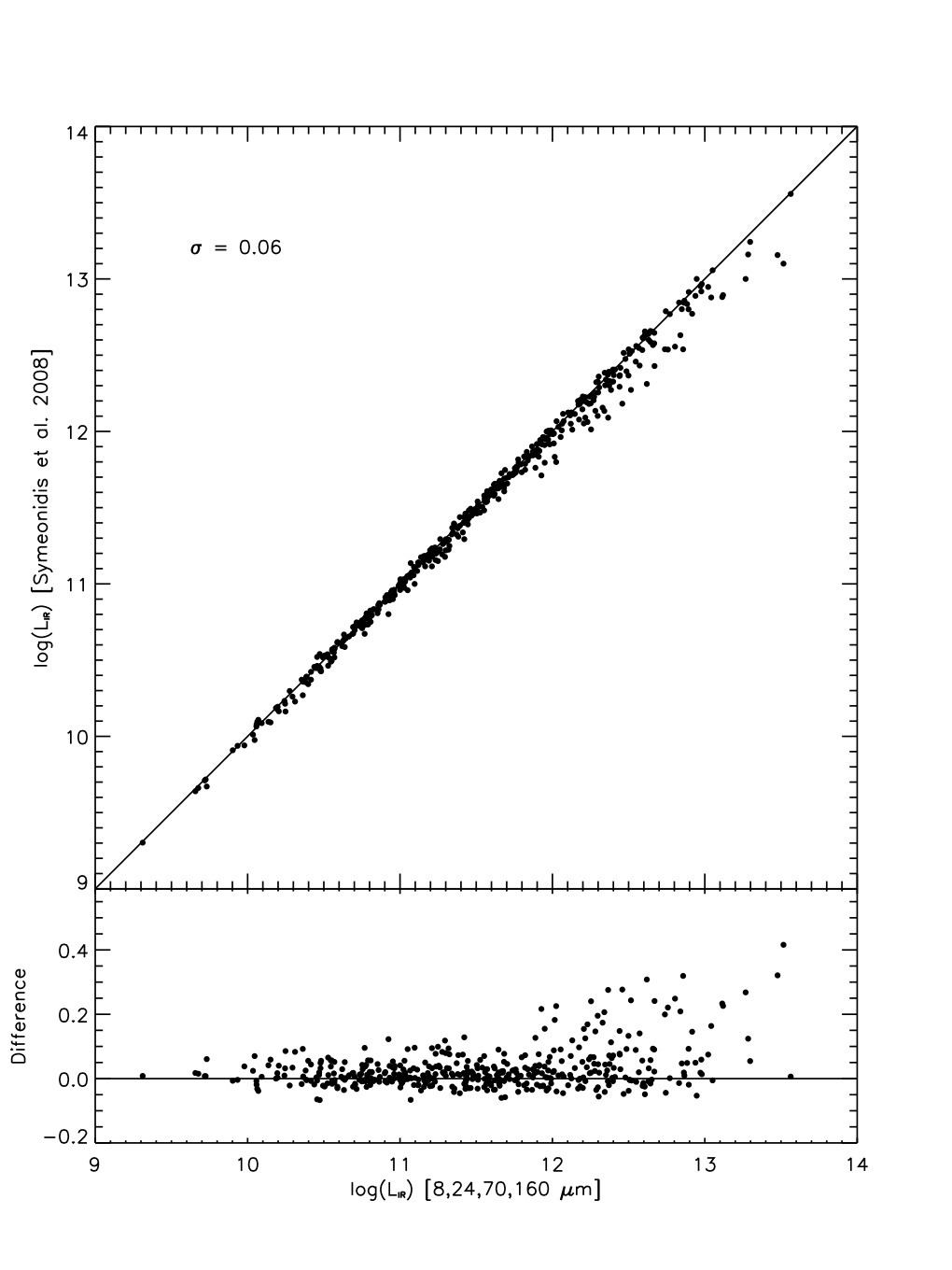}
\caption{Comparison of the total infrared luminosity obtained with the fits to the four infrared data points to that obtained with equation 1 of Symeonidis \etal 2008. The agreement between our estimates and the equation is excellent although the dispersion increases at $L_{\rm IR}>10^{12}\ts L_{\odot}$. At these higher luminosities, and consequently higher redshifts, the 160\ts${\mu}$m data point is not long enough to precisely constrain the peak and thus there is a larger spread among the best fit templates.}
\label{lir_sym}
\end{figure}

\cite{Symeonidis:2008p2625} performed a similar comparison with 43 sources from the Extended Groth Strip. They derived an empirical relation (Equation 1 in \citealt{Symeonidis:2008p2625}) to obtain $L_{\rm IR}$ directly from the 24, 70, and 160\ts${\mu}$m flux measurements without the need to fit templates or apply $k$-corrections. This relation was found by summing the approximate contributions to the total infrared luminosity obtained using triangles and polygons of heights determined by the fluxes at the three MIPS wavelengths. They then add in a correction for the systematic offset found between this approximation and their total $L_{\rm IR}$ found through template fitting to arrive at the final equation. They find that this equation agrees well with their 43 EGS sources at all $L_{\rm IR}$ with a very small scatter (0.06 dex). Figure~\ref{lir_sym} shows a comparison of the $L_{\rm IR}$ derived here for the COSMOS sources and that using the \cite{Symeonidis:2008p2625} equation.  We also find excellent agreement and a small scatter (0.06 dex overall), however, we find that the scatter increases beyond $10^{12}\ts L_{\odot}$ ($\sim$ 0.1 dex). A closer inspection of the sources with the largest deviation from the equation indicates that at these redshifts, the 160\ts${\mu}$m point is not long enough to constrain the peak of the emission and therefore, a wider range of templates fit the points than at lower redshifts.

\begin{figure}[b]
\plotone{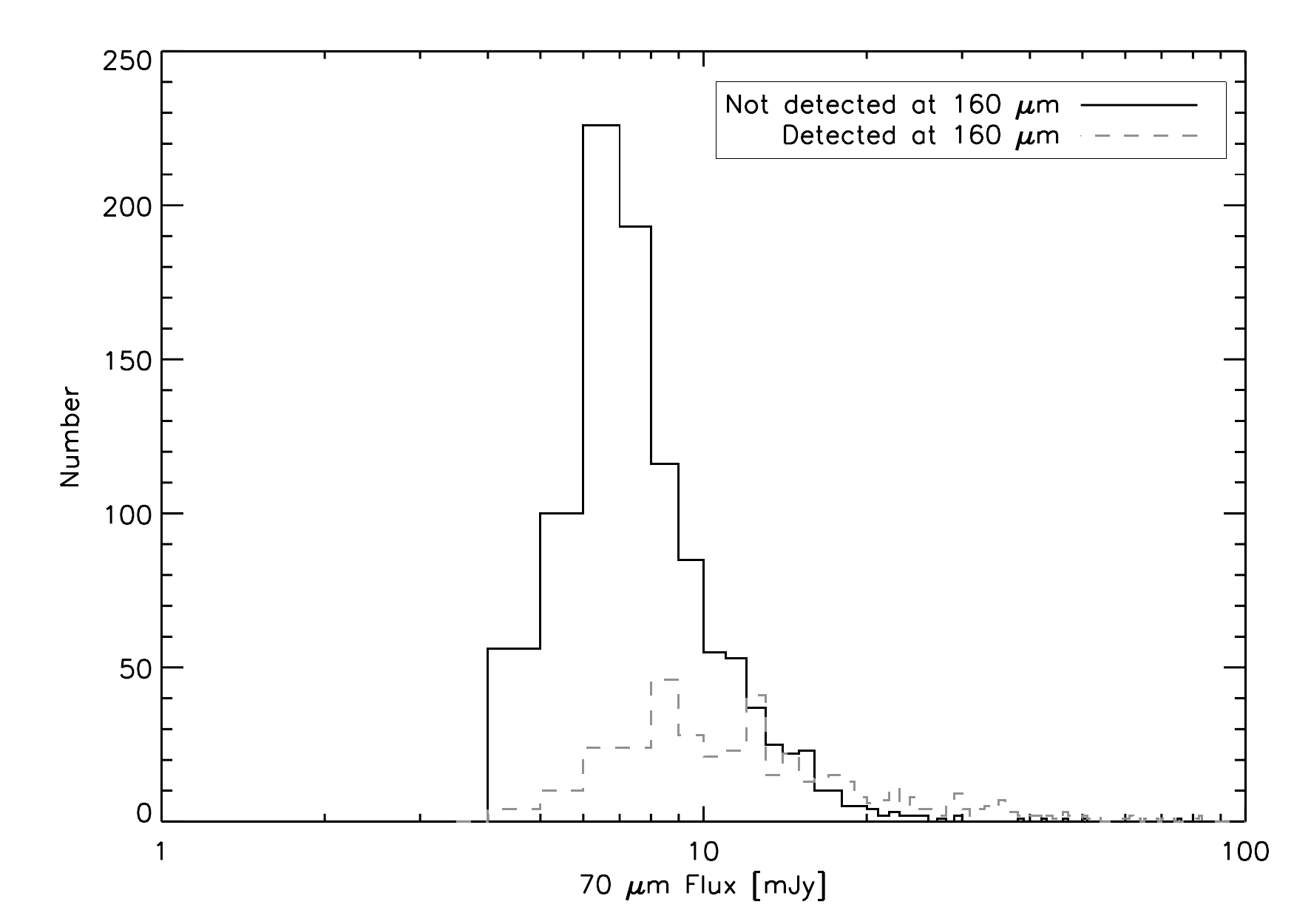}
\caption{70\ts${\mu}$m flux distribution of the 463 sources detected at 160\ts${\mu}$m (gray dashed) and the 1042 not detected at 160\ts${\mu}$m. On average, those detected at 160\ts${\mu}$m have higher 70\ts${\mu}$m fluxes. }
\label{160}
\end{figure}

\begin{figure*}
\vspace*{-0.9in}
\plotone{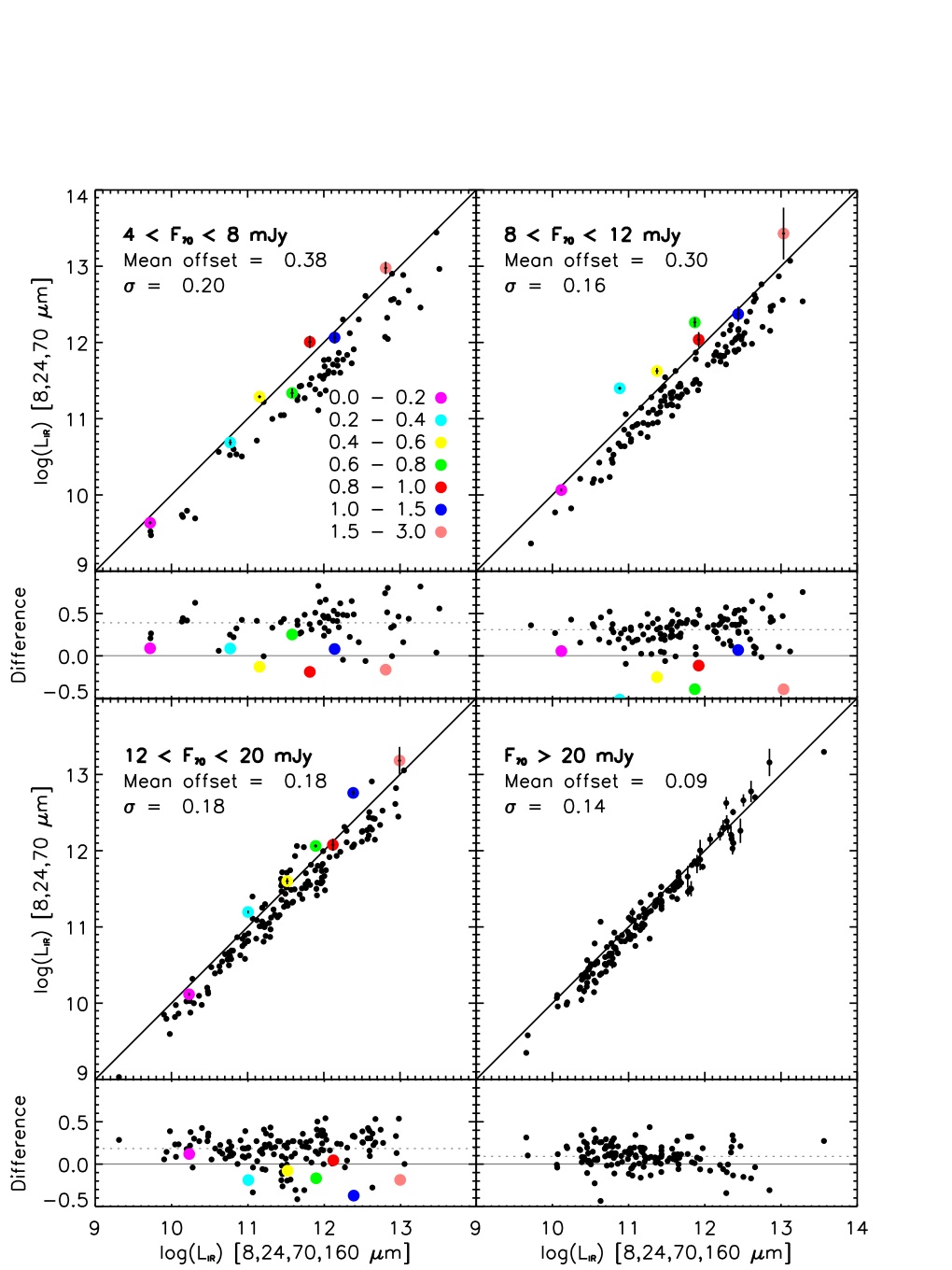}
\caption{Same as Figure~\ref{L_flux} but with the luminosities obtained by fitting the mean SED for the 1042 sources not detected at 160\ts${\mu}$m with stacked 160\ts${\mu}$m flux measurements divided into several flux and redshift bins (shown as colored points). Note that the while the stacked points show the same amount of scatter overall between the two luminosity estimates, they do not have the same systematic offset as the 160\ts${\mu}$m detected sources. This suggests that there is an intrinsic difference in the shape of the SEDs between sources with a 160\ts${\mu}$m detection and those without.}
\label{stack}
\end{figure*}

\subsection{Sources not detected at 160\ts${\mu}$m}

For the remaining 1042 sources without a detection at 160\ts${\mu}$m,  the relationship shown in Figures~\ref{lir}--\ref{L_z} indicates that an empirical correction may be needed. However, such a correction factor may not be warranted if those sources detected at 160\ts${\mu}$m are intrinsically different from those not detected. Figure~\ref{160} shows the distribution of 70\ts${\mu}$m fluxes for these two subsamples. The sources detected at 160\ts${\mu}$m are clearly weighted toward the bright end of the flux distribution. 

\begin{figure*}
\epsscale{1}
\plotone{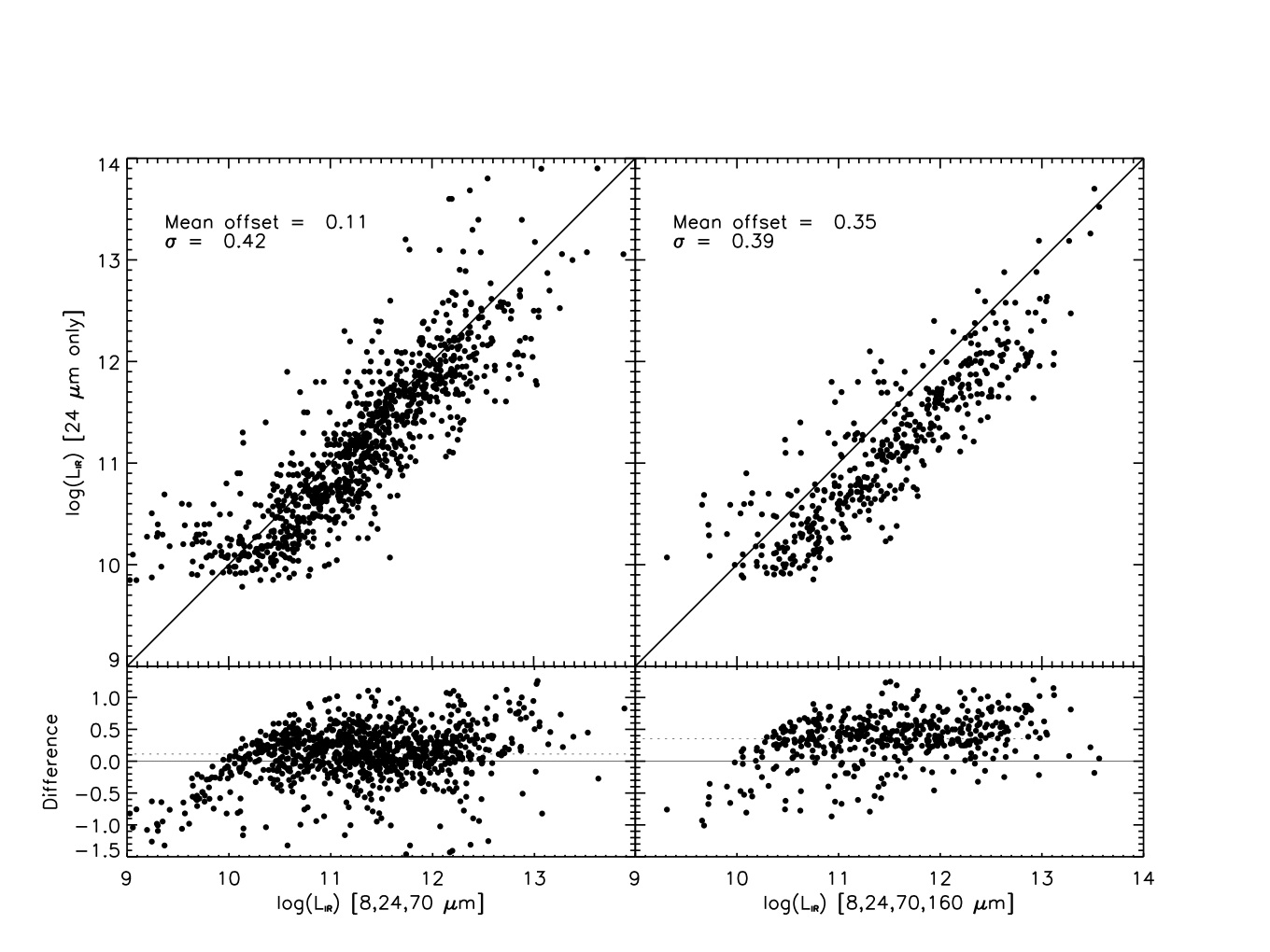}
\caption{Comparison of luminosities obtained with all of the available bands for the sources without 160\ts$\mu$m detections (left) and those with 160\ts$\mu$m detections (right) to the luminosities obtained from just  the 24\ts$\mu$m data point. The dotted line represents the mean difference between the two and the $\sigma$ value is the internal dispersion about that mean. The sources detected at 160\ts${\mu}$m show a similar offset in their luminosities as they did in the 70/160\ts${\mu}$m luminosity comparison (see Fig.~\ref{lir}. This is further indication that the sources detected at 160\ts${\mu}$m are intrinsically different (i.e., have a colder dust component) than those that are not.}
\label{L24}
\end{figure*}

\begin{figure*}
\plotone{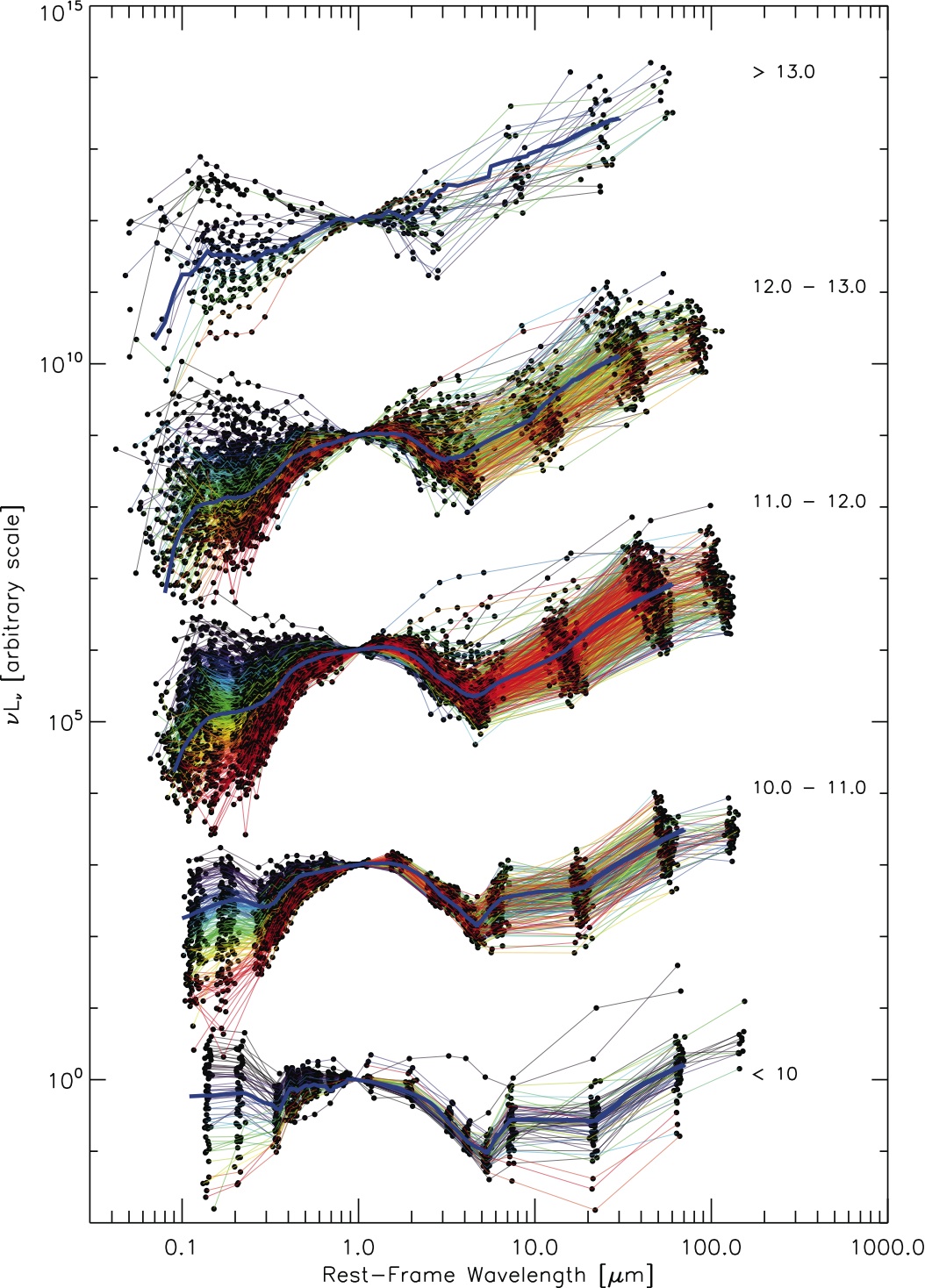}
\caption{Spectral energy distributions of all 1503 70\ts$\mu$m sources binned by luminosity and normalized at 1\ts$\mu$m. The SEDs are color coded such that the ones with the bluest slope in the UV-optical are purple/blue and the reddest are red and the spectrum in between shows the range of SED colors. The median SED for each bin is overplotted in blue. Note the large variety in SED shapes, particularly in the UV, and the trends with $L_{\rm IR}$. At the highest luminosities there is an obvious trend toward power law SEDs all the way from the optical through the FIR.}
\label{sed_L}
\end{figure*}

To further examine the properties of the sources not detected at 160\ts${\mu}$m, we fit templates to their mean SED using a stacked 160\ts${\mu}$m data point in several redshift and flux bins in the same manner as the detected sources. The resulting comparison is shown as the colored points in Figure~\ref{stack}. While these stacked points show a similar amount of scatter as the 160\ts${\mu}$m detected sources, they do not show the same systematic offset. For the stacked sources, the luminosity obtained with the 160\ts${\mu}$m point is in agreement with the luminosity obtained without it. This suggests that the sources detected at 160\ts${\mu}$m are indeed biased toward colder sources. Therefore, a correction to the $L_{\rm IR}$ for the rest of the 70\ts${\mu}$m sample is not necessary. Instead, we include the scatter in the relationship as an additional source of uncertainty and include this in our error bars in Figure~\ref{lir_all}. Thus the typical uncertainty for sources detected at 160\ts${\mu}$m is 0.05 dex while for those not detected it is $\sim 0.2$ dex.
	
\begin{figure*}
\plotone{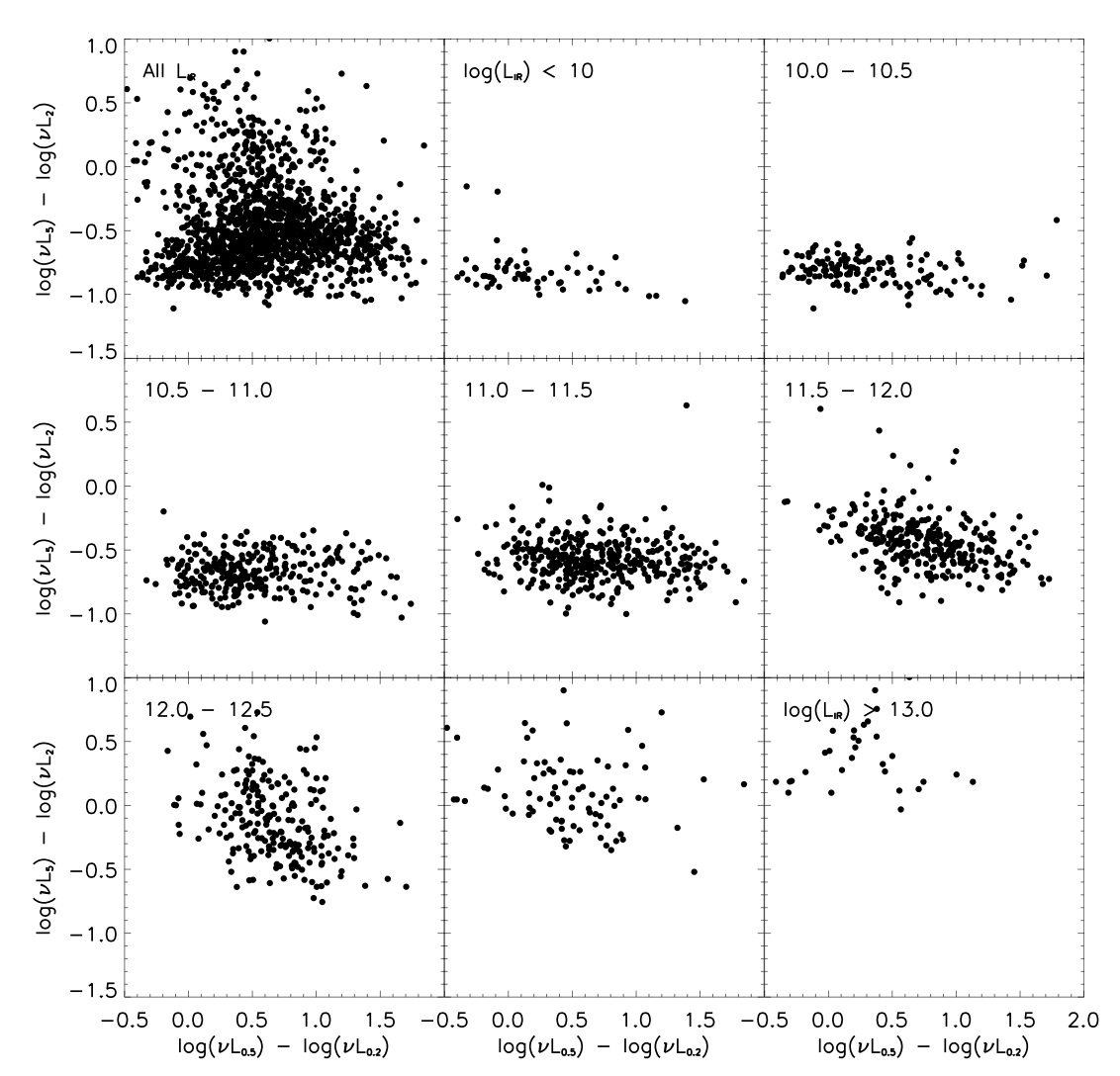}
\caption{Rest frame NIR spectral index ($\alpha^{5}_{2}$) versus UV/optical spectral index ($\alpha^{0.5}_{0.2}$) divided into $L_{\rm IR}$ bins. Note the large range spanned by both indices. $\alpha^{5}_{2}$ increases with $L_{\rm IR}$ as more sources become dominated by a power-law SED while $\alpha^{0.5}_{0.2}$ spans a large range at all luminosities.}
\label{color1}
\end{figure*}

\begin{figure*}
\plotone{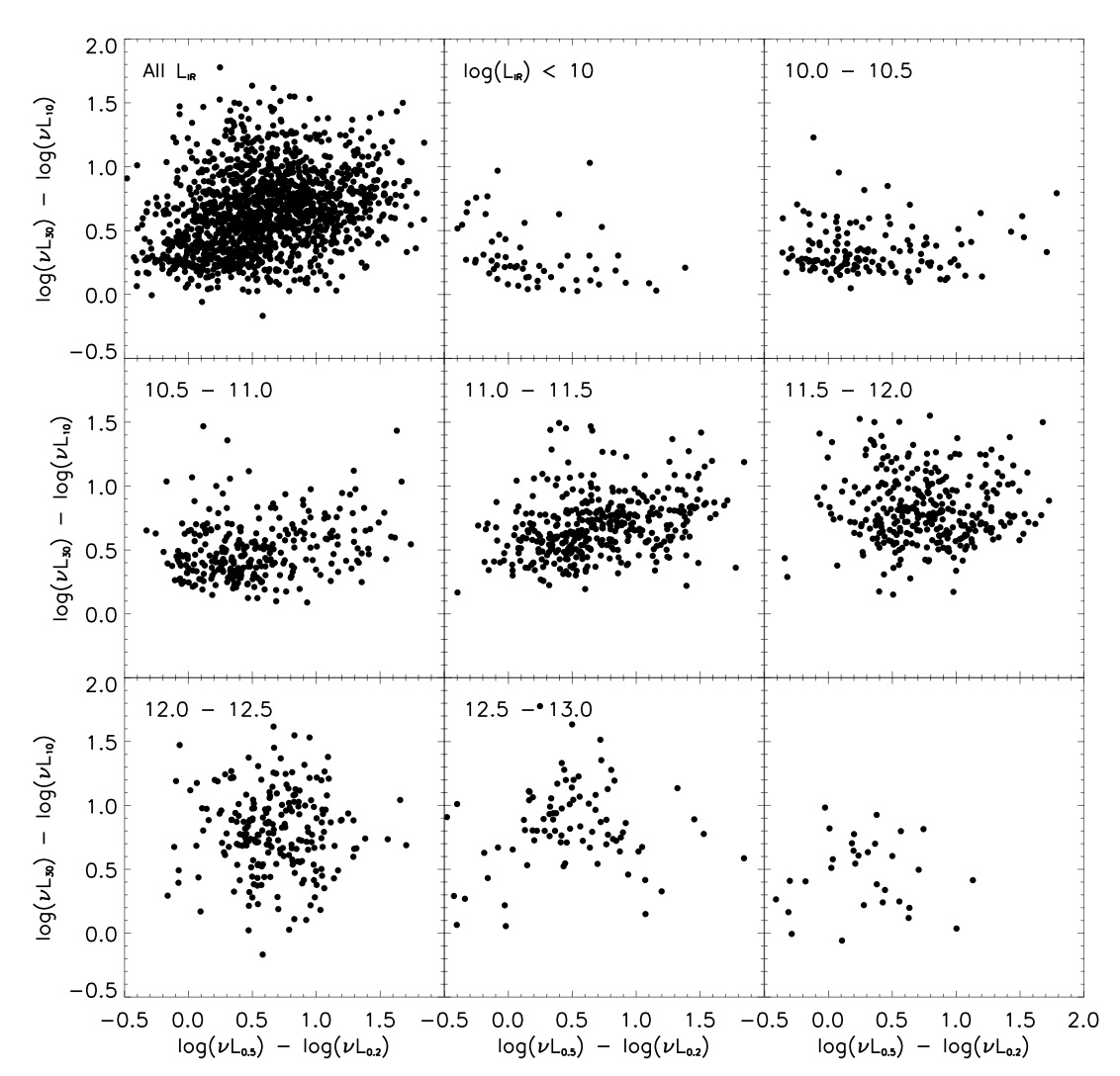}
\caption{Rest frame MIR spectral index ($\alpha^{30}_{10}$) versus UV/optical spectral index ($\alpha^{0.5}_{0.2}$) divided into $L_{\rm IR}$ bins. Note that $\alpha^{30}_{10}$ appears to span a larger range in values with increasing $L_{\rm IR}$.}
\label{color2}
\end{figure*}

\begin{figure*}
\plotone{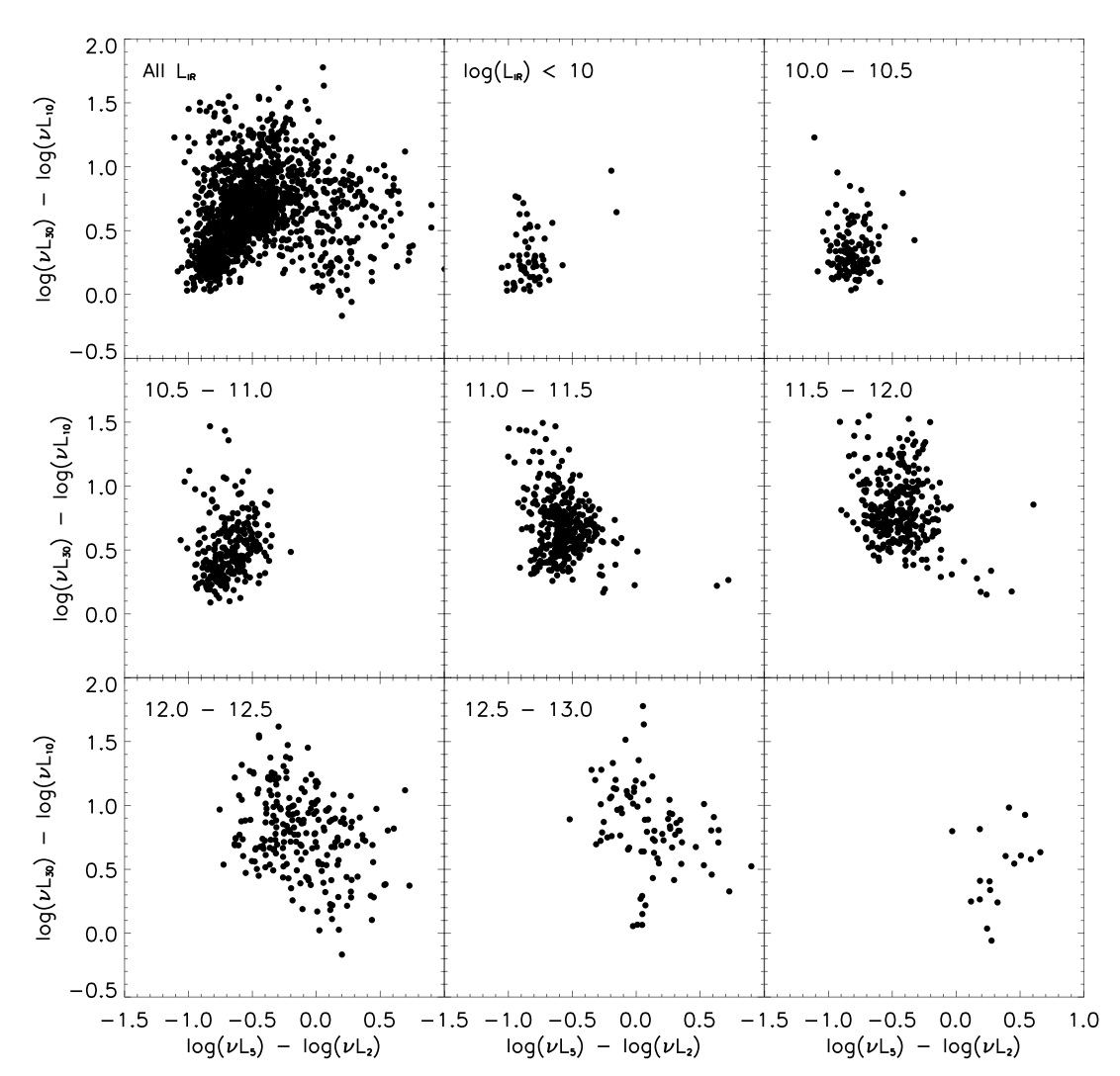}
\caption{Rest frame MIR spectral index ($\alpha^{30}_{10}$) versus NIR spectral index ($\alpha^{5}_{2}$) divided into $L_{\rm IR}$ bins. Notice that at higher luminosities ($L_{\rm IR} > 10^{11}\ts L_{\odot}$) a second branch appears with large values of $\alpha^{5}_{2}$ (power-law SEDs) and intermediate values of $\alpha^{30}_{10}$.}
\label{color3}
\end{figure*}

\subsection{Estimates based on 24\ts${\mu}$m}

Since MIPS is the most sensitive at 24\ts${\mu}$m, samples selected at this wavelength are very useful and will continue to be used in the future. Therefore, it is essential to understand how well their total infrared luminosities can be determined with just this data point. To test this, we obtained the best fit template using just the 24\ts${\mu}$m data point. Since there is only one point, the templates are not really fit but instead the best match to the 24\ts${\mu}$m flux at the source redshift is obtained from each of the template libraries without allowing for rescaling. For the final luminosity estimate we chose to use the \citeauthor{Siebenmorgen:2007p2697} library since it was the best match for the vast majority of sources using 70 and 160\ts${\mu}$m and has the largest variety of SED shapes. Figure~\ref{L24} shows the comparison between luminosity estimates using all available bands and using only the 24\ts${\mu}$m point for both 160\ts${\mu}$m detected and non-detected sources. The scatter in this relationship is much larger ($\sigma \sim 0.4$ dex) and the difference can be up to 1\ts dex for some sources. A larger offset is also seen for the sources detected at 160\ts$\mu$m than those not, consistent  with the results discussed in the previous section. The large scatter indicates that luminosity estimates for samples based solely on 24\ts${\mu}$m detection alone can be off by a significant amount and many LIRG and ULIRG samples in the literature based on 24\ts${\mu}$m may actually be contaminated by lower luminosity sources since their number densities are higher though it is possible that this effect is mitigated by the slight offset ($\sim 0.1$ dex on average) obtained without the 70\ts$\mu$m data point.

\section{Discussion}
In this paper we have presented a sample of 1503 70\ts${\mu}$m sources in the COSMOS field and determined a reliable measure of their total infrared luminosity. We believe that this is the largest sample of 70\ts${\mu}$m sources to date. When the source is detected at 160\ts${\mu}$m we obtain an estimate of $L_{\rm IR}$ accurate to within 0.05 dex on average and 0.2 dex for those without a 160\ts${\mu}$m detection. This result is much better than can be obtained from a 24\ts${\mu}$m data point alone ($\sim 0.5$ dex). In this section, we discuss in more detail the UV-to-FIR SEDs and the variations of the mean SED shape with $L_{\rm IR}$. In particular we explore three spectral regions (the UV, NIR, and MIR) since these have been found to be the most sensitive to the underlying power source (i.e., starbursts and AGN). We make use of the multiwavelength properties of the sample to identify a large population of AGN candidates. 

\subsection{Analysis of SEDs}

Figures~\ref{sed_z} and \ref{sed_L} show the full range of spectral energy distributions that we observe for the 70\ts ${\mu}$m selected sample of galaxies binned by redshift and $L_{\rm IR}$ with the median SED in each bin over plotted as a blue line.  The median SEDs are similar to sources observed locally for sources at low luminosity (e.g., Dale \etal 2007: SINGS, $L_{\rm IR} < 10^{11}\ts L_{\odot}$) and at high luminosity  (e.g., U \etal 2009, in preparation: RBGS, $L_{\rm IR} > 10^{11}\ts L_{\odot}$). All of these studies have shown that the median UV-to-FIR SEDs have two prominent thermal peaks (the optical stellar peak and the FIR dust peak) and, in addition, a range of spectral shapes, particularly in the UV (0.2--0.5\ts${\mu}$m), NIR (2--5\ts${\mu}$m), and MIR (10--30\ts${\mu}$m). With the large number of sources in our 70\ts${\mu}$m sample we can finally begin to quantify the full extent of these variations and explore how different parts of the SED correlate with each other and with $L_{\rm IR}$.  As should already be known from studies of smaller samples of local objects (e.g., RBGS: \citealt{Sanders:2003p1575}, 1 Jy ULIRG Survey: \citealt{Kim:1998p3280}), it is clear that a single archetypical template for a given infrared luminosity (e.g., Arp 220 or Mrk 231 for ULIRGs) would not be applicable for all objects with that luminosity.

The median SEDs clearly show several general features. First, most of them show the 1.6\ts ${\mu}$m peak of the stellar bump. This feature becomes less pronounced at higher luminosities. The SEDs also display a prominent 5\ts ${\mu}$m dip at low luminosities which becomes less clear at high luminosities.  The 7.7\ts ${\mu}$m PAH emission feature is also evident at $L_{\rm IR} <10^{11}\ts L_{\odot}$ near 8\ts${\mu}$m but becomes less evident in the higher luminosity bins as rest frame 7.7\ts${\mu}$m is no longer probed. However, in the highest luminosity bins, 24\ts${\mu}$m approaches rest-frame 7.7\ts${\mu}$m and this enhancement is still not observed, possibly due to the median SED being dominated by AGN at those luminosities. Many of the SEDs show an increase in emission toward the UV, indicative of the big blue bump \citep{Malkan:1982p3097, Sanders:1989p3181,Elvis:1994p5581} seen in many AGN. The GALEX data points, particularly at high redshift where they probe further into the FUV, point to a broad maximum for many of these sources.

To provide a more quantitative measure of the spread in SED shapes, we compute three spectral indices (logarithmic rest-frame $\nu L_{\nu}$ ratios) probing three different wavelength regimes: UV/optical from 0.2 to 0.5\ts$\mu$m ($\alpha^{0.5}_{0.2}$), NIR from 2 to 5\ts$\mu$m ($\alpha^{5}_{2}$), and MIR from 10 to 30\ts$\mu$m ($\alpha^{30}_{10}$). These spectral indices are plotted against each other in Figures~\ref{color1}--\ref{color3} binned by $L_{\rm IR}$.  All three of these plots show strong trends with $L_{\rm IR}$.   $\alpha^{5}_{2}$ increases with $L_{\rm IR}$ as sources become dominated by power-law SEDs (see \S5.2). This is also evident in Figure~\ref{color3} where the sources with high  $\alpha^{5}_{2}$ form a separate distinct branch at intermediate values of  $\alpha^{30}_{10}$. For galaxies with low values of  $\alpha^{5}_{2}$,  $\alpha^{30}_{10}$ is low at low luminosities and steadily increases with $L_{\rm IR}$.  $\alpha^{0.5}_{0.2}$ spans a wide range at all $L_{\rm IR}$ (a range of nearly 3 orders of magnitude) and has lower values on average at the most extreme luminosities.

\begin{figure*}
\epsscale{1.1}
\hspace*{-0.5in}
\plotone{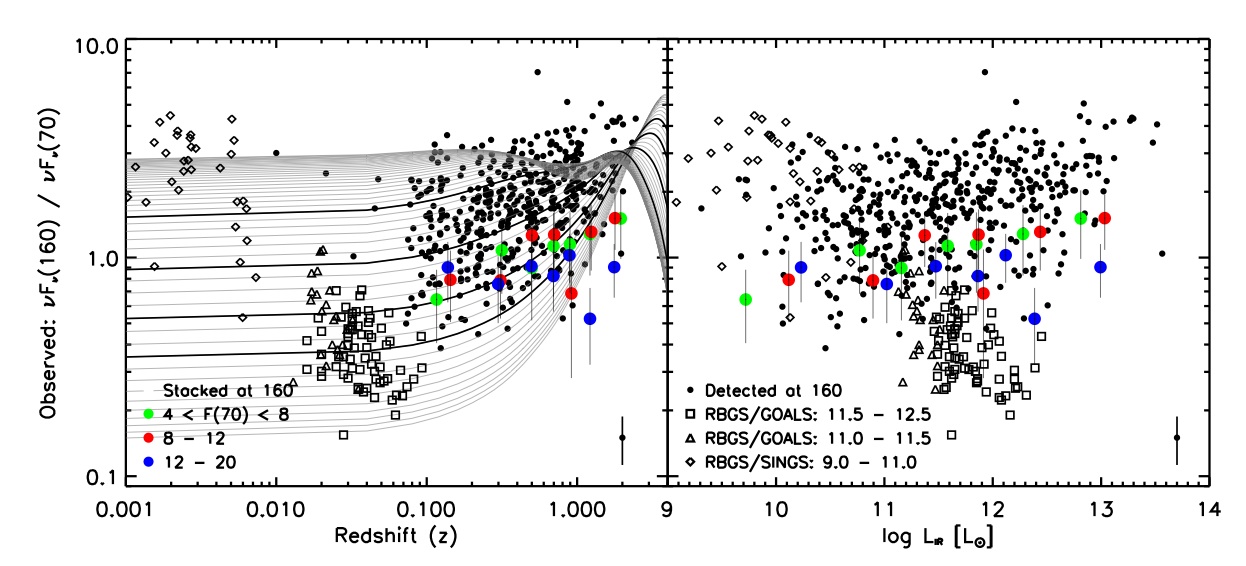}
\caption{Observed far-infrared colors ($\nu F_{\nu}(160)/ \nu F_{\nu}(70)$) as a function of redshift (left) and $L_{\rm IR}$ (right). A typical $1\ts\sigma$ error bar is shown in the corner. The colored points represent the stacked data points (described in \S4.4) for those sources not detected at 160\ts $\mu$m divided into several flux and redshift bins. Local infrared sources from the RBGS are over plotted with their FIR colors obtained from MIPS observations from SINGS ($10^{9}<L_{\rm IR}<10^{11}\ts L_{\odot}$: open diamonds) and GOALS ($10^{11}<L_{\rm IR}<10^{11.5}\ts L_{\odot}$: open triangles; $10^{11.5}<L_{\rm IR}<10^{12.5}\ts L_{\odot}$: open squares). The gray lines represent the redshift evolution of empirical templates of local objects from \cite{Dale:2002p2130} in the luminosity range $10^{8.3}<L_{\rm IR}<10^{13.5}\ts L_{\odot}$. while the black lines highlight the luminosities at $10^{9}, 10^{10}, 10^{11}$, and $10^{12}\ts L_{\odot}$ from top to bottom.}
\label{fir_obs}
\end{figure*}

\begin{figure*}
\epsscale{1.1}
\hspace*{-0.5in}
\plotone{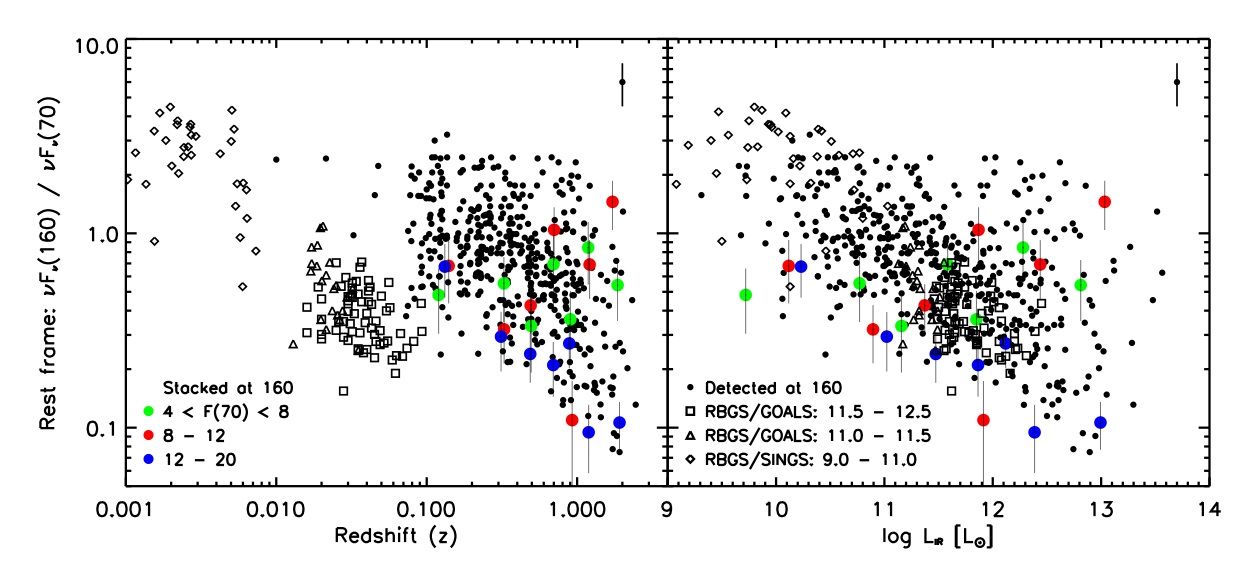}
\caption{Same as Fig.~\ref{fir_obs} but converted into rest-frame colors using the best fit template to the infrared. Overall, the COSMOS sources span the same range in FIR colors as the local sources, however, at high luminosities there are a number of sources with colder SEDs than would be expected based on local sources. This may be evidence for a population of sources with colder SEDs at high redshift, however, at these redshifts the 160\ts${\mu}$m data point is not long enough to constrain the peak of the SED nor to obtain a precise dust temperature. The large range in SED templates that fit these data points is indicated by the fact that these sources are the same ones with a large scatter in Fig.~\ref{lir_sym}.}
\label{fir_rf}
\end{figure*}

\subsubsection{160/70\ts${\mu}$m Flux Ratios}

In addition to the SED shapes described above, our large 70\ts${\mu}$m sample and the substantial number of detections at 160\ts${\mu}$m makes it particularly useful to investigate the 160/70\ts${\mu}$m flux ratios as a crude measure of the characteristic dust temperature for the FIR peak emission. Figure~\ref{fir_obs} shows the observed FIR colors ($\nu F_{\nu}(160)/ \nu F_{\nu}(70)$) as a function of redshift and $L_{\rm IR}$ for all 70\ts${\mu}$m sources detected at 160\ts${\mu}$m. For those not detected at 160\ts${\mu}$m, the colored points on the plot represent the colors obtained from their stacked 160\ts${\mu}$m fluxes. To put these values in context we also over plot the flux ratios for local IRAS 60\ts${\mu}$m selected sources from the RBGS over a wide range of $L_{\rm IR}$ ($10^8-10^{12.5}\ts L_{\odot}$) with MIPS observations obtained as a part of GOALS or SINGS (the Spitzer Infrared Nearby Galaxies Survey: \citealt{Kennicutt:2003p5199}; \citealt{Dale:2007p5190}). A typical error bar for the sources detected at 160\ts${\mu}$m is shown in the corner. The overall scatter of the data points remains the same even with a cut at larger $S/N$, indicating that the observed trends are likely real. The local sources span a large range in $\nu F_{\nu}(160) / \nu F_{\nu}(70)$ from 0.2--5 and show a strong trend with $L_{\rm IR}$. Also over plotted in gray are a set of values obtained from the \cite{Dale:2002p2130} templates spanning an $L_{\rm IR}$ range of  $10^{8.3}-10^{13.5}$ from top to bottom as a function of redshift. These templates span most of the range occupied by the COSMOS 70\ts${\mu}$m selected sources (the full set of libraries covers the entire range of colors observed at each redshift but we show just this one library for illustration). The sources not detected at 160\ts${\mu}$m span the lower portion of the range occupied by the sources detected at 160\ts${\mu}$m. This result is expected if the 160\ts${\mu}$m detection preferentially selects colder sources than the rest of the sample.

We obtained rest-frame $\nu F_{\nu}(160) /  \nu F_{\nu}(70)$ values by using the 70 and 160\ts${\mu}$m fluxes determined from the best-fit IR template used to estimate $L_{\rm IR}$. These values are plotted in Figure~\ref{fir_rf} as a function of redshift and luminosity. As a note of caution, since these values are dependent on the best-fit SED template this adds an extra source of uncertainty, particularly at the high redshift end (where shorter rest-frame wavelengths are observed). We find that the COSMOS sources span the same range of colors (and thus dust temperatures) as the local sources but that there is an apparent excess of colder sources at higher luminosities, similar to the result of \cite{Symeonidis:2009p5203}.  However, our interpretation is that this excess is likely a result of the longer wavelength selection (rest-frame 80\ts${\mu}$m at $z\sim1$ versus 60\ts${\mu}$m locally) and the fact that the peak of emission is not well constrained at these redshifts and thus a wide range of templates can fit the data points. Future surveys at FIR and submillimeter wavelengths with {\it Herschel} and SCUBA2 will be able to disentangle the temperature dependent selection effects and investigate the dust temperature distributions of high redshift galaxies.

\subsection{AGN Diagnostics}

Both X-ray and radio detections are traditional methods used to identify potential AGN in high-redshift galaxies.  The vast majority of the 70\ts${\mu}$m selected galaxies that are detected in the X-ray (86\%) have an X-ray luminosity greater than $10^{42} \ts \rm erg\ts s^{-1}$ and are therefore bona fide AGN. The 22 lower luminosity sources are all at low redshifts ($z<0.35$).  Most of the X-ray detected AGN have power law SEDs (though not all), particularly at high $L_{\rm IR}$. Many sources that are not detected as AGN in the X-ray also have a power-law SED. These are possibly heavily obscured (Compton thick) AGN that are opaque even to hard X-ray emission \citep{Polletta:2006p5580}.

\begin{figure*}
\epsscale{0.9}
\plotone{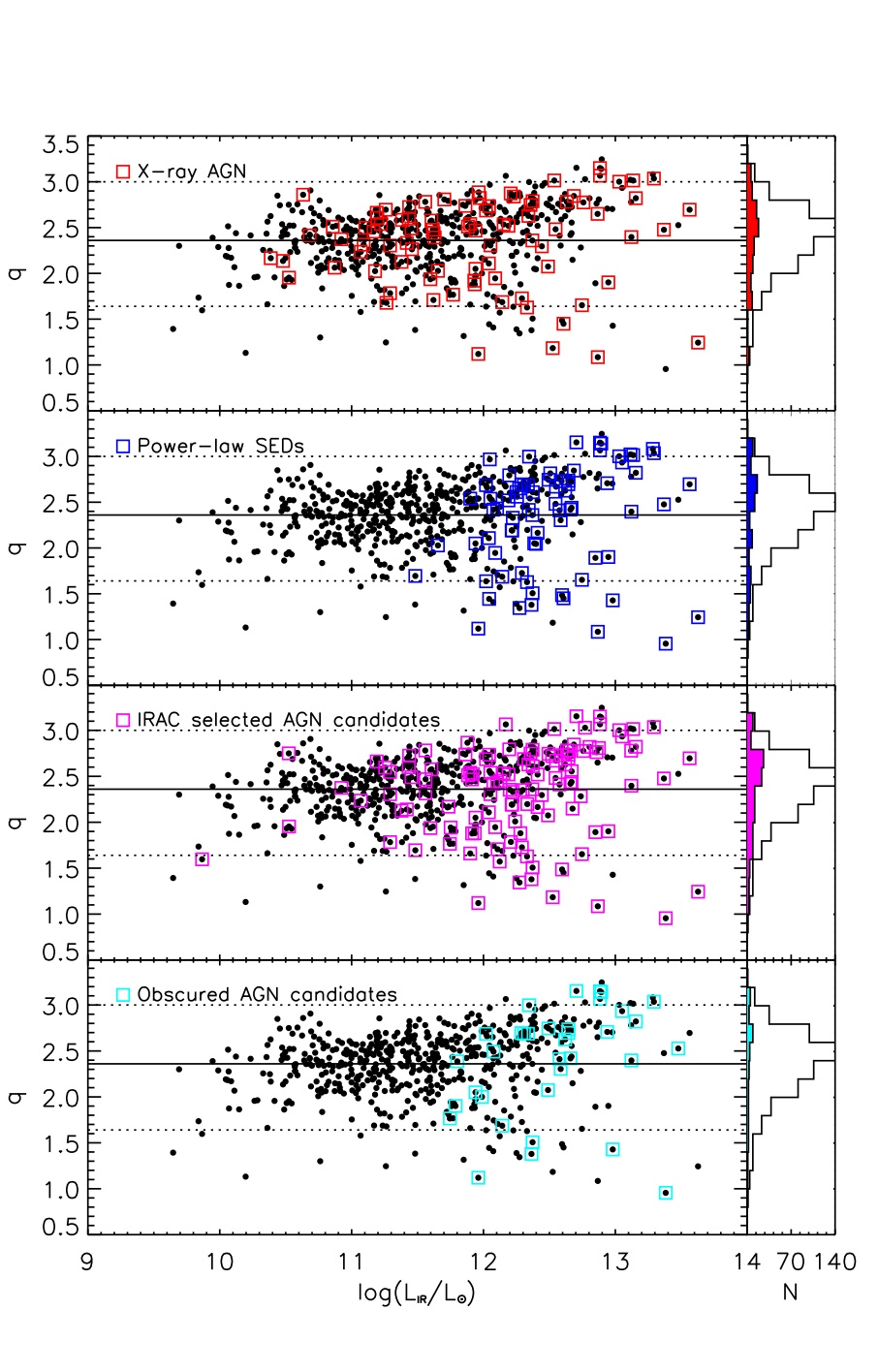}
\caption{Radio/infrared flux ratio ($q$) vs. $L_{\rm IR}$ (same as Figure~\ref{q}) with X-ray detected AGN (red), objects with power-law SEDs (blue), IRAC color selected AGN candidates (magenta), and obscured AGN candidates (cyan) over plotted from top to bottom. The solid line indicates the median value of $q$ (2.34) for the entire sample and the dotted lines represent the dividing lines for radio ($q<1.64$) and infrared ($q>3.0$) excess sources following Yun \etal (2001). The panels on the right show the distribution of the entire sample with radio detections and the different AGN selections over plotted in color.}
\label{q_compare}
\end{figure*}

Since $\alpha^{5}_{2}$ is a measure of the $\nu\ts F_{\nu}$ slope between 2 and 5\ts${\mu}$m it is therefore a good way to select power law galaxies. Anything above zero on the y-axis of Figure~\ref{color1} or the x-axis of Figure~\ref{color3} has the shape of a power-law in the NIR, i.e., these objects do not have a strong 1.6\ts${\mu}$m bump nor a dip at 5\ts${\mu}$m. The  $\alpha^{5}_{2}>0$ cutoff here corresponds to a $F_{\nu}$ spectral slope of 0.4 (similar to what is used as a power law slope cutoff in the literature, e.g., \citealt{AlonsoHerrero:2006p2013} and \citealt{Donley:2007p3092}). X-ray detected AGN span almost the entire area of the spectral index plots rather than being clearly separated the way the power law galaxies are. However, they do on average have lower $\alpha^{30}_{10}$ as AGN in the local universe do. 

We select all sources with $\alpha^{5}_{2}> 0$ to be power-law AGN candidates. A total of 166 (11.2\%) objects satisfy this criterion. Forty-five of these are detected in X-ray ($\sim 27\%$; a slightly lower fraction than found by \citealt{AlonsoHerrero:2006p2013} and \citealt{Donley:2007p3092} but consistent with the shallower COSMOS X-ray depths).  A few differences between power-law galaxies detected in the X-ray and those not detected in the X-ray can be seen in their SEDs. Those sources detected in the X-ray tend to display the full range of properties in the UV where only one or two sources that are not X-ray detected appear bright in the UV. The prominence of the UV emission in the X-ray detected power-law galaxies could be evidence that these sources tend to be less obscured than the non X-ray detected sources and hence are transparent to UV (and X-ray) emission. The SEDs of the non-X-ray detected power-law galaxies also tend to have a steeper slope.

We use the infrared-radio correlation as another way to investigate the energy source. Using the limits defined in \cite{Yun:2001p2885} (5 times larger radio or IR flux than expected from the infrared-radio correlation) for radio excess ($q<1.64$) and infrared excess ($q>3.0$) objects we find many objects that fall into these two categories. In total there are 27 radio excess sources, 6 of which (22\%) are also detected as X-ray AGN (as shown in Fig.~\ref{q_compare}). The fraction of objects that this population represents increases with $L_{\rm IR}$ from 4 (2\%) at $<10^{11}\ts L_{\odot}$, to 5 (2\%)  LIRGs and 16 (14\%) ULIRGs.  Radio excess sources are believed to be potential AGN hosts where the extra radio emission seen comes from either a compact radio core or from associated radio jets or lobes (Sanders \& Mirabel 1996). 

All 14 of the infrared excess objects are ULIRGs or HyLIRGs and represent 11\% of the sample at these luminosities. 6 (43\%) of them are X-ray detected AGN. A total of 25 of the obscured AGN candidates (64\%) described in \S5.2.2 are detected in the radio and have $<q>=2.54\pm0.63$. 5 of these (2 X-ray AGN) are infrared excess objects.  It is interesting to note that the IR excess objects contain a higher fraction of X-ray detected AGN than the radio excess objects. It is also interesting to note that many of the radio and infrared excess sources also have a power-law SED (see Fig.~\ref{q_compare}).

\subsubsection{IRAC Color Selected AGN}

\begin{figure*}
\epsscale{0.8}
\plotone{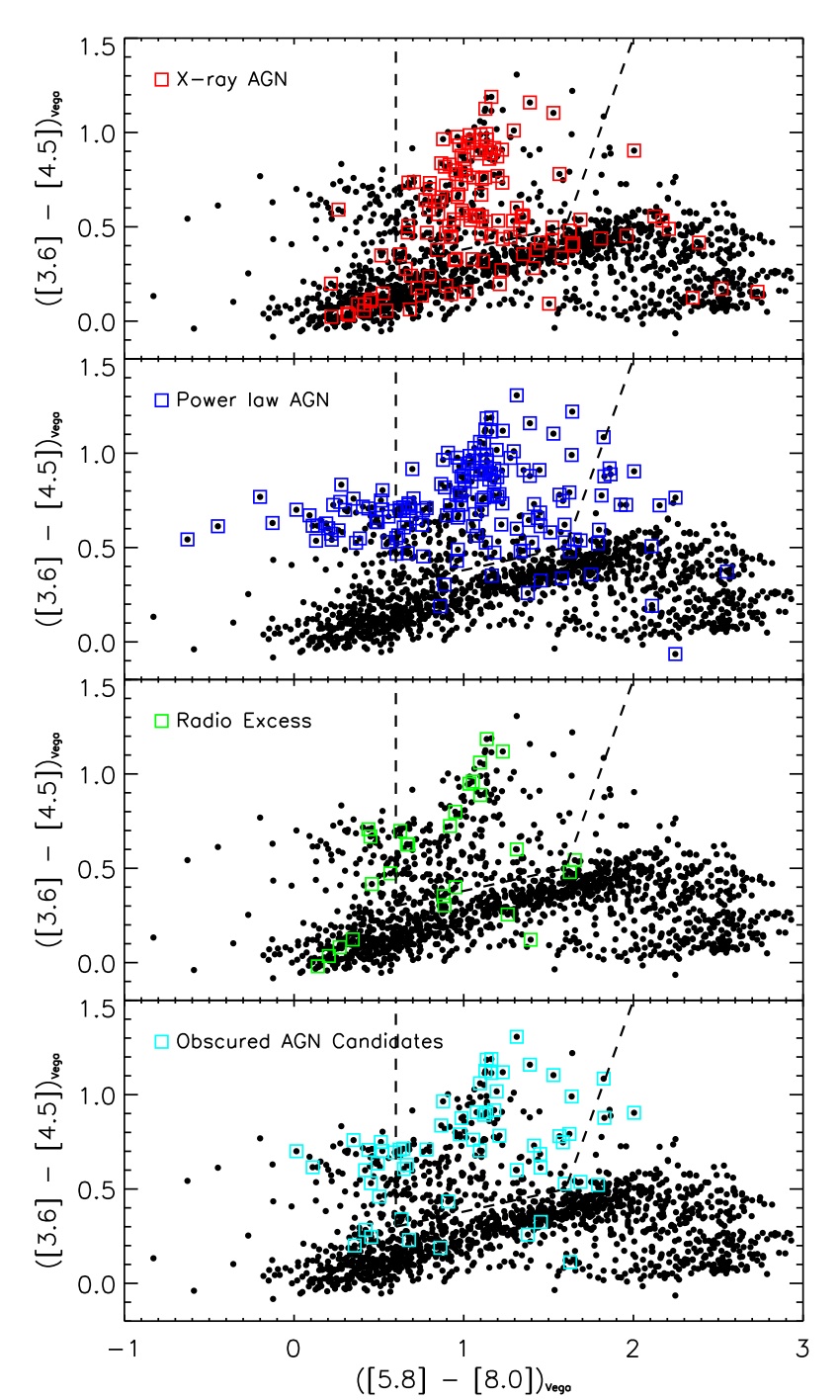}
\caption{\cite{Stern:2005p3000} IRAC color-color plot for selecting AGN candidateswith X-ray detected AGN (red), objects with power-law SEDs (blue), radio excess objects (green), and obscured AGN candidates (cyan) over plotted from top to bottom. The \cite{Stern:2005p3000} selection box is represented by the dashed lines.}
\label{irac}
\end{figure*}

Color selection in the mid-infrared has also been shown to be an effective way to select potential AGN \citep{Stern:2005p3000,Lacy:2004p3062} as these colors can separate star-forming galaxies with a stellar bump at 1.6\ts${\mu}$m from those with a power-law like SED. Also, this method can potentially find dust obscured AGN since the MIR wavelengths are not as affected by extinction as the UV and optical. Following the criteria of \cite{Stern:2005p3000} we used the color-color IRAC plot (see Fig.~\ref{irac}) to select potential AGN and compare these sources with those found using four other methods. A total of 232 sources meet the Stern \etal criteria, representing $\sim 15\%$ of the entire 70\ts${\mu}$m sample. Of these 232, 82 (35\%) are detected as AGN in the X-ray. This is 62\% of the total X-ray AGN population, which means the other 50 X-ray AGN are not selected by this method. According to Stern \etal, this method is good at selecting broad line AGN (selects $\sim 91\%$) and not as good at selecting narrow line AGN ($\sim 40\%$) with an overall contamination rate of $\sim 20\%$. \cite{Donley:2007p3092} show that this region of the plot can be contaminated by starburst dominated ULIRGs (like Arp 220) and normal star-forming galaxies. If we assume that 20\% of the 232 objects are potentially contaminants, this leaves us with 186 actual AGN, or about 1.5 times the number detected in the X-ray.  Of the potentially obscured objects described in the next section ($F(24\ts {\mu} {\rm m})/F(R) > 10^3$), 36 out of the 60 match these color criteria as do 33 out of the 56 that also have $R-K>4.5$. Most of the power law galaxies (62\%) fall in a narrow region in the \cite{Stern:2005p3000} selection box but many ($\sim 35\%$) fall outside of this region. Very few of these objects fall in the lower left corner of the selection region, indicating that this is possibly the region that contributes most of the contamination in the selection (consistent with the results of \citealt{Donley:2008p3069}).  13 of the radio excess sources are also selected by the Stern \etal criteria.

\begin{figure*}
\epsscale{1}
\plotone{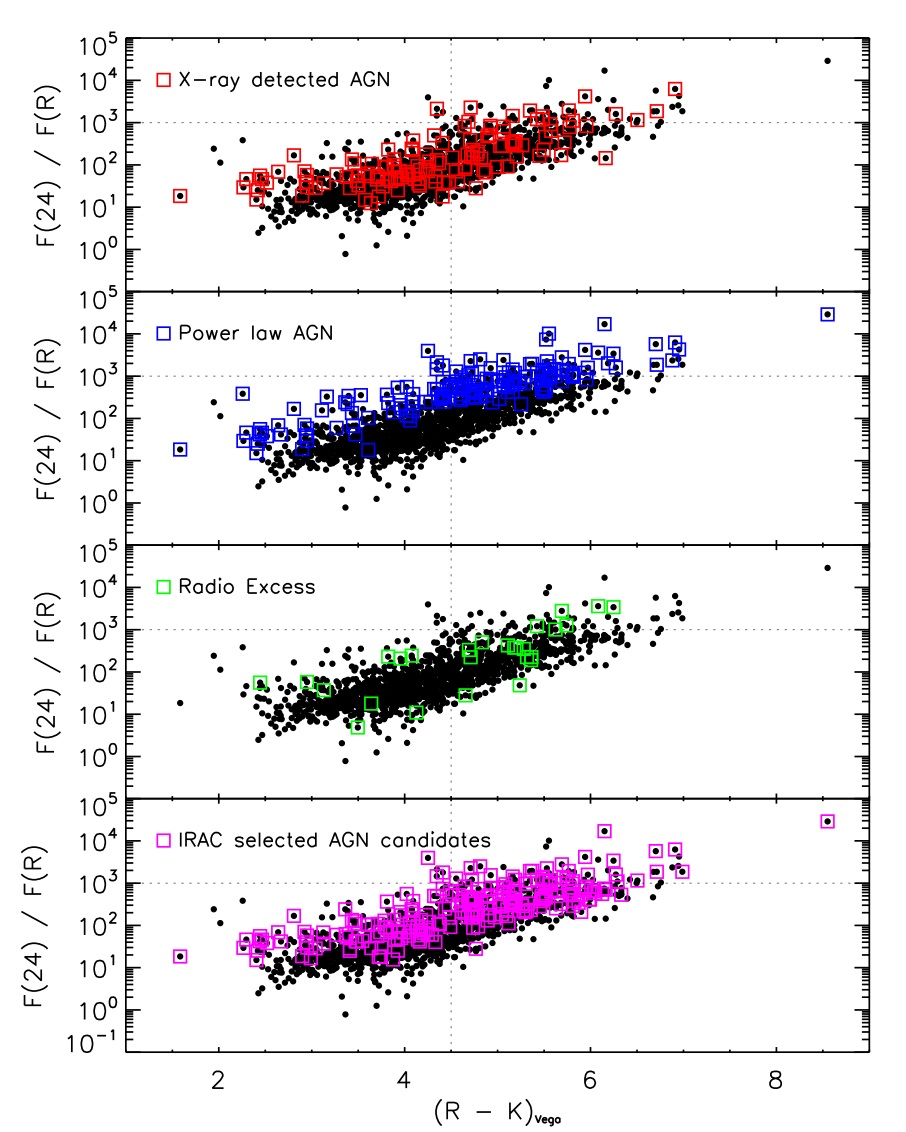}
\caption{MIR (24\ts$\mu$m) to optical (R-band) flux ratio as a function of $R-K$ color with X-ray detected AGN (red), objects with power-law SEDs (blue), radio excess objects (green), and IRAC color selected AGN candidates (magenta) over plotted from top to bottom. The dotted lines represent the $F(24)/F(R)>10^{3}$ and $R-K>4.5$ selection cuts from Fiore \etal 2008 for obscured AGN candidates.}
\label{obscure}
\end{figure*}

\subsubsection{Obscured AGN Candidates}

Fiore \etal (\citeyear{Fiore:2008p2961}, \citeyear{Fiore:2009p2143}) have shown that obscured AGN tend to have high MIR to optical flux ratios and red $R-K$ colors. The 24\ts${\mu}$m to R-band flux ratio for each of the 70\ts${\mu}$m sources in the sample is shown in Figure~\ref{obscure} as a function of $R-K$ color with AGN selected through four other methods over plotted. The dotted lines show the cuts of $(R-K)_{\rm Vega} > 4.5$ and $\rm F(24\ts {\mu} m)/F(R) > 10^3$ defined in \cite{Fiore:2008p2961} for selecting candidate obscured AGN at high redshift.  56 of the galaxies in our sample make both of these cuts and 14 of them are detected in the X-rays.  Three are detected only in the hard band and one only in the soft band. The remaining sources have a median hardness ratio of 0.2. The redshifts of these galaxies range from 0.7 to 2.5 with the median being $z=1.5$, confirming the results of \cite{Fiore:2008p2961}. Additionally, four galaxies match the MIR to optical flux ratio criteria but are bluer than the $(R-K)_{\rm Vega} > 4.5$ magnitude cutoff. Of these, one is detected as an AGN in the X-ray with a detection only in the hard band. These four sources have redshifts that range from $z=1.2$ to 2.4 with a median redshift of 2.1 -- a similar distribution to the sources with red $R-K$ colors. The 24\ts${\mu}$m to R-band flux ratio correlates with the total infrared luminosity (as expected for obscured AGN where the luminosity from the nuclear regions is greatly reduced by the obscuring material) and those above the cutoff of $10^3$ are predominantly ULIRGs and HyLIRGs. We consider all 61 of these galaxies to be obscured AGN candidates. 

It is also worth noting that this selection is similar to other IR/optical color selections in the literature, such as dust obscured galaxies \citep{Dey:2008p2965} and infrared bright, optically faint galaxies \citep{Yan:2004p2979,Houck:2005p2986, Weedman:2006p2991}. The dust obscured galaxy color cut ($R-[24] > 14$) is actually the same as the one used here and is expected to select high redshift ($z\sim2$) extremely luminous galaxies with large column densities of dust.

\subsubsection{AGN Fraction Among Luminous and Ultraluminous Infrared Galaxies}

In the local universe, the AGN fraction of galaxies is known to increase very strongly with $L_{\rm IR}$. Using spectroscopic diagnostics of a sample of 200 luminous infrared galaxies from IRAS, \cite{Veilleux:1995p2081} found that 62\% of objects at the highest luminosities were AGN. A similar study of the 1\ts Jy sample of ULIRGs \citep{Veilleux:1999p2073} found that above $L_{\rm IR} > 10^{12.3}$ $\sim 50\%$ of objects are AGN.  \cite{Tran:2001p3271} found a similar transition luminosity of $L_{\rm IR} > 10^{12.4}\ts L_{\odot}$ between objects whose luminosities are dominated by star-burst emission and those dominated by AGN. Above this transition they find that ``most ULIRGs appear AGN-like". With the large sample of objects over the wide range of luminosities probed by our 70\ts${\mu}$m selected sample, we can address the question of whether this trend holds out to higher redshifts.

Table~\ref{xray_area} shows the total number of galaxies in each $L_{\rm IR}$ bin along with the number of X-ray detected sources over the full area, in the area covered by {\it Chandra}, and in the area with the deepest {\it Chandra} coverage. The total fraction of X-ray detected AGN increases with increasing $L_{\rm IR}$, from 5\% at $L_{\rm IR} < 10^{10}\ts L_{\odot}$ to $\sim 15\%$ for ULIRGs and $\sim 42\%$ for HyLIRGs, however, these numbers are lower limits due to the uneven X-ray coverage across the field. The fraction increases when limited to the smaller areas with deeper X-ray coverage, though the small numbers make a statistical analysis difficult. This indicates that as many as 12\% of LIRGs, 30\% of ULIRGs, and 43\% of HyLIRGs are AGN with detectable X-ray emission. For the rest of the discussion here, we use the fractions obtained over the entire area.

\begin{figure}
\epsscale{1.2}
\plotone{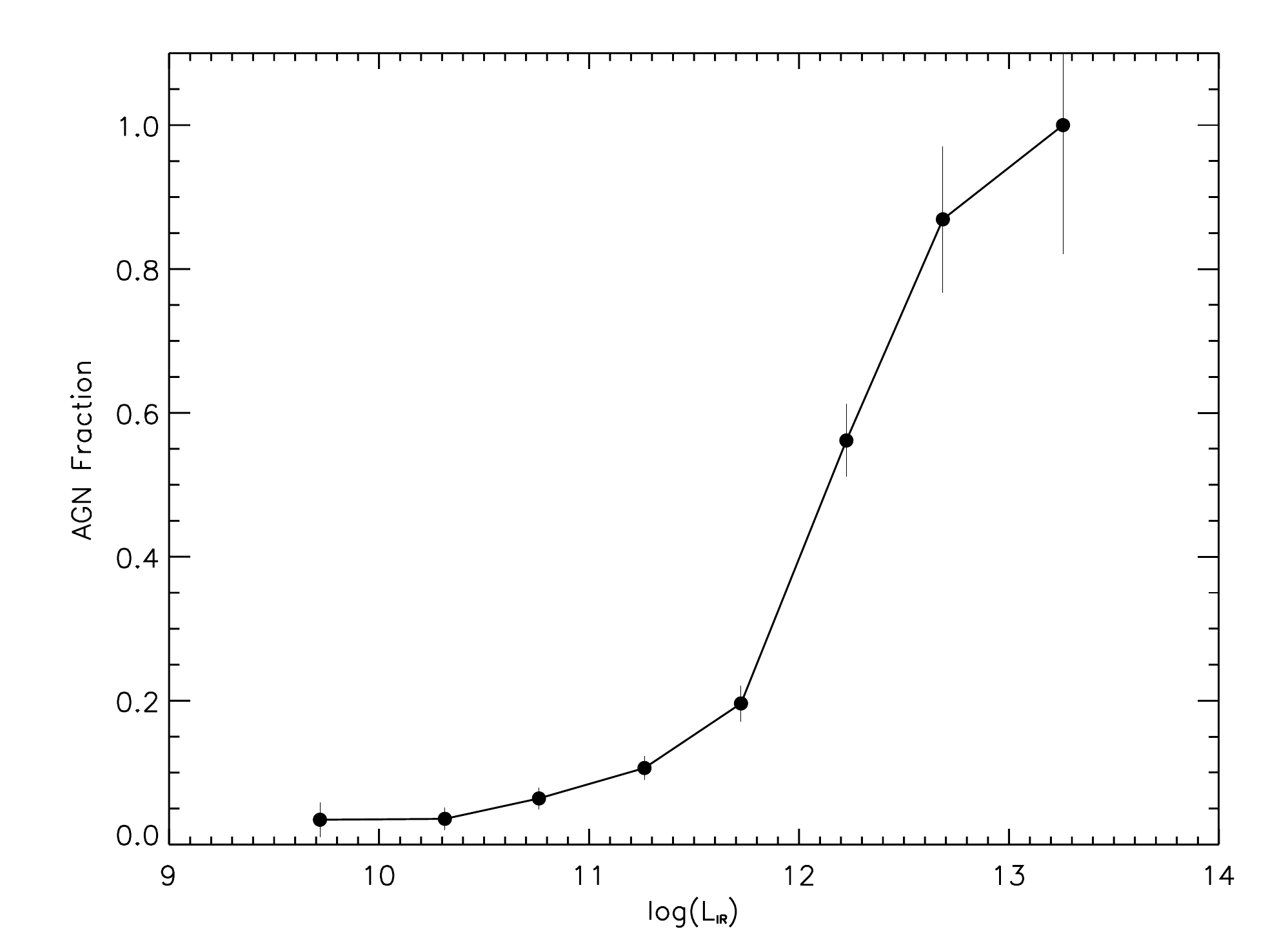}
\caption{Fraction of 70\ts${\mu}$m sources selected as AGN candidates by any one of the five different selection methods illustrated in Figs.~\ref{q_compare}--\ref{obscure} as given in Column 9 of Table~\ref{frac}. $1\ts\sigma$ Poissonian error bars are shown. Many of these sources are selected by more than one method though each source is counted only once. The fraction goes from $<5\%$ at $L_{\rm IR} < 10^{11}\ts L_{\odot}$ to 100\% at $L_{\rm IR} > 10^{13}\ts L_{\odot}$.}
\label{fraction}
\end{figure}

Table~\ref{frac} shows the fraction of candidate AGN found by each method (including X-ray detected, radio selected, IRAC color selected, power-law galaxies, and potentially obscured objects). If we add up all of the galaxies that fall into any of the possible AGN candidate categories (total column in Table~\ref{frac}) then we find that the AGN fraction ranges from 3\% at $L_{\rm IR} < 10^{10}\ts L_{\odot}$ to $\sim 70-100\%$ for ULIRGs and HyLIRGs (illustrated in Fig.~\ref{fraction}). There is likely to be some contamination in this estimate, particularly from the IRAC color selection method.  If we leave out the IRAC selected objects, the AGN fraction goes from 2\% for objects with $L_{\rm IR} < 10^{10}\ts L_{\odot}$ to 10\% for LIRGs, 51\% for ULIRGs, and 97\% for HyLIRGs. This result is in excellent agreement with the AGN fraction in the local universe, indicating that high redshift ULIRGs are indeed similar to local sources in this respect.

\begin{figure*}
\plotone{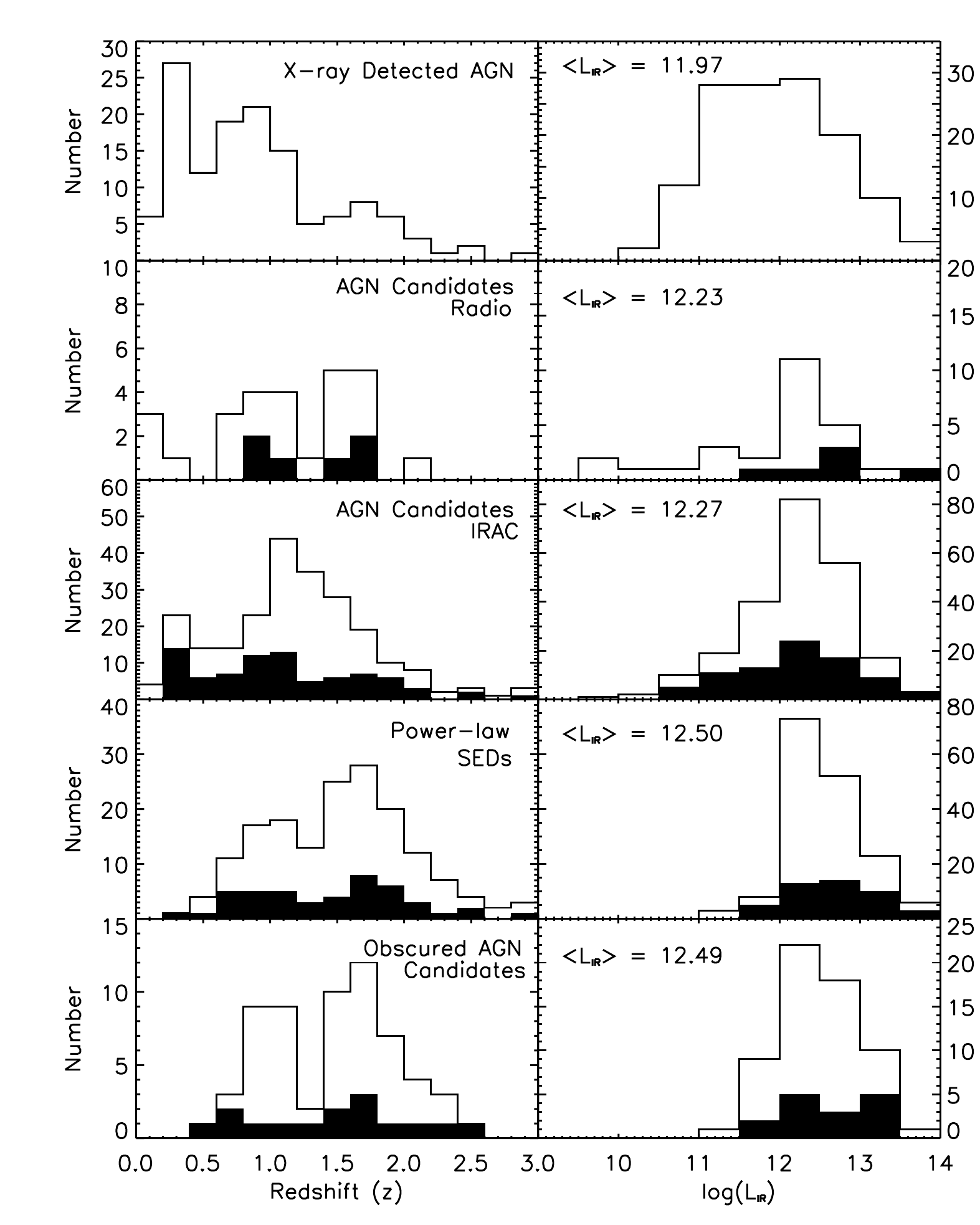}
\caption{Redshift (left) and infrared luminosity (right) histograms for AGN selected by each of the five different selection methods: (from top down) X-ray selected, radio excess sources, IRAC color selected sources, sources with power law SEDs, and obscured AGN candidates. The filled histograms represent the objects in each method that are also detected in the X-ray. The median $L_{\rm IR}$ is plotted for each method.}
\label{hists}
\end{figure*}

The redshift and $L_{\rm IR}$ distribution for the AGN candidates found using each method is shown in Figure~\ref{hists}. Those objects that are also selected in the X-ray are shown by the filled histogram.  It is interesting to note that while the X-ray selection finds objects at a wide range in redshifts and $L_{\rm IR}$, each of the other methods has a narrower range and typically identifies objects at higher luminosities.  The IRAC color selection finds two peaks in the redshift distribution, one at $z\sim0.35$ and the other at $z\sim1.15$. For the radio, IRAC, and power-law selections the X-ray seems to detect objects over the full range of redshifts and $L_{\rm IR}$.  The majority of the power-law objects are ULIRGs or HyLIRGs with only a few ($\sim 10$) with LIRG luminosities.

\section{Summary}

We have presented a large robust sample of 1503 unconfused 70\ts${\mu}$m selected galaxies with $S/N>3$ (approximate flux limit of 6.5\ts mJy)  from the 1.8 deg$^{2}$ ACS-COSMOS field. These sources span a redshift range of $0.01<z<3.5$ with a median redshift of 0.5 and an infrared luminosity range of $10^{8} < L_{\rm IR} < 10^{14}\ts L_{\odot}$ with a median luminosity of $10^{11.4}\ts L_{\odot}$ This is the largest sample of 70\ts${\mu}$m sources to date. The large multiwavelength data set available for the full COSMOS field has allowed us to compile the complete spectral energy distributions for each object  from the ultraviolet to the far-infrared and obtain an estimate of the total infrared luminosity. Spectroscopic redshifts are available for $\sim 40\%$ of the sample and for the remainder we use photometric redshifts with a precision of $0.02\times(1+z)$. The coverage in the X-ray, radio, and IRAC bands have allowed us to identify a large sample of potential AGN. From the analysis of the multiwavelength properties of these objects, the following conclusions can be drawn:

\begin{enumerate}

\item{The long wavelength {\it Spitzer}-MIPS 70 and 160\ts${\mu}$m data are essential for obtaining a more accurate and reliable estimate of the total infrared luminosity  than was previously possible using 24\ts$\mu$m data alone. The typical uncertainty in $L_{\rm IR}$ when 160\ts$\mu$m is detected is $\sim 0.05$ dex while when it is not detected it is $\sim 0.2$ dex. Using only the 24\ts${\mu}$m data point we find the dispersion is significantly increased to $\sim 0.5$ dex.}

\item{The shape of the mean SED for the 70\ts$\mu$m selected sample including 18 different bands from the FUV through the FIR is similar to what has been observed for local LIRGs and ULIRGs. In particular, the maximum IR to optical ratio ($\sim 100$) and the maximum $L_{\rm IR}$ ($\sim 10^{14}\ts L_{\odot}$) is the same as seen locally. Our large sample presents a more complete picture of the range of SED shapes, particularly in the UV where the ratio of the optical (0.5\ts$\mu$m) to ultraviolet (0.2\ts$\mu$m) luminosities varies by nearly a factor of 1000. Similarly, the ratio of luminosities in the near (2 to 5\ts$\mu$m) and mid-infrared (10 to 30\ts$\mu$m) varies by more than a factor of 100. Although our selection at 70\ts$\mu$m is biased toward a warmer sample at higher redshift it still appears that the dispersion of shapes is similar at all redshifts. Clearly, one needs to consider a wider range of spectral templates as opposed to the typical cold (e.g., Arp 220) and warm (e.g., Mrk 231) templates used to represent infrared galaxies.}

\item{A comparison of our luminosity estimates using model fits with and without the 160\ts$\mu$m data points indicates that the sources detected at 160\ts$\mu$m have a colder SED than those not detected. This is confirmed by the difference in the FIR colors between the two subsamples. A comparison with local infrared selected sources finds the same range in FIR colors overall but also presents evidence that high redshift and high luminosity sources detected at 160\ts$\mu$m in our sample are colder than local sources of comparable luminosity. This is likely a combination of a selection effect due to the longer wavelength 160\ts$\mu$m detection and the poor constraints that 160\ts${\mu}$m places on the peak of emission and dust temperature at these redshifts ($z>1.5$). Future FIR and submillimeter studies with {\it Herschel} and SCUBA2 will be necessary to disentangle these possible causes.}

\item{We use five of the most commonly used multiwavelength selection techniques to identify potential AGN in our sample, including 132 X-ray detected AGN, 27 sources with radio fluxes five times greater than that expected from the infrared-radio correlation, 166 sources with power-law SEDs, 232 IRAC color selected galaxies, and 61 potentially obscured ($\rm F(24\ts {\mu} m)/F(R) > 10^3$) AGN. We find that there is significant overlap among these different methods resulting in a total of 354 sources with signatures of an AGN.}

\item{We find that the total fraction of AGN increases strongly with $L_{\rm IR}$ similarly to the behavior of galaxies in the local universe. Our estimate indicates that nearly 70\% of 70\ts$\mu$m selected ULIRGs and all HyLIRGs contain a powerful AGN. Less than half of the candidate AGN selected based on several different methods are detected in the X-ray providing evidence for a significant population of obscured AGN among this infrared selected sample. Since a 70\ts$\mu$m selection selects warmer SEDs at higher redshifts it remains to be seen whether this AGN fraction is representative of all ULIRGs at these redshifts. }

\end{enumerate}

The complete 70\ts$\mu$m selected catalog of all 1503 sources including their coordinates, redshifts, and total infrared luminosities (Table~\ref{catalog}) will be made publicly available with the online version of this paper.

 \acknowledgments

Support for this work was provided in part by NASA through contracts 1282612, 1298213 and 1344920 issued by the Jet Propulsion Laboratory. We would also like to recognize the contributions from all of the members of the COSMOS Team who helped in obtaining and reducing the large amount of multi-wavelength data that are now publicly available through the NASA Infrared Science Archive (IRSA) at http://irsa.ipac.caltech.edu/Missions/cosmos.html. The analysis pipeline used to reduce the DEIMOS data was developed at UC Berkeley with support from NSF grant AST-0071048. This research has made use of the NASA/IPAC Extragalactic Database (NED) which is operated by the Jet Propulsion Laboratory, California Institute of Technology, under contract with the National Aeronautics and Space Administration. This research has also made use of data from the Sloan Digital Sky Survey (SDSS-DR7). Funding for the SDSS and SDSS-II has been provided by the Alfred P. Sloan Foundation, the Participating Institutions, the National Science Foundation, the U.S. Department of Energy, the National Aeronautics and Space Administration, the Japanese Monbukagakusho, the Max Planck Society, and the Higher Education Funding Council for England. The SDSS Web Site is http://www.sdss.org/.


\begin{deluxetable}{lcccc}
\tablewidth{0pt}
 \tabletypesize{\small}
\tablecaption{Summary of COSMOS Multiwavelength Data Sets \label{data}}
\tablehead{\colhead{Instrument} & \colhead{Band} & \colhead{Date of Observations} & \colhead{Depth (5\ts$\sigma$)}  & \colhead{\# in 70\ts$\mu$m sample}}
\startdata
Spitzer/IRAC   & 3.6\ts$\mu$m & 2006             & 0.9\ts $\mu$Jy  & 1503 \\
Spitzer/IRAC   & 4.5\ts$\mu$m & 2006             & 1.7\ts $\mu$Jy  & 1503 \\
Spitzer/IRAC   & 5.6\ts$\mu$m & 2006             & 11.3\ts $\mu$Jy  & 1501 \\
Spitzer/IRAC   & 8.0\ts$\mu$m & 2006             & 14.6\ts $\mu$Jy  & 1503 \\
Spitzer/MIPS   & 24\ts$\mu$m      & 2006--2008  & 60\ts$\mu$Jy    &  1503 \\
Spitzer/MIPS   & 70\ts$\mu$m      & 2006--2008  & 4.5\ts mJy          & 1503 \\
Spitzer/MIPS   & 160\ts$\mu$m   & 2006--2008  &  65\ts mJy          & 463\\
HST/ACS         & F814W                & 2003--2005  & 27.2\ts mag      &1503 \\
HST/NICMOS & F160W                & 2003--2005  & 25.9\ts mag      & 74 \\
Galex             & FUV,NUV             & 2004               & 25.7, 26.0\ts mag  & 892, 1284\\
XMM               &  0.5--10\ts keV    &  2004--2005  &  $5\times 10^{-16}\rm\ts erg\ts cm^{-2}\ts s^{-1}$   & 119 \\
Chandra        & 0.5--10\ts keV     & 2006--2007    & $5.7\times 10^{-16}\rm\ts erg\ts cm^{-2}\ts s^{-1}$ & 103 \\
CFHT/MEGACAM & $u^*$            & 2003--2007   & 26.5\ts mag       & 1502 \\
Subaru/SuprimeCam & $B_JV_Jg^+r^+i^+z^+$ &2004--2005 & $i^+\sim26.2\ts$ mag & 1503 \\
UKIRT/WFCAM & $J$                 &  2004--2007  & 23.7\ts mag & 1498 \\
CFHT/WIRCAM & $K_{\rm S}$        &  2005--2007  & 23.5\ts mag & 1503 \\
VLA                 &  1.4 GHz            & 2003--2005   & 55\ts $\mu$Jy & 562 
\enddata
\end{deluxetable}

\begin{deluxetable}{llccc}
 \tabletypesize{\small}
\tablecaption{Spectroscopic Redshift Surveys \label{spectra}}
\tablehead{\colhead{Survey} & \colhead{Reference} & \colhead{Instrument} & \colhead{Wavelength Range [\AA]}  & \colhead{\# in 70\ts$\mu$m sample}}
\startdata

24\ts$\mu$m sources	& Kartaltepe \etal, in prep. 	& Keck II/DEIMOS   		& 4000--10000 	& 62 \\   
\nodata				& Capak, Salvato \etal, in prep	& Keck II/DEIMOS		& 4000--10000		& 18  \\
zCOSMOS                   	& Lilly \etal 2009                  	& VLT/VIMOS          		& 5550--9650  		& 355 \\
XMM sources               	& Trump \etal 2007, 2008  	& Magellan/IMACS 		& 5600--9200 		& 223 \\
Quasar candidates		& \citealt{Prescott:2006p2333}	& MMT/Hectospec		&  3100--9000		& 43 \\
\nodata				& Anguita \etal 2009			& VLT/FORS1			& 3300-11000		& 4 \\
SDSS                             	& \citealt{Abazajian:2009p5536} 	& SDSS Spectrograph	&  3900--9100 	& 65 \\
2dFGRS		   		& \citealt{Colless:2001p5237}	& 2dF Spectrograph		& 3500--10000		& 34 

\enddata
\end{deluxetable}

\begin{deluxetable}{lccccccccc}
  \tablewidth{0pt}
  \tabletypesize{\scriptsize}
  \tablecaption{COSMOS 70 $\mu$m Selected Source Catalog \label{catalog}}
  \setlength{\tabcolsep}{0.05in}
\tablehead{\colhead{ID} & \colhead{Name} & \colhead{RA} & \colhead{Dec} & 
              \colhead{24 $\mu$m flux\tablenotemark{a}} & \colhead{70 $\mu$m flux} & \colhead{160 $\mu$m flux} & 
              \colhead{Redshift} & \colhead{Flag\tablenotemark{b}} & \colhead{log($L_{\rm IR}/L_{\odot})$\tablenotemark{c}} 
              }
\startdata
  1017  & COSMOS J100046.07+013439.89 &   150.19197  &  1.57775  &  $0.69\pm 0.02$  &  $7.25\pm2.13$  & \nodata  &  0.518 & I  & $11.16\pm0.25$ \\
 1077  & COSMOS J100034.22+013543.18  &   150.14259  &   1.59533  & $1.08\pm0.29$  &  $16.03\pm2.03$  & \nodata &   0.23 & P &$10.87\pm0.22$   \\
 1081  & COSMOS J100133.37+013547.59  &   150.38905  &   1.59655 &  $0.75\pm0.04$ & $10.82\pm2.00$ &  \nodata & 0.23 & P & $10.44\pm0.22 $ \\
 1082  & COSMOS J100041.55+013552.68  &   150.17314  &  1.59797 & $1.62\pm0.35$  &   $22.26\pm2.45$ &  \nodata & 0.37 & P &  $11.26\pm0.23 $ \\
 1083  & COSMOS J100241.78+013550.07  &   150.15190   & 1.59699   & $0.14\pm0.02$  &  $18.61\pm2.12$  & \nodata &  1.20 & P &   $12.62\pm0.32 $
\enddata
\\
\tablenotetext{a}{All fluxes and errors are in mJy. Fluxes and errors for all of the bands discussed in this paper are presented in the online version of the paper. Here, only the MIPS data points are shown as an example.}
\tablenotetext{b}{Redshift Flag --- D: DEIMOS spectroscopy on Keck II, Z: zCOSMOS spectroscopy from VLT/VIMOS, I: IMACS spectroscopy from Magellan, S: SDSS spectroscopy, 2dF: 2dFGRS spectroscopy, F: FORS1 spectroscopy, M: Hectospec spectroscopy from MMT, P: Photometric redshift from Ilbert \etal 2009 or Salvato \etal 2009}
\tablenotetext{c}{Total infrared luminosity and uncertainties estimated from the template fits described in \S4.}
\end{deluxetable}

\begin{deluxetable}{llccccc}
\tablewidth{0pt}
\tablecaption{Offset in $L_{\rm IR}$ Estimate Without 160\ts$\mu$m Data Point \label{offsets}}
\tablehead{\colhead{Flux bin [mJy]} &\colhead{z bin} & \colhead{med(flux) [mJy]} & \colhead{med(z)} & \colhead{\# of sources}  & 
\colhead{Mean Offset } & \colhead{Dispersion}}
\startdata

$0.4-0.8$       & \nodata 	& 6.6  	& 0.84 	&  61    	& 0.41 	& 0.20 \\
$0.8-12.0$    &\nodata 	&  9.4  	& 0.58 	& 117    	& 0.33 	& 0.16 \\
$12.0-20.0$  & \nodata 	&14.7 	& 0.41 	& 140  	& 0.20 	& 0.18 \\
$> 20.0$    &\nodata 		& 32.8	& 0.22 	& 141 	& 0.09 	& 0.14 \\ \\
\hline \\
\nodata & $0.0-0.2$ & 24.8  	& 0.12 	& 106 	& 0.17 	& 0.16 \\
\nodata & $0.2-0.4$ & 15.9 	& 0.27 	& 124 	&  0.14 	& 0.15 \\
\nodata & $0.4-0.6$ & 12.8 	& 0.49 	& 69 		& 0.19 	& 0.22\\
\nodata & $0.6-0.8$ & 12.2 	& 0.69 	& 49 		& 0.28 	& 0.20\\
\nodata & $0.8-1.0$ & 11.4 	& 0.90 	& 50 		& 0.30 	& 0.19 \\
\nodata & $1.0-1.5$ & 10.6 	& 1.15 	& 39 		& 0.31 	& 0.25\\
\nodata & $1.5-3.0$ &  8.3 	& 1.78 	& 22 		& 0.32 	& 0.24 

\enddata
\end{deluxetable}

\begin{deluxetable}{lccccc}
\tablewidth{0pt}
\tabletypesize{\small}
\tablecaption{Fraction of 70\ts$\mu$m Sources Detected in the X-ray \label{xray_area}}
\tablehead{\colhead{log$(L_{\rm IR}/L_{\odot})$} & \colhead{Median} &  \colhead{Number} & \colhead{X-ray: Full area\tablenotemark{a}} & \colhead{X-ray: {\it Chandra} area\tablenotemark{b}} & \colhead{X-ray: {\it Chandra} deep area\tablenotemark{c}} \\  
\colhead{ } & \colhead{Redshift} & \colhead { } & \colhead{\# (\%)} & \colhead{\# (\%)} & \colhead{\# (\%)} } 
\startdata
$< 9.0$ 	    	& 0.02    	&   3   	&   0		   	&  0	 		& 0 \\
$9.0-10.0$     	& 0.10    	&   58   	&   3 (5.2)   	&  1 (3.6) 		& 0 \\
$10.0-10.5$	&  0.17 	&  140 	&  8 (5.7)     	& 4 (5.2) 		&  2 (4.9) \\
$10.5-11.0$  	& 0.27 	&  281  	& 18 (6.4)    	&  14 (8.0) 	& 9 (9.6) \\
$11.0-11.5$  	& 0.47 	&  376  	& 35 (9.3)    	&   27 (12.0) 	& 15 (13.6) \\
$11.5-12.0$ 	&  0.77 	&   311 	&  28 (9.0)    	&  27 (14.1) 	& 14 (14.7) \\
$12.0-12.5$  	&  1.05 	&  219 	&  29 (13.2)  	&  19 (13.0)	&  14 (16.3) \\
$12.5-13.0$  	&  1.47 	&  84  	&  20 (23.8)   	&  13 (29.5) 	&  9 (34.6) \\
$>13.0$     	&  2.00    	&   31   	&   13 (41.9 )	&  6  (46.2) 	&  3  (42.9) \\
\hline
All $L_{\rm IR}$ & 0.50 	& 1503 	& 154 (10.2) 	& 111 (12.3) 	& 66 (14.0)          
\enddata
\tablenotetext{a}{Percentage of 70\ts$\mu$m sources detected in the X-ray by XMM and {\it Chandra} over the full ACS area of the COSMOS field. The XMM observations cover 100\% of this area.}
\tablenotetext{b}{Percentage of 70\ts$\mu$m sources detected in the X-ray by XMM and {\it Chandra} in the area of the field observed by {\it Chandra} (55\% of the ACS area) as illustrated in  Figure~\ref{xray_dist}.}
\tablenotetext{c}{Percentage of 70\ts$\mu$m sources detected in the X-ray by XMM and {\it Chandra} in the area of the field with deep {\it Chandra} coverage (30\% of the ACS area).}
\end{deluxetable}

\begin{deluxetable}{lccccccccc}
\tablewidth{0pt}
  \tabletypesize{\small}
\tablecaption{AGN Fraction by Luminosity Bin \label{frac}}

\tablehead{\colhead{log$(\rm L_{\rm IR}/L_{\odot})$} & \colhead{Median} &  \colhead{Number} & 
\colhead{X-ray AGN\tablenotemark{a}} & \colhead{PL\tablenotemark{b}} & \colhead{IRAC\tablenotemark{c}} & \colhead{Obscured\tablenotemark{d}} & \colhead{Radio Excess\tablenotemark{e}} & \colhead{Total\tablenotemark{f}}\\  
\colhead{ } & \colhead{Redshift} & \colhead { } & \colhead{\# (\%)} & \colhead{\# (\%)} & \colhead{\# (\%)} & \colhead{Candidates} & \colhead{\# (\%)} & \colhead{\# (\%)}} 
\startdata
$< 9$     		& 0.02    	&   3   	&  0			&  0		    	& 1 (33)  		&    0   		&  0		&  1 (33)\\
$9.0-10.0$     	& 0.10    	&   58   	&   0			&  0 		    	& 1 (1.7)  		&    0   		&  2 (3.4)	&  2 (3.4)\\
$10.0-10.5$	&  0.17 	&  140 	&  2 (1.4)     	&  0		       	& 2 (1.4)  		&    0   		&  1 (0.7)	&  5 (3.6) \\
$10.5-11.0$  	& 0.27 	&  281  	& 12 (4.3)    	&  0            	& 10 (3.6)		&    0 		&  1 (0.4)	& 18 (6.4) \\
$11.0-11.5$  	& 0.47 	&  376  	& 28 (7.4)    	&  3 (0.8)       	&  19 (5.1)		&    1 (0.3)		&  3 (0.8) 	& 40 (10.6) \\
$11.5-12.0$ 	&  0.77 	&   311	&  28 (9.0)    	& 8 (2.6)	  	& 40 (12.9)	&    9 (2.9)		& 2 (0.6)	& 61 (19.6) \\
$12.0-12.5$  	&  1.05 	&  219 	&  29 (13.2)  	& 73 (33.3)    	& 82 (37.4)  	&    22 (10.0)	& 11 (5.0)	& 123 (56.2) \\
$12.5-13.0$  	&  1.47	&  84  	&  20 (23.8)   	& 52 (61.9)	& 56 (66.7)  	&    18 (21.4)	& 5 (6.0)  	& 73 (86.9) \\
$>13.0$     	&  2.00    	&   31   	&   13 (41.9)	& 30 (96.8)    	& 21 (67.7)   	&    11 (35.5)	& 2 (6.5)  	& 31 (100.0)\\
\hline
All $L_{\rm IR}$ & 0.50 	& 1503 	& 132 (8.8) 	& 166 (11.0) 	& 232 (15.4)	&  61 (4.1))	& 27 (1.8)	& 354 (23.6)         
\enddata
\tablenotetext{a}{Number and percentage of objects that are detected as AGN in the X-ray (see \S6.4 for details).}
\tablenotetext{b}{Number and percentage of objects that have power-law SEDs as defined in \S6.4}
\tablenotetext{c}{Number and percentage of objects whose colors match the AGN candidate criteria of \cite{Stern:2005p3000}}
\tablenotetext{d}{Number and percentage of objects whose infrared to optical flux ratios ($F(24\ts\mu m)/F(R) > 10^3$) match the Fiore \etal 2008 criteria of potentially obscured AGN. }
\tablenotetext{e}{Number and percentage of objects that with a radio excess (5 times more radio flux than expected from the radio-IR correlation).}
\tablenotetext{f}{Total number and percentage of objects that fall into any of the AGN candidate categories.}
\end{deluxetable}


\begin{thebibliography}{78}
\expandafter\ifx\csname natexlab\endcsname\relax\def\natexlab#1{#1}\fi

\bibitem[{Abazajian {et~al.}(2009)Abazajian, Adelman-McCarthy, Ag{\"u}eros,
  Allam, Prieto, An, Anderson, Anderson, Annis, Bahcall, Bailer-Jones,
  Barentine, Bassett, Becker, Beers, Bell, Belokurov, Berlind, Berman,
  Bernardi, Bickerton, Bizyaev, Blakeslee, Blanton, Bochanski, Boroski,
  Brewington, Brinchmann, Brinkmann, Brunner, Budav{\'a}ri, Carey, Carliles,
  Carr, Castander, Cinabro, Connolly, Csabai, Cunha, Czarapata, Davenport,
  de~Haas, Dilday, Doi, Eisenstein, Evans, Evans, Fan, Friedman, Frieman,
  Fukugita, G{\"a}nsicke, Gates, Gillespie, Gilmore, Gonzalez, Gonzalez,
  Grebel, Gunn, Gy{\"o}ry, Hall, Harding, Harris, Harvanek, Hawley, Hayes,
  Heckman, Hendry, Hennessy, Hindsley, Hoblitt, Hogan, Hogg, Holtzman, Hyde,
  ichi Ichikawa, Ichikawa, Im, Ivezi{\'c}, Jester, Jiang, Johnson, Jorgensen,
  Juri{\'c}, Kent, Kessler, Kleinman, Knapp, Konishi, Kron, Krzesinski,
  Kuropatkin, Lampeitl, Lebedeva, Lee, Lee, Leger, L{\'e}pine, Li, Lima, Lin,
  Long, Loomis, Loveday, Lupton, Magnier, Malanushenko, Malanushenko,
  Mandelbaum, Margon, Marriner, Mart{\'\i}nez-Delgado, Matsubara, McGehee,
  McKay, Meiksin, Morrison, Mullally, Munn, Murphy, Nash, Nebot, Neilsen,
  Newberg, Newman, Nichol, Nicinski, Nieto-Santisteban, Nitta, Okamura,
  Oravetz, Ostriker, Owen, Padmanabhan, Pan, Park, Pauls, Peoples, Percival,
  Pier, Pope, Pourbaix, Price, Purger, Quinn, Raddick, Fiorentin, Richards,
  Richmond, Riess, Rix, Rockosi, Sako, Schlegel, Schneider, Scholz, Schreiber,
  Schwope, Seljak, Sesar, Sheldon, Shimasaku, Sibley, Simmons, Sivarani, Smith,
  Smith, Smol{\v c}i{\'c}, Snedden, Stebbins, Steinmetz, Stoughton, Strauss,
  Rao, Suto, Szalay, Szapudi, Szkody, Tanaka, Tegmark, Teodoro, Thakar,
  Tremonti, Tucker, Uomoto, Berk, Vandenberg, Vidrih, Vogeley, Voges, Vogt,
  Wadadekar, Watters, Weinberg, West, White, Wilhite, Wonders, Yanny, Yocum,
  York, Zehavi, Zibetti, \& Zucker}]{Abazajian:2009p5536}
Abazajian, K.~N., {et~al.} 2009, ApJS, 182, 543

\bibitem[{Alonso-Herrero {et~al.}(2006)Alonso-Herrero, P{\'e}rez-Gonz{\'a}lez,
  Alexander, Rieke, Rigopoulou, Floc'h, Barmby, Papovich, Rigby, Bauer, Brandt,
  Egami, Willner, Dole, \& Huang}]{AlonsoHerrero:2006p2013}
Alonso-Herrero, A., {et~al.} 2006, ApJ, 640, 167

\bibitem[{Appleton {et~al.}(2004)Appleton, Fadda, Marleau, Frayer, Helou,
  Condon, Choi, Yan, Lacy, Wilson, Armus, Chapman, Fang, Heinrichson, Im,
  Jannuzi, Storrie-Lombardi, Shupe, Soifer, Squires, \&
  Teplitz}]{Appleton:2004p5559}
Appleton, P.~N., {et~al.} 2004, ApJS, 154, 147

\bibitem[{Armus {et~al.}(2009)Armus, Mazzarella, Evans, Surace, Sanders,
  Iwasawa, Frayer, Howell, Chan, Petric, Vavilkin, Kim, Haan, Inami, Murphy,
  Appleton, Barnes, Bothun, Bridge, Charmandaris, Jensen, Kewley, Lord, Madore,
  Marshall, Melbourne, Rich, Satyapal, Schulz, Spoon, Sturm, U, Veilleux, \&
  Xu}]{Armus:2009p5172}
Armus, L., {et~al.} 2009, PASP, 121, 559

\bibitem[{Bavouzet {et~al.}(2008)Bavouzet, Dole, Floc'h, Caputi, Lagache, \&
  Kochanek}]{Bavouzet:2008p2620}
Bavouzet, N., Dole, H., Floc'h, E.~L., Caputi, K.~I., Lagache, G., \& Kochanek,
  C.~S. 2008, A\&A, 479, 83

\bibitem[{Bertin(2006)}]{Bertin:2006p2670}
Bertin, E. 2006, Astronomical Data Analysis Software and Systems XV ASP
  Conference Series, 351, 112

\bibitem[{Bertin \& Arnouts(1996)}]{Bertin:1996p322}
Bertin, E., \& Arnouts, S. 1996, A\&AS, 117, 393

\bibitem[{Bertin {et~al.}(2002)Bertin, Mellier, Radovich, Missonnier, Didelon,
  \& Morin}]{Bertin:2002p2651}
Bertin, E., Mellier, Y., Radovich, M., Missonnier, G., Didelon, P., \& Morin,
  B. 2002, Astronomical Data Analysis Software and Systems XI, 281, 228

\bibitem[{Bondi {et~al.}(2008)Bondi, Ciliegi, Schinnerer, Smol{\v c}i{\'c},
  Jahnke, Carilli, \& Zamorani}]{Bondi:2008p68}
Bondi, M., Ciliegi, P., Schinnerer, E., Smol{\v c}i{\'c}, V., Jahnke, K.,
  Carilli, C., \& Zamorani, G. 2008, ApJ, 681, 1129

\bibitem[{Brusa {et~al.}(2009)Brusa, Comastri, Gilli, Hasinger, Iwasawa,
  Mainieri, Mignoli, Salvato, Zamorani, Bongiorno, Cappelluti, Civano, Fiore,
  Merloni, Silverman, Trump, Vignali, Capak, Elvis, Ilbert, Impey, \&
  Lilly}]{Brusa:2009p2165}
Brusa, M., {et~al.} 2009, ApJ, 693, 8

\bibitem[{Cappelluti {et~al.}(2009)Cappelluti, Brusa, Hasinger, Comastri,
  Zamorani, Finoguenov, Gilli, Puccetti, Miyaji, Salvato, Vignali, Aldcroft,
  B{\"o}hringer, Brunner, Civano, Elvis, Fiore, Fruscione, Griffiths, Guzzo,
  Iovino, Koekemoer, Mainieri, Scoville, Shopbell, Silverman, \&
  Urry}]{Cappelluti:2009p2159}
Cappelluti, N., {et~al.} 2009, A\&A, 497, 635

\bibitem[{Caputi {et~al.}(2007)Caputi, Lagache, Yan, Dole, Bavouzet, Floc'h,
  Choi, Helou, \& Reddy}]{Caputi:2007p2597}
Caputi, K.~I., {et~al.} 2007, ApJ, 660, 97

\bibitem[{Chary \& Elbaz(2001)}]{Chary:2001p2083}
Chary, R., \& Elbaz, D. 2001, ApJ, 556, 562

\bibitem[{Colless {et~al.}(2001)Colless, Dalton, Maddox, Sutherland, Norberg,
  Cole, Bland-Hawthorn, Bridges, Cannon, Collins, Couch, Cross, Deeley,
  Propris, Driver, Efstathiou, Ellis, Frenk, Glazebrook, Jackson, Lahav, Lewis,
  Lumsden, Madgwick, Peacock, Peterson, Price, Seaborne, \&
  Taylor}]{Colless:2001p5237}
Colless, M., {et~al.} 2001, MNRAS, 328, 1039

\bibitem[{Condon(1992)}]{Condon:1992p2865}
Condon, J.~J. 1992, ARAA, 30, 575

\bibitem[{Condon {et~al.}(1991{\natexlab{a}})Condon, Anderson, \&
  Helou}]{Condon:1991p1977}
Condon, J.~J., Anderson, M.~L., \& Helou, G. 1991{\natexlab{a}}, ApJ, 376, 95

\bibitem[{Condon {et~al.}(1991{\natexlab{b}})Condon, Huang, Yin, \&
  Thuan}]{Condon:1991p1982}
Condon, J.~J., Huang, Z.-P., Yin, Q.~F., \& Thuan, T.~X. 1991{\natexlab{b}},
  ApJ, 378, 65

\bibitem[{Dale {et~al.}(2007)Dale, de~Paz, Gordon, Hanson, Armus, Bendo,
  Bianchi, Block, Boissier, Boselli, Buckalew, Buat, Burgarella, Calzetti,
  Cannon, Engelbracht, Helou, Hollenbach, Jarrett, Kennicutt, Leitherer, Li,
  Madore, Martin, Meyer, Murphy, Regan, Roussel, Smith, Sosey, Thilker, \&
  Walter}]{Dale:2007p5190}
Dale, D.~A., {et~al.} 2007, ApJ, 655, 863

\bibitem[{Dale \& Helou(2002)}]{Dale:2002p2130}
Dale, D.~A., \& Helou, G. 2002, ApJ, 576, 159

\bibitem[{Dale {et~al.}(2001)Dale, Helou, Contursi, Silbermann, \&
  Kolhatkar}]{Dale:2001p2742}
Dale, D.~A., Helou, G., Contursi, A., Silbermann, N.~A., \& Kolhatkar, S. 2001,
  ApJ, 549, 215

\bibitem[{Desert {et~al.}(1990)Desert, Boulanger, \& Puget}]{Desert:1990p2747}
Desert, F.-X., Boulanger, F., \& Puget, J.~L. 1990, A\&A, 237, 215

\bibitem[{Dey {et~al.}(2008)Dey, Soifer, Desai, Brand, Floc'h, Brown, Jannuzi,
  Armus, Bussmann, Brodwin, Bian, Eisenhardt, Higdon, Weedman, \&
  Willner}]{Dey:2008p2965}
Dey, A., {et~al.} 2008, ApJ, 677, 943

\bibitem[{Donley {et~al.}(2008)Donley, Rieke, P{\'e}rez-Gonz{\'a}lez, \&
  Barro}]{Donley:2008p3069}
Donley, J.~L., Rieke, G.~H., P{\'e}rez-Gonz{\'a}lez, P.~G., \& Barro, G. 2008,
  ApJ, 687, 111

\bibitem[{Donley {et~al.}(2007)Donley, Rieke, P{\'e}rez-Gonz{\'a}lez, Rigby, \&
  Alonso-Herrero}]{Donley:2007p3092}
Donley, J.~L., Rieke, G.~H., P{\'e}rez-Gonz{\'a}lez, P.~G., Rigby, J.~R., \&
  Alonso-Herrero, A. 2007, ApJ, 660, 167

\bibitem[{Dunne {et~al.}(2000)Dunne, Eales, Edmunds, Ivison, Alexander, \&
  Clements}]{Dunne:2000p2799}
Dunne, L., Eales, S., Edmunds, M., Ivison, R., Alexander, P., \& Clements,
  D.~L. 2000, MNRAS, 315, 115

\bibitem[{Dunne \& Eales(2001)}]{Dunne:2001p2816}
Dunne, L., \& Eales, S.~A. 2001, MNRAS, 327, 697

\bibitem[{Elvis {et~al.}(2009)Elvis, Civano, Vignali, Puccetti, Fiore,
  Cappelluti, Aldcroft, Fruscione, Zamorani, Comastri, Brusa, Gilli, Miyaji,
  Damiani, Koekemoer, Finoguenov, Brunner, Urry, Silverman, Mainieri, Hasinger,
  Griffiths, Carollo, Hao, Guzzo, Blain, Calzetti, Carilli, Capak, Ettori,
  Fabbiano, Impey, Lilly, Mobasher, Rich, Salvato, Sanders, Schinnerer,
  Scoville, Shopbell, Taylor, Taniguchi, \& Volonteri}]{Elvis:2009p2154}
Elvis, M., {et~al.} 2009, eprint arXiv, 0903, 2062

\bibitem[{Elvis {et~al.}(1994)Elvis, Wilkes, McDowell, Green, Bechtold,
  Willner, Oey, Polomski, \& Cutri}]{Elvis:1994p5581}
---. 1994, ApJS, 95, 1

\bibitem[{Fiore {et~al.}(2008)Fiore, Grazian, Santini, Puccetti, Brusa,
  Feruglio, Fontana, Giallongo, Comastri, Gruppioni, Pozzi, Zamorani, \&
  Vignali}]{Fiore:2008p2961}
Fiore, F., {et~al.} 2008, ApJ, 672, 94

\bibitem[{Fiore {et~al.}(2009)Fiore, Puccetti, Brusa, Salvato, Zamorani,
  Aldcroft, Aussel, Brunner, Capak, Cappelluti, Civano, Comastri, Elvis,
  Feruglio, Finoguenov, Fruscione, Gilli, Hasinger, Koekemoer, Kartaltepe,
  Ilbert, Impey, LeFloc'h, Lilly, Mainieri, Martinez-Sansigre, McCracken,
  Menci, Merloni, Miyaji, Sanders, Sargent, Schinnerer, Scoville, Silverman,
  Smolcic, Steffen, Santini, Taniguchi, Thompson, Trump, Vignali, Urry, \&
  Yan}]{Fiore:2009p2143}
---. 2009, ApJ, 693, 447

\bibitem[{Frayer {et~al.}(2009)Frayer, Sanders, Surace, Aussel, Salvato,
  Floc'h, Huynh, Scoville, Afonso-Luis, Bhattacharya, Capak, Fadda, Fu, Helou,
  Ilbert, Kartaltepe, Koekemoer, Lee, Murphy, Sargent, Schinnerer, Sheth,
  Shopbell, Shupe, \& Yan}]{Frayer:2009p2141}
Frayer, D.~T., {et~al.} 2009, eprint arXiv, 0902, 3273

\bibitem[{Hasinger {et~al.}(2007)Hasinger, Cappelluti, Brunner, Brusa,
  Comastri, Elvis, Finoguenov, Fiore, Franceschini, Gilli, Griffiths, Lehmann,
  Mainieri, Matt, Matute, Miyaji, Molendi, Paltani, Sanders, Scoville, Tresse,
  Urry, Vettolani, \& Zamorani}]{Hasinger:2007p2291}
Hasinger, G., {et~al.} 2007, ApJS, 172, 29

\bibitem[{Helou {et~al.}(1985)Helou, Soifer, \&
  Rowan-Robinson}]{Helou:1985p2892}
Helou, G., Soifer, B.~T., \& Rowan-Robinson, M. 1985, ApJ, 298, L7

\bibitem[{Houck {et~al.}(2005)Houck, Soifer, Weedman, Higdon, Higdon, Herter,
  Brown, Dey, Jannuzi, Floc'h, Rieke, Armus, Charmandaris, Brandl, \&
  Teplitz}]{Houck:2005p2986}
Houck, J.~R., {et~al.} 2005, ApJ, 622, L105

\bibitem[{Huynh {et~al.}(2007)Huynh, Frayer, Mobasher, Dickinson, Chary, \&
  Morrison}]{Huynh:2007p62}
Huynh, M.~T., Frayer, D.~T., Mobasher, B., Dickinson, M., Chary, R.-R., \&
  Morrison, G. 2007, ApJ, 667, L9

\bibitem[{Ibar {et~al.}(2008)Ibar, Cirasuolo, Ivison, Best, Smail, Biggs,
  Simpson, Dunlop, Almaini, McLure, Foucaud, \& Rawlings}]{Ibar:2008p5546}
Ibar, E., {et~al.} 2008, MNRAS, 386, 953

\bibitem[{Ilbert {et~al.}(2009)Ilbert, Capak, Salvato, Aussel, McCracken,
  Sanders, Scoville, Kartaltepe, Arnouts, Floc'h, Mobasher, Taniguchi,
  Lamareille, Leauthaud, Sasaki, Thompson, Zamojski, Zamorani, Bardelli,
  Bolzonella, Bongiorno, Brusa, Caputi, Carollo, Contini, Cook, Coppa,
  Cucciati, de~la Torre, de~Ravel, Franzetti, Garilli, Hasinger, Iovino,
  Kampczyk, Kneib, Knobel, Kovac, LeBorgne, LeBrun, F{\`e}vre, Lilly, Looper,
  Maier, Mainieri, Mellier, Mignoli, Murayama, Pell{\`o}, Peng,
  P{\'e}rez-Montero, Renzini, Ricciardelli, Schiminovich, Scodeggio, Shioya,
  Silverman, Surace, Tanaka, Tasca, Tresse, Vergani, \&
  Zucca}]{Ilbert:2009p2146}
Ilbert, O., {et~al.} 2009, ApJ, 690, 1236

\bibitem[{Kennicutt {et~al.}(2003)Kennicutt, Armus, Bendo, Calzetti, Dale,
  Draine, Engelbracht, Gordon, Grauer, Helou, Hollenbach, Jarrett, Kewley,
  Leitherer, Li, Malhotra, Regan, Rieke, Rieke, Roussel, Smith, Thornley, \&
  Walter}]{Kennicutt:2003p5199}
Kennicutt, R.~C., {et~al.} 2003, PASP, 115, 928

\bibitem[{Kim \& Sanders(1998)}]{Kim:1998p3280}
Kim, D.-C., \& Sanders, D.~B. 1998, ApJS, 119, 41

\bibitem[{Koekemoer {et~al.}(2007)Koekemoer, Aussel, Calzetti, Capak,
  Giavalisco, Kneib, Leauthaud, F{\`e}vre, McCracken, Massey, Mobasher, Rhodes,
  Scoville, \& Shopbell}]{Koekemoer:2007p2299}
Koekemoer, A.~M., {et~al.} 2007, ApJS, 172, 196

\bibitem[{Kov{\'a}cs {et~al.}(2006)Kov{\'a}cs, Chapman, Dowell, Blain, Ivison,
  Smail, \& Phillips}]{Kovacs:2006p2929}
Kov{\'a}cs, A., Chapman, S.~C., Dowell, C.~D., Blain, A.~W., Ivison, R.~J.,
  Smail, I., \& Phillips, T.~G. 2006, ApJ, 650, 592

\bibitem[{Kruegel(2003)}]{Kruegel:2003p2827}
Kruegel, E. 2003, The physics of interstellar dust

\bibitem[{Lacy {et~al.}(2004)Lacy, Storrie-Lombardi, Sajina, Appleton, Armus,
  Chapman, Choi, Fadda, Fang, Frayer, Heinrichsen, Helou, Im, Marleau, Masci,
  Shupe, Soifer, Surace, Teplitz, Wilson, \& Yan}]{Lacy:2004p3062}
Lacy, M., {et~al.} 2004, ApJS, 154, 166

\bibitem[{Lagache {et~al.}(2003)Lagache, Dole, \& Puget}]{Lagache:2003p1825}
Lagache, G., Dole, H., \& Puget, J.-L. 2003, MNRAS, 338, 555

\bibitem[{Le F{\`e}vre {et~al.}(2003)Le F{\`e}vre, Vettolani, Maccagni, Picat, Adami,
  Arnaboldi, Bardelli, Bondi, Bottini, Bolzonella, Busarello, Cappi, Ciliegi,
  Contini, Charlot, Foucaud, Franzetti, Garilli, Gavignaud, Guzzo, Ilbert,
  Iovino, Brun, Marano, Marinoni, McCracken, Mathez, Mazure, Mellier, Meneux,
  Merluzzi, Paltani, Pell{\`o}, Pollo, Pozzetti, Radovich, Rizzo, Scaramella,
  Scodeggio, Tresse, Zamorani, Zanichelli, \& Zucca}]{LeFevre:2003p2676}
Le F{\`e}vre, O.~L., {et~al.} 2003, eprint arXiv, 11475

\bibitem[{Le Floc'h {et~al.}(2005)Le Floc'h, Papovich, Dole, Bell, Lagache, Rieke,
  Egami, P{\'e}rez-Gonz{\'a}lez, Alonso-Herrero, Rieke, Blaylock, Engelbracht,
  Gordon, Hines, Misselt, Morrison, \& Mould}]{LeFloch:2005p2544}
Le Floc'h, E., {et~al.} 2005, ApJ, 632, 169

\bibitem[{Lilly {et~al.}(2007)Lilly, F{\`e}vre, Renzini, Zamorani, Scodeggio,
  Contini, Carollo, Hasinger, Kneib, Iovino, Brun, Maier, Mainieri, Mignoli,
  Silverman, Tasca, Bolzonella, Bongiorno, Bottini, Capak, Caputi, Cimatti,
  Cucciati, Daddi, Feldmann, Franzetti, Garilli, Guzzo, Ilbert, Kampczyk,
  Kovac, Lamareille, Leauthaud, Borgne, McCracken, Marinoni, Pello,
  Ricciardelli, Scarlata, Vergani, Sanders, Schinnerer, Scoville, Taniguchi,
  Arnouts, Aussel, Bardelli, Brusa, Cappi, Ciliegi, Finoguenov, Foucaud,
  Franceschini, Halliday, Impey, Knobel, Koekemoer, Kurk, Maccagni, Maddox,
  Marano, Marconi, Meneux, Mobasher, Moreau, Peacock, Porciani, Pozzetti,
  Scaramella, Schiminovich, Shopbell, Smail, Thompson, Tresse, Vettolani,
  Zanichelli, \& Zucca}]{Lilly:2007p2297}
Lilly, S.~J., {et~al.} 2007, ApJS, 172, 70

\bibitem[{Magnelli {et~al.}(2009)Magnelli, Elbaz, Chary, Dickinson, Borgne,
  Frayer, \& Willmer}]{Magnelli:2009p2619}
Magnelli, B., Elbaz, D., Chary, R.~R., Dickinson, M., Borgne, D.~L., Frayer,
  D.~T., \& Willmer, C. N.~A. 2009, A\&A, 496, 57

\bibitem[{Makovoz \& Khan(2005)}]{Makovoz:2005p2643}
Makovoz, D., \& Khan, I. 2005, Astronomical Data Analysis Software and Systems
  XIV ASP Conference Series, 347, 81

\bibitem[{Malkan \& Sargent(1982)}]{Malkan:1982p3097}
Malkan, M.~A., \& Sargent, W. L.~W. 1982, ApJ, 254, 22

\bibitem[{Polletta {et~al.}(2006)del Carmen~Polletta, Wilkes, Siana,
  Lonsdale, Kilgard, Smith, Kim, Owen, Efstathiou, Jarrett, Stacey,
  Franceschini, Rowan-Robinson, Babbedge, Berta, Fang, Farrah,
  Gonz{\'a}lez-Solares, Morrison, Surace, \& Shupe}]{Polletta:2006p5580}
Polletta, M., {et~al.} 2006, ApJ, 642, 673

\bibitem[{Prescott {et~al.}(2006)Prescott, Impey, Cool, \&
  Scoville}]{Prescott:2006p2333}
Prescott, M. K.~M., Impey, C.~D., Cool, R.~J., \& Scoville, N.~Z. 2006, ApJ,
  644, 100

\bibitem[{Sajina {et~al.}(2008)Sajina, Yan, Lutz, Steffen, Helou, Huynh,
  Frayer, Choi, Tacconi, \& Dasyra}]{Sajina:2008p2894}
Sajina, A., {et~al.} 2008, ApJ, 683, 659

\bibitem[{Salvato {et~al.}(2009)Salvato, Hasinger, Ilbert, Zamorani, Brusa,
  Scoville, Rau, Capak, Arnouts, Aussel, Bolzonella, Buongiorno, Cappelluti,
  Caputi, Civano, Cook, Elvis, Gilli, Jahnke, Kartaltepe, Impey, Lamareille,
  LeFloch, Lilly, Mainieri, McCarthy, McCracken, Mignoli, Mobasher, Murayama,
  Sasaki, Sanders, Schiminovich, Shioya, Shopbell, Silverman, Smol{\v c}i{\'c},
  Surace, Taniguchi, Thompson, Trump, Urry, \& Zamojski}]{Salvato:2009p2142}
Salvato, M., {et~al.} 2009, ApJ, 690, 1250

\bibitem[{Sanders {et~al.}(2003)Sanders, Mazzarella, Kim, Surace, \&
  Soifer}]{Sanders:2003p1575}
Sanders, D.~B., Mazzarella, J.~M., Kim, D.-C., Surace, J.~A., \& Soifer, B.~T.
  2003, AJ, 126, 1607

\bibitem[{Sanders {et~al.}(1989)Sanders, Phinney, Neugebauer, Soifer, \&
  Matthews}]{Sanders:1989p3181}
Sanders, D.~B., Phinney, E.~S., Neugebauer, G., Soifer, B.~T., \& Matthews, K.
  1989, ApJ, 347, 29

\bibitem[{Sanders {et~al.}(2007)Sanders, Salvato, Aussel, Ilbert, Scoville,
  Surace, Frayer, Sheth, Helou, Brooke, Bhattacharya, Yan, Kartaltepe, Barnes,
  Blain, Calzetti, Capak, Carilli, Carollo, Comastri, Daddi, Ellis, Elvis,
  Fall, Franceschini, Giavalisco, Hasinger, Impey, Koekemoer, F{\`e}vre, Lilly,
  Liu, McCracken, Mobasher, Renzini, Rich, Schinnerer, Shopbell, Taniguchi,
  Thompson, Urry, \& Williams}]{Sanders:2007p11}
Sanders, D.~B., {et~al.} 2007, ApJS, 172, 86

\bibitem[{Schinnerer {et~al.}(2007)Schinnerer, Smol{\v c}i{\'c}, Carilli,
  Bondi, Ciliegi, Jahnke, Scoville, Aussel, Bertoldi, Blain, Impey, Koekemoer,
  Fevre, \& Urry}]{Schinnerer:2007p2300}
Schinnerer, E., {et~al.} 2007, ApJS, 172, 46

\bibitem[{Scoville {et~al.}(2007{\natexlab{a}})Scoville, Abraham, Aussel,
  Barnes, Benson, Blain, Calzetti, Comastri, Capak, Carilli, Carlstrom,
  Carollo, Colbert, Daddi, Ellis, Elvis, Ewald, Fall, Franceschini, Giavalisco,
  Green, Griffiths, Guzzo, Hasinger, Impey, Kneib, Koda, Koekemoer, Lefevre,
  Lilly, Liu, McCracken, Massey, Mellier, Miyazaki, Mobasher, Mould, Norman,
  Refregier, Renzini, Rhodes, Rich, Sanders, Schiminovich, Schinnerer,
  Scodeggio, Sheth, Shopbell, Taniguchi, Tyson, Urry, Waerbeke, Vettolani,
  White, \& Yan}]{Scoville:2007p1769}
Scoville, N., {et~al.} 2007{\natexlab{a}}, ApJS, 172, 38

\bibitem[{Scoville {et~al.}(2007{\natexlab{b}})Scoville, Aussel, Brusa, Capak,
  Carollo, Elvis, Giavalisco, Guzzo, Hasinger, Impey, Kneib, Lefevre, Lilly,
  Mobasher, Renzini, Rich, Sanders, Schinnerer, Schminovich, Shopbell,
  Taniguchi, \& Tyson}]{Scoville:2007p1776}
---. 2007{\natexlab{b}}, ApJS, 172, 1

\bibitem[{Siebenmorgen \& Kr{\"u}gel(2007)}]{Siebenmorgen:2007p2697}
Siebenmorgen, R., \& Kr{\"u}gel, E. 2007, A\&A, 461, 445

\bibitem[{Silva {et~al.}(1998)Silva, Granato, Bressan, \&
  Danese}]{Silva:1998p2723}
Silva, L., Granato, G.~L., Bressan, A., \& Danese, L. 1998, ApJ, 509, 103

\bibitem[{Soifer {et~al.}(1989)Soifer, Boehmer, Neugebauer, \&
  Sanders}]{Soifer:1989p2523}
Soifer, B.~T., Boehmer, L., Neugebauer, G., \& Sanders, D.~B. 1989, AJ, 98, 766

\bibitem[{Stern {et~al.}(2005)Stern, Eisenhardt, Gorjian, Kochanek, Caldwell,
  Eisenstein, Brodwin, Brown, Cool, Dey, Green, Jannuzi, Murray, Pahre, \&
  Willner}]{Stern:2005p3000}
Stern, D., {et~al.} 2005, ApJ, 631, 163

\bibitem[{Stetson(1987)}]{Stetson:1987p2648}
Stetson, P.~B. 1987, PASP, 99, 191

\bibitem[{Stickel {et~al.}(2000)Stickel, Lemke, Klaas, Beichman,
  Rowan-Robinson, Efstathiou, Bogun, Kessler, \& Richter}]{Stickel:2000p2754}
Stickel, M., {et~al.} 2000, A\&A, 359, 865

\bibitem[{Symeonidis {et~al.}(2009)Symeonidis, Page, Seymour, Dwelly, Coppin,
  McHardy, Rieke, \& Huynh}]{Symeonidis:2009p5203}
Symeonidis, M., Page, M.~J., Seymour, N., Dwelly, T., Coppin, K., McHardy, I.,
  Rieke, G.~H., \& Huynh, M. 2009, eprint arXiv, 0905, 854

\bibitem[{Symeonidis {et~al.}(2008)Symeonidis, Willner, Rigopoulou, Huang,
  Fazio, \& Jarvis}]{Symeonidis:2008p2625}
Symeonidis, M., Willner, S.~P., Rigopoulou, D., Huang, J.-S., Fazio, G.~G., \&
  Jarvis, M.~J. 2008, MNRAS, 385, 1015

\bibitem[{Tran {et~al.}(2001)Tran, Lutz, Genzel, Rigopoulou, Spoon, Sturm,
  Gerin, Hines, Moorwood, Sanders, Scoville, Taniguchi, \&
  Ward}]{Tran:2001p3271}
Tran, Q.~D., {et~al.} 2001, ApJ, 552, 527

\bibitem[{Trump {et~al.}(2009)Trump, Impey, Elvis, McCarthy, Huchra, Brusa,
  Salvato, Capak, Cappelluti, Civano, Comastri, Gabor, Hao, Hasinger, Jahnke,
  Kelly, Lilly, Schinnerer, Scoville, \& Smol{\v c}i{\'c}}]{Trump:2009p5204}
Trump, J.~R., {et~al.} 2009, ApJ, 696, 1195

\bibitem[{Trump {et~al.}(2007)Trump, Impey, McCarthy, Elvis, Huchra, Brusa,
  Hasinger, Schinnerer, Capak, Lilly, \& Scoville}]{Trump:2007p2229}
---. 2007, ApJS, 172, 383

\bibitem[{Veilleux {et~al.}(1999)Veilleux, Kim, \&
  Sanders}]{Veilleux:1999p2073}
Veilleux, S., Kim, D.-C., \& Sanders, D.~B. 1999, ApJ, 522, 113

\bibitem[{Veilleux {et~al.}(1995)Veilleux, Kim, Sanders, Mazzarella, \&
  Soifer}]{Veilleux:1995p2081}
Veilleux, S., Kim, D.-C., Sanders, D.~B., Mazzarella, J.~M., \& Soifer, B.~T.
  1995, ApJS, 98, 171

\bibitem[{Weedman {et~al.}(2006)Weedman, Polletta, Lonsdale, Wilkes, Siana,
  Houck, Surace, Shupe, Farrah, \& Smith}]{Weedman:2006p2991}
Weedman, D., {et~al.} 2006, ApJ, 653, 101

\bibitem[{Yan {et~al.}(2004)Yan, Helou, Fadda, Marleau, Lacy, Wilson, Soifer,
  Drozdovsky, Masci, Armus, Teplitz, Frayer, Surace, Storrie-Lombardi,
  Appleton, Chapman, Choi, Fan, Heinrichsen, Im, Schmitz, Shupe, \&
  Squires}]{Yan:2004p2979}
Yan, L., {et~al.} 2004, ApJS, 154, 60

\bibitem[{Yang {et~al.}(2007)Yang, Greve, Dowell, \& Borys}]{Yang:2007p3462}
Yang, M., Greve, T.~R., Dowell, C.~D., \& Borys, C. 2007, ApJ, 660, 1198

\bibitem[{Younger {et~al.}(2009)Younger, Omont, Fiolet, Huang, Fazio, Lai,
  Polletta, Rigopoulou, \& Zylka}]{Younger:2009p2901}
Younger, J.~D., {et~al.} 2009, MNRAS, 394, 1685

\bibitem[{Yun {et~al.}(2001)Yun, Reddy, \& Condon}]{Yun:2001p2885}
Yun, M.~S., Reddy, N.~A., \& Condon, J.~J. 2001, ApJ, 554, 803

\bibitem[{Zamojski {et~al.}(2007)Zamojski, Schiminovich, Rich, Mobasher,
  Koekemoer, Capak, Taniguchi, Sasaki, McCracken, Mellier, Bertin, Aussel,
  Sanders, F{\`e}vre, Ilbert, Salvato, Thompson, Kartaltepe, Scoville, Barlow,
  Forster, Friedman, Martin, Morrissey, Neff, Seibert, Small, Wyder, Bianchi,
  Donas, Heckman, Lee, Madore, Milliard, Szalay, Welsh, \&
  Yi}]{Zamojski:2007p28}
Zamojski, M.~A., {et~al.} 2007, ApJS, 172, 468

\end{thebibliography}
\end{document}